%% file: main.tex
\documentclass[aps,prxquantum,twocolumn,superscriptaddress,floatfix,nofootinbib]{revtex4-2}
\usepackage[T1]{fontenc}
\usepackage{amsmath,amssymb,amsthm,graphicx,booktabs,bm}
\usepackage{xcolor}\usepackage{placeins}
\usepackage[colorlinks=true,linkcolor=blue!50!black,citecolor=blue!50!black,urlcolor=blue!50!black]{hyperref}
\newtheorem{theorem}{Theorem}\newtheorem{proposition}{Proposition}\newtheorem{lemma}{Lemma}\newtheorem{corollary}{Corollary}
\theoremstyle{definition}\newtheorem{definition}{Definition}\theoremstyle{remark}\newtheorem{remark}{Remark}
\newcommand{\Fid}{F}\renewcommand{\Pr}{\mathbb{P}}\newcommand{\E}{\mathbb{E}}
\newcommand{\Umax}{U_{\max}}\newcommand{\Lmax}{L_{\max}}

\begin{document}
\title{What can a quantum network promise about the entangled pairs it delivers?}

\author{Tanish Singh Rajpal}
\email{trajpal@cmu.edu}
\affiliation{Carnegie Mellon University, Pittsburgh, Pennsylvania 15213, USA}
\author{Michael Mattarock}
\email{mattarock@cmu.edu}
\affiliation{Carnegie Mellon University, Pittsburgh, Pennsylvania 15213, USA}

\begin{abstract}
Quantum networks will give users entangled pairs of qubits, the resource behind teleportation and quantum-secure keys. But a user cannot check a pair without destroying it, so networks promise the quality of each pair, its fidelity, from a physical model rather than from a measurement. If the model is too optimistic, the delivery rule alone may miss the error. We ask: what can a network actually guarantee? Our main result has two sides. By testing some pairs at random, in secret, users can put a lower bound on the average fidelity of the pairs they keep, without any model and at every moment. But when they test only a small fraction and keep most pairs, the tests cannot reliably show that every kept pair meets a fidelity threshold: a rare bad pair can still slip through. A promise about every single pair therefore rests on a model, and we give a rule that bounds each user's risk of ever receiving a bad pair under any data-driven routing, as long as the model is not optimistic or a large enough margin is declared in advance. Real data show calibration noise beyond chance, and in simulated networks a model error of the size seen in one experiment breaks the per-pair promise unless the margin was declared. On real data from trapped-ion and diamond links, the number of pairs decided how strong the certificate was: with about 3{,}000 pairs of a kilohertz link and $20\%$ tested, the median certificate was $0.818$, against the $0.817$ we predicted in advance. We then ran the protocol ourselves on a 36-qubit trapped-ion processor in 50 pre-registered jobs. The certificate came within $0.003$ of the measured fidelity, dropped by the predicted amount when we secretly damaged a third of the pairs, and, without changes, certified three- and four-qubit entangled states. The same tests also catch an optimistic model, but slowly. We pre-registered every confirmatory analysis and report the failures too.
\end{abstract}
\maketitle

\section{Introduction}\label{sec:intro}
Quantum networks are being built to connect quantum devices across a laboratory, a city and, one day, a continent~\cite{Wehner2018,Craddock2024}. What they deliver is an \emph{entangled pair}: two quantum bits (qubits), one at each end of a connection, whose joint state is correlated in a way that no two ordinary bits can be. A pair can teleport an unknown quantum state from one place to another, seed a secret key whose security comes from physics, or join two parts of a distributed computation. Every one of these uses consumes the pair. And here is the catch: you cannot check how good a pair is without disturbing it, because a single copy of an unknown state cannot be characterised and kept intact.

This creates a problem of trust (Box~1 explains our answer without mathematics). The quality of a pair is its \emph{fidelity}, a number between 0 and 1 that says how close the pair is to the ideal entangled state (Table~\ref{tab:terms} defines this and the other terms we use). A user who needs a fidelity of at least $0.9$ cannot check the pair in hand, because checking would disturb it. Network protocols for routing and purification answer with a promise: every delivered pair clears the threshold~\cite{EFiRAP2022,QPATH2022,QGUARD2026}. But that promise is computed, not observed. A physical model turns estimated properties of each link, and the time each qubit waited in memory, into a predicted fidelity, and the network delivers the pairs whose prediction clears the threshold. In other words, the protocol grades its own paper with its own answer key. If the model or the estimates are wrong, the promise is wrong, and nothing in the protocol notices. We came to this question from routing: while building routers for simulated networks, we saw that every promise our routers made was checked only by the model that made it. A promise about each pair matters whenever a pair is used for something that cannot be repeated, such as teleporting a state that cannot be prepared again, or feeding an error-correction or distillation protocol (which turns several weaker pairs into fewer, better ones) that assumes a minimum input quality. An average over many pairs limits the share of bad pairs only weakly (Remark~\ref{rem:markov}) and does not certify the pair at hand.

\begin{figure}[t]
\fbox{\parbox{\dimexpr\columnwidth-2\fboxsep-2\fboxrule\relax}{\small\leftskip=0pt\rightskip=0pt\parfillskip=0pt plus 1fil\parindent=0pt
\textbf{Box 1 \textbar\ How it works, in plain words.}\par\smallskip
\textbf{The problem.} A network hands users entangled pairs. Measuring a pair destroys it, so nobody can check the pair they are about to use. Instead, networks predict each pair's quality from a model of their own hardware, which is a bit like grading your own paper with your own answer key.\par\smallskip
\textbf{Layer 1: a guard.} The network delivers a pair only if the model, fed with cautious estimates of each link, says it is good enough. If the model is right, every user is protected: with high probability they never get a bad pair. If the model is too optimistic, this protection breaks, unless a safety margin was set in advance.\par\smallskip
\textbf{Layer 2: hidden tests.} After each pair arrives, the two users secretly flip a biased coin. On heads (say one time in ten) they measure the pair along a random direction and note whether it fails; on tails they keep it. The network cannot tell which pairs will be tested, so it cannot cheat. From the failures alone, a simple formula gives a number that the average quality of the kept pairs stays above, at every moment, without trusting any model.\par\smallskip
\textbf{What tests cannot do.} No test like this can promise that \emph{every} kept pair is good: a rare bad pair can hide among the kept ones. Tests can only vouch for the average.\par\smallskip
\textbf{Closing the loop.} The same failures also show when the model is too optimistic, and then the guard raises its margin. This works, but slowly.\par\smallskip
\textbf{The evidence.} We checked each claim on data from real experiments, in simulated networks and on two quantum processors. We fixed the analyses in advance and report the failures too.}}
\end{figure}

There is a second, opposite way of thinking, which uses no model at all. In quantum certification, one measures a random sample of copies and infers the quality of the rest~\cite{Kliesch2021,Pallister2018,Takeuchi2018,ZhuHayashi2019,Gocanin2022,BoumanFehr2010}; the same idea is behind the security proofs of quantum key distribution~\cite{BoumanFehr2010,Tomamichel2012}. Modern \emph{anytime-valid} statistics make such inferences hold at every moment, so a user may look at the data as often as they like and stop whenever they want~\cite{ZhangGlancyKnill2011,Knill2020,Ramdas2023,Cumitini2026}. These tools certify \emph{averages} over the copies that were not measured.

In this paper we ask what a network can really guarantee about the pairs it delivers, and what that costs, when both ways of thinking are available. We answer in four steps. First, we show that when most pairs are kept, testing a random fraction cannot reliably prove that every kept pair is good, whatever the test; it can prove the average. Second, we give a model-based rule that bounds each user's risk of ever receiving a bad pair, but only as far as its model and its declared safety margins are right. Third, we check where these conditions hold and where they fail, on data from eight experiments, in simulated networks and on two quantum processors. Fourth, we show that the tests users run anyway can also measure how optimistic the model is, and feed that measurement back into the model-based rule.

\begin{figure*}[t]
\includegraphics[width=\textwidth]{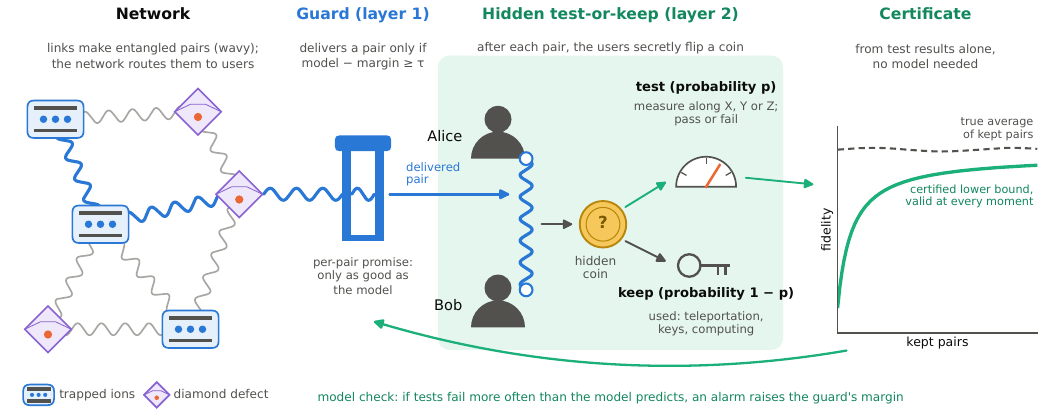}
\caption{How certified delivery works (a schematic; the inset curve shows the shape of the certificate, not data). The links of a quantum network, here drawn with trapped-ion and diamond-defect nodes, make entangled pairs (wavy lines) that the network routes to its users. The guard (layer 1) delivers a pair only if a physics model, fed with cautious, always-valid bounds on the quality of each link, predicts a fidelity of at least $\tau$ after a declared margin is subtracted; its per-pair promise holds only as far as the model is right. After each delivered pair the two users draw a hidden random bit (layer 2): with probability $p$ they test the pair along a random Pauli axis, $X$, $Y$ or $Z$, and otherwise they keep it for use. Because the network cannot see this choice, the test results certify, with trustworthy measurements and whatever the network does, the average quality of the kept pairs (the fraction that would pass a check against the ideal state $\Phi^+$, and their average fidelity), at every moment. The same tests run a model check on layer 1; an alarm raises the guard's margin (Sec.~\ref{sec:loop}).}
\label{fig:layers}
\end{figure*}

What we found is that \emph{certified delivery has two layers} (Fig.~\ref{fig:layers}). A model-based layer makes the per-pair promises, and it needs a large enough safety margin in force from the start. A model-free layer of hidden random tests certifies the average whatever the network does, and, given enough tests, can tell the first layer when its model is wrong. To be clear about what is new: the statistical tools we use are classical, and the certificate is of a known type (Sec.~\ref{sec:related}). What is new is the analysis of delivered pairs in an adaptive network, the measurement of where each guarantee holds or fails at the noise levels of real hardware, and one technical step that lets the tests bound the model's optimism (Theorem~\ref{thm:post}).

We pre-registered every confirmatory analysis: the hypotheses, pass marks and analysis code were fixed in a commit before we computed the reported statistic, with the exceptions listed in Table~\ref{tab:prereg}, and we report every outcome, including the failures. For every pre-registered study from S26 on, GitHub's server-side push records show that the pre-registration was pushed to the project's public GitHub repository before the first run or data read; for earlier analyses only the commit is recorded, and a few were specified after we had partly seen the data (Table~\ref{tab:prereg}). To keep the text readable, the running text does not name studies; figure captions, the summary table and the Methods do, and Table~\ref{tab:keys} (Appendix~\ref{app:prov}) lists every study key (for example S30), its pre-registration and its outcome. All network results are simulations; the real-data results re-analyse published or author-provided data.

\begin{table*}[!t]
\caption{\label{tab:terms}Key terms in plain language. Formal definitions are in Methods (Sec.~\ref{sec:methods}).}
\begin{ruledtabular}
\begin{tabular}{p{3.6cm}p{13.4cm}}
Entangled pair & Two qubits at different places, sharing one quantum state. The target here is the ideal state $\Phi^+$ (or another named Bell state).\\
Bell states & The four maximally entangled two-qubit states ($\Phi^\pm$, $\Psi^\pm$). Each is a perfect pair for some purpose, but a wrong Bell state counts as a bad pair.\\
Pauli test & Both users measure their qubit along the same axis, $X$, $Y$ or $Z$, and compare results. The ideal state always gives a known pattern.\\
Readout error; readout correction & A measurement that reports the wrong result; a correction that undoes such errors on average, using a separately measured error rate.\\
Fidelity $\Fid$ & How close a pair is to the target state, from 0 to 1. $\Fid=1$ is perfect; for two qubits, $\Fid>1/2$ proves entanglement.\\
Threshold $\tau$; bad pair & The minimum fidelity a user needs (mostly $0.9$ here). A pair below it is \emph{bad}.\\
Link; herald & A physical connection between neighbouring nodes. A herald is the signal that a link has just made a pair.\\
Calibration & Measurements a network makes on some of its link pairs to estimate the quality of each link.\\
Physics model & A formula that predicts a delivered pair's fidelity from the estimated link qualities and from how long each qubit was stored.\\
Model offset $\Delta$ & The model's predicted fidelity minus the true fidelity. A positive offset means the model is \emph{optimistic}.\\
Margin & An allowance subtracted from the prediction before it is compared with the threshold, to cover model error.\\
Guard & The network's rule for deciding whether a pair is good enough to deliver.\\
Per-user guarantee & With probability at least $1-\alpha$ (for example 95\%), a user \emph{never} receives a bad pair.\\
Test-or-keep & After a pair arrives, the two users draw a shared random bit the network cannot see. With probability $p$ they test the pair (measure it, using it up); otherwise they keep it.\\
Certificate & A number that, with stated confidence, lies below the average fidelity of the pairs kept so far (strictly, their average fidelity given the past, which equals the plain average when pairs are independent of each other and of the network).\\
CHSH test & A test of entanglement that scores how strongly two measurement results are correlated. Even a perfect pair fails it $14.6\%$ of the time.\\
Device-independent; self-testing & Conclusions that do not trust the measuring devices to work as designed; self-testing infers the state from test statistics alone.\\
Anytime-valid & Holding at every moment at once, so that watching the data continuously and stopping at any time does not spoil the statement.\\
Confidence sequence & A range of plausible values, updated as data arrive, that contains the true value at all times with stated confidence.\\
Model check; alarm & A running test of whether the test results are worse than the model allows. An alarm says the model looks too optimistic.\\
Throughput; oracle & Good pairs delivered, as a share of those delivered by an oracle that knows every pair's true fidelity.\\
One-sided upper bound & A value that the true quantity stays below with stated confidence (here 95\%).\\
Pre-registration & Fixing and time-stamping the hypothesis, the pass mark and the analysis code before analysing the data.\\
GHZ state & An entangled state of $n\ge3$ qubits, the many-qubit version of a Bell state: $(|0\cdots0\rangle+|1\cdots1\rangle)/\sqrt2$.\\
Genuinely multipartite entangled & Entangled across all qubits at once, not only within smaller groups. A GHZ fidelity above $1/2$ proves it.\\
Shot; job & One preparation and measurement of a processor's qubits; a batch of shots of one circuit sent to a processor at once.\\
Tomography & Measuring many copies of a state in all combinations of settings to reconstruct it in full.\\
$p$-value & The probability of a result at least as extreme as the one observed if only chance were at work.\\
Clopper--Pearson interval & An exact confidence interval for a success rate estimated from counts.
\end{tabular}
\end{ruledtabular}
\end{table*}

\section{Results}\label{sec:results}
\subsection{Tests on a few pairs cannot vouch for every pair}\label{sec:nogo-res}
Suppose the users of a network test a random fraction $p$ of their pairs, say one in ten, and keep the rest. Can the test results prove that no kept pair is bad? Not when most pairs are kept, whatever measurement is used as the test (Proposition~\ref{prop:N}). Compare two networks. One always delivers perfect pairs. The other, very rarely, swaps in a pair of the wrong kind, so rarely that only about one is expected over the whole run. The tests seldom land on the rare bad pairs, so the two networks produce almost the same test records, yet the second network usually hands a bad pair to the user. With one test in ten and the random tests described below, any rule that approves the good network 95\% of the time also approves the bad one at least 88\% of the time (at least 85\% for any test), even though the bad network hands over a bad pair with probability at least 59\%; such a rule therefore makes a false all-clear with probability at least 48\%. The argument covers test fractions up to about $0.46$ ($0.37$ for any test; Sec.~\ref{sec:nogo}); if almost every pair is tested, simple rules can succeed, but then almost nothing is left to use. Here ``bad'' refers to the pair's actual state, including the network's own record of what it sent. A user who sees only the test results would assign every pair of the second network a fidelity just below 1, and that is the point: averages can be certified, the individual pair cannot (Sec.~\ref{sec:setting}).

What tests can certify is an average, and an average does limit the share of bad pairs, but only weakly. If the average fidelity is $\bar\Fid$, at most a fraction $(1-\bar\Fid)/(1-\tau)$ of the pairs can lie below $\tau$, and this cannot be improved (Remark~\ref{rem:markov}). To be sure that at most 1\% of pairs fall below $0.9$, one would have to certify an average of $0.999$. In this regime, a promise about every pair, or that a user \emph{never} receives a bad pair, must therefore rest on explicit model assumptions. Our simulations show what this means in practice: a network whose true average fidelity was above the threshold while a quarter of its pairs were below it (Sec.~\ref{sec:cert-res}).

\subsection{A guard can protect every user, if its model is right}\label{sec:guard-res}
\begin{figure*}[t]
\includegraphics[width=\textwidth]{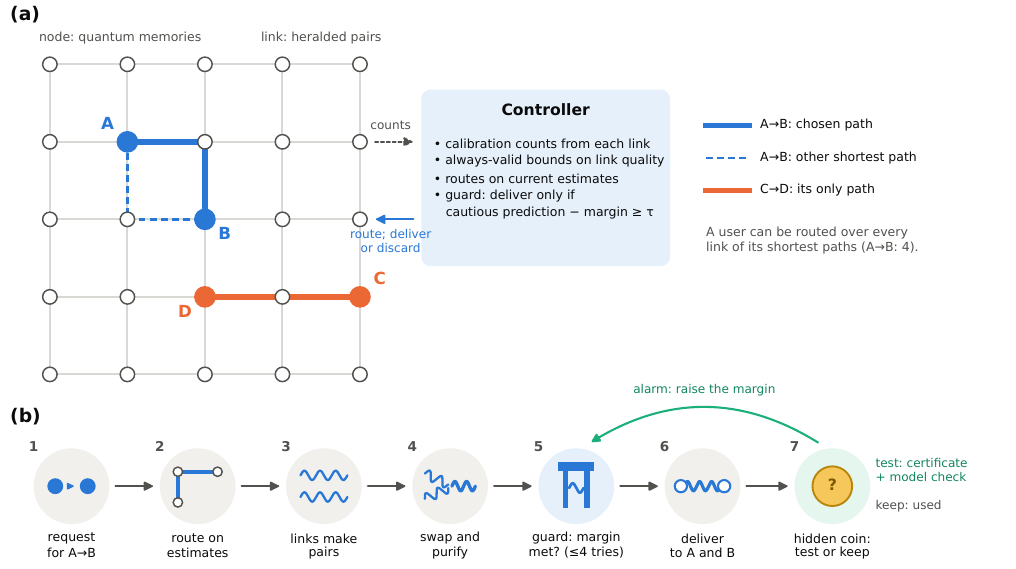}
\caption{How the simulated network serves its users. (a) The $5\times5$ grid. Each node holds quantum memories and each link makes heralded pairs. A user is an ordered pair of end nodes within $\Lmax$ hops: 204 users for $\Lmax=2$ and 340 for $\Lmax=3$ (1{,}004 and 1{,}860 on $10\times10$; 10{,}204 and 19{,}940 on $30\times30$). User A$\to$B has two shortest paths, so this user can be routed over four links ($|U_{A\to B}|=4$); user C$\to$D has one. The controller keeps always-valid bounds on each link's quality from calibration counts, routes each request and applies the guard. Requests form one stream for the whole network, each for a user drawn at random. In the network studies they are served one at a time, so two users never compete for a link at the same moment; only the finite-memory cells of S26 served disjoint paths in parallel. Each delivered pair is recorded for the user whose request it served, which in a real network is the job of the entanglement identifiers that a link-layer protocol attaches to every pair~\cite{Dahlberg2019,Pompili2022}. (b) The life of one pair in seven steps: a request arrives, is routed on the current estimates, the links make pairs for two chains, which are swapped and combined by one purification round, the guard (layer 1) approves or discards the result (a request that fails is retried, up to four attempts in total), the pair is delivered, and the users make the hidden test-or-keep choice (layer 2). Tested pairs feed the certificate and the model check, whose alarm raises the guard's margin (Sec.~\ref{sec:loop}).}
\label{fig:arch}
\end{figure*}
The model-based layer is a \emph{guard}: it delivers a pair only if the model, fed with cautious estimates of the link qualities, predicts a fidelity above the threshold (Fig.~\ref{fig:arch}). Each link keeps a confidence sequence for its quality, built from its calibration counts~\cite{Robbins1970,Howard2021}, and the guard uses the lower end of that range. Any other controller, for example a learned routing policy, can sit on top of the guard but may only remove deliveries.

The subtle point is routing. A network that chooses paths from data will tend to choose links whose estimates happen to look good, and those are the links whose estimates are most likely to be too high. A safety budget designed for one fixed path therefore no longer guarantees its level for a user served by adaptive routing: its bound grows with the number of links the user could be routed over. It is enough to give each link the error budget $\alpha$ divided by the largest number of links that any user could ever be routed over. Then, under any routing, the user receives a bad pair with probability at most $\alpha$ (Theorem~\ref{thm:guard}). On a $5\times5$ grid with paths of at most two links, a user can be routed over up to four links (seven for paths of at most three). Spending the budget over all links of the network protects all users at once (Corollary~\ref{cor:NW}).

The theorem has two quiet conditions: the model must not be optimistic, and the calibration counts must behave like independent coin flips. If either fails, the guard must declare a margin and a noise allowance in advance, estimated from calibration data (Sec.~\ref{sec:guard}). A common alternative does not give this kind of guarantee. Delivering whenever an estimate clears the threshold by some multiple of its error bar controls the error of each estimate, but not the share of delivered pairs that are bad, which depends on how many borderline links are on offer (Remark~\ref{rem:baserate})~\cite{Storey2003}.

\subsection{Real links strain the guard's assumptions}\label{sec:real}
We re-analysed data from eight experiments, each with a pre-registered analysis (Sec.~\ref{sec:methods-real}):
\begin{itemize}
\item ICFO: entanglement between two solid-state memories, heralded by telecom photons~\cite{LagoRivera2021};
\item USTC: remote entanglement between trapped ions~\cite{USTC2026};
\item Innsbruck: ion--photon entanglement from a 230\,m network~\cite{Krutyanskiy2023};
\item Delft: entanglement between two diamond-spin (NV) nodes in Delft and The Hague, 10\,km apart~\cite{Stolk2024};
\item Oxford: the shot-by-shot record of a run of a device-independent key-distribution experiment between two trapped-ion nodes about 2\,m apart~\cite{Nadlinger2022}, provided by its authors;
\item Harvard: electron- and nuclear-spin entanglement between two silicon-vacancy (SiV) nodes in a Boston-area telecom network~\cite{Knaut2024}, the per-shot records provided by its authors;
\item IonQ: kilohertz-rate entanglement between a trapped ion and an SiV node, from its public data and code~\cite{Hartung2026};
\item Oxford Sr--Sr: a 200{,}000-event characterisation run of the Oxford ion-trap network~\cite{Main2025b,Stephenson2020} in which each node draws its measurement setting at random for every event, provided by its authors.
\end{itemize}

\begin{figure*}[t]
\includegraphics[width=\textwidth]{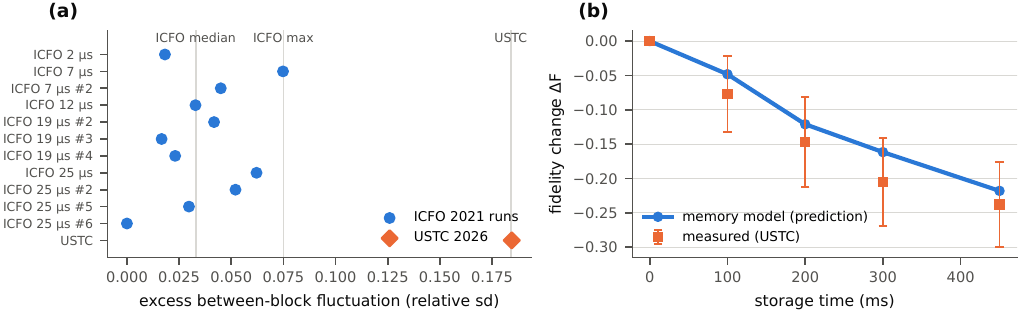}
\caption{Real links. (a) Fluctuation of calibration counts between time blocks, beyond what binomial (coin-flip) noise explains: the echo probability per herald at ICFO~\cite{LagoRivera2021} (every run) and the error rate at USTC~\cite{USTC2026}. Vertical lines mark the three levels used in the simulations (Sec.~\ref{sec:sim}). (b) USTC remote-entanglement fidelity change against storage time (squares; uncertainties propagated from the published values assuming independent errors), compared with a memory model built from single-ion coherence (line). The point estimates are optimistic at every storage time, but the offset is not significant (text).}
\label{fig:real}
\end{figure*}

\paragraph*{Calibration counts are noisier than chance.} In 7 of 11 ICFO runs (5 after correcting for the number of runs tested), the probability that a stored photon was re-emitted (the memory echo) fluctuated between time blocks more than coin-flip noise allows, by up to $7.5\%$ in relative terms (median $3.3\%$) [Fig.~\ref{fig:real}(a)]. Across 200 USTC data sets the error rate (mean $3.6\%$) fluctuated $18\%$ beyond chance; unequal data-set sizes, which the published data do not let us check, would inflate this figure. %
In the Harvard runs, which lasted hours, the herald rate varied between time blocks beyond chance (smallest p-value $6\times10^{-6}$ after correcting for 13 tests), while the error rate showed no excess where it could be tested (3 tests; 8 had too few errors). In the IonQ runs, each about 9\,s long, none of 28 tests found excess fluctuation, as predicted.

\paragraph*{A standard memory model is optimistic, but not significantly.} From USTC's single-ion coherence measurements alone, we built a standard model of how stored entanglement fades (pure dephasing, a slow loss of the phase relation between the two qubits). It predicts the remote-entanglement fidelity over 450\,ms of storage within the published error bars, but its point estimates are optimistic at all four storage times, by $0.020$--$0.044$ [Fig.~\ref{fig:real}(b)]. The four gaps share one reference point, and a joint fit gives $0.029\pm0.044$, which is not significant. A safe declaration must therefore use the one-sided 95\% upper bound (Table~\ref{tab:terms}), $0.10$. In simulation we use the largest single gap, $0.044$, as a stress level calibrated to these data, not as a measured offset.

\paragraph*{At Delft the authors' model agrees with the data.} Their fidelity model has no free parameters but uses calibrations interleaved with the runs~\cite{Stolk2024}. For the post-selected states of their Fig.~3 (selected after the fact, not delivered in real time) it matched the data (offset $-0.004\pm0.008$; upper bound $0.009$). For the delivered states of their Fig.~4 (every heralded pair) it showed no optimism ($-0.007\pm0.015$), but offsets up to $0.018$ are not excluded, and the test had a 63\% chance of detecting an offset of $0.03$. At Harvard the published memory model (pure dephasing with the measured nuclear-spin coherence, $T_2\approx2.1$\,s, anchored at 250\,ms of storage) was, if anything, pessimistic at 500--1000\,ms: pooled offset $-0.077$ (95\% interval $[-0.207,0.050]$). %
The experiments therefore give no common offset to rely on: a margin must be declared from data for each link.

\subsection{In simulated networks, undeclared noise and model errors break the promise}\label{sec:sim}
We simulated a $5\times5$ grid network with best-path routing, link qualities that vary from link to link, calibration and memory decay. We used the parameters of four kinds of hardware: two trapped-ion sets (ion2020, anchored to a demonstrated link, and ionproj, a projected one) and two kinds of diamond colour centre, atom-sized defects that act as qubits (NV and SiV; we call these low-rate hardware because their links make pairs slowly), as detailed in Sec.~\ref{sec:methods-sim}. Every ordered pair of nodes within the path-length limit is a user, which gives 204 users for paths of at most two links and 340 for three; requests arrive at a fixed total rate for the whole network, each for a user drawn at random, and are served one at a time (Fig.~\ref{fig:arch}). Each condition ran on 32 networks. A rule counted as \emph{safe} if, with 95\% confidence, at most 5\% of users ever received a bad pair and at most 1\% of delivered pairs were bad. Table~\ref{tab:results} collects the numbers.

\paragraph*{With an exact model, the guard is safe at a small cost.} With the per-user budget, 0 of 91{,}392 users received a bad pair, in $1.01$ million deliveries. Users, and runs of the same network under different rules, are not independent, so we also count networks: none of the 128 simulated networks delivered any bad pair in any of its 384 runs, which bounds the chance that a network does so at $2.3\%$ (conservative 95\% bound over the four hardware types, $8.9\%$ within each; Table~\ref{tab:cluster}). Compared with a budget designed for fixed paths, it cost at most $0.75$ percentage points of throughput on ions, and up to $2.4$ points on SiV, above the pre-registered 2-point bar. An earlier pre-registered study ran a guard of the same construction on a $30\times30$ grid with 900 nodes and 1{,}740 links, with a per-path rather than a per-user budget, an exact model and stable links, so it tests scale, not model error. It delivered no bad pair in about $2.11$ million deliveries, at a throughput cost of $6$--$9\%$ (G3). By contrast, our re-implementation of a plug-in rule that trusts point estimates, representing the class of fidelity-guaranteed routing protocols~\cite{EFiRAP2022,QPATH2022,QGUARD2026}, gave $17$--$19\%$ of ion users a bad pair. With a margin tuned on ground truth from separate networks it was safe on ions, but such a margin carries no guarantee beyond its tuning data.

\begin{figure*}[t]
\includegraphics[width=\textwidth]{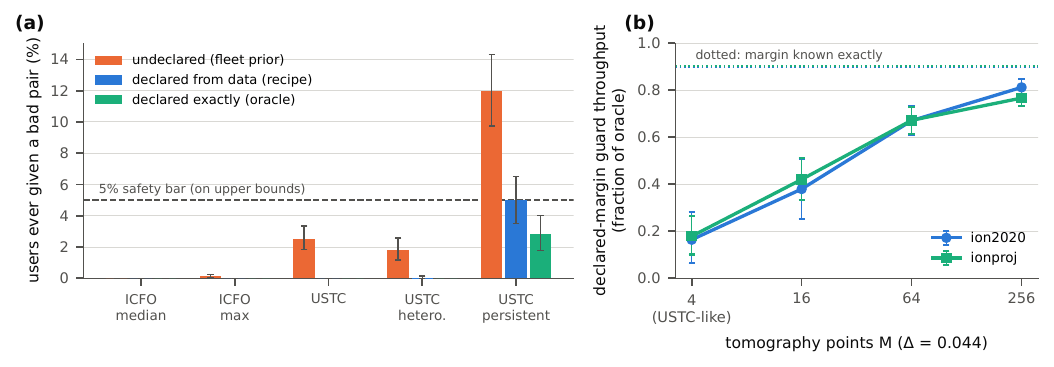}
\caption{The model-based layer at measured hardware noise levels (S26, 32 networks per bar or point). (a) ion2020: fraction of users that ever received a below-threshold pair (the safety bar applies to one-sided upper bounds, Table~\ref{tab:results}), for a guard that ignores excess noise, one that declares it from a 30-period calibration campaign, and one told the true value. Noise levels are those of Fig.~\ref{fig:real}(a); ``hetero.'' varies the noise between nodes; ``persistent'' has a period-to-period correlation of $0.9$. (b) Throughput of the declared-margin guard when a model offset $\Delta=0.044$ is declared from $M$ tomography points of USTC precision; dotted lines (overlapping): margin known exactly. Under persistent noise every rule failed the safety bar through delivered-pair risk (Table~\ref{tab:results}), which panel (a) does not show. Error bars: 95\% intervals from a bootstrap over simulated networks (4{,}000 resamples within hardware type); they have zero width when no network has an event (see Table~\ref{tab:cluster}).}
\label{fig:control}
\end{figure*}

\paragraph*{Measured noise can be declared; in our tests, persistent noise could not.} Noise at the ICFO level did no harm [Fig.~\ref{fig:control}(a)]. At the USTC level, guards that ignored it gave $1.1$--$5.5\%$ of ion2020 users a bad pair. Declaring a noise allowance from a calibration campaign restored safety (at most 67 bad pairs in $10^5$ deliveries), at a cost of up to $4.5$ points of throughput on SiV. When the same noise persisted from one calibration period to the next (correlation $0.9$), every model-based rule on ion2020 failed, even one told the true noise level: $1.4$--$7.3\%$ of delivered pairs were bad. In our tests, an allowance declared one period at a time did not cover noise whose effect builds up over time.

\paragraph*{An optimistic model voids the guarantee unless a margin is declared from data.} With an undeclared offset of $0.044$, $51\%$ of ion2020 users received a bad pair. Declaring the margin from $M$ independent fidelity measurements (tomography points), each as precise as one USTC data point, restored the per-user guarantee: at most $0.9\%$ of users received a bad pair (upper bounds at most $2.7\%$), against a guaranteed level of 10\%, which adds the 5\% chance that the declared margin itself is too small. With only four points, up to $5\%$ of delivered pairs were still bad, because some networks declared too small a margin (24 of 512 across all $M$). The price is set by calibration precision [Fig.~\ref{fig:control}(b)]: throughput was $16\%$ of that of an oracle that knows the truth with $M=4$ points, $67\%$ with 64 and $81\%$ with 256, against $90\%$ with the exact margin.

\paragraph*{Small memories do not break the guard.} With 2, 4 or unlimited qubits per node at four times the load, 0 of $4.3$ million deliveries were bad, although two qubits per node raised the end-to-end delay by 20--60\% on NV and SiV.

\subsection{Hidden random tests certify the average, whatever the network does}\label{sec:cert-res}
The model-free layer works as follows. After a pair arrives and its final correction has been applied, the two users draw a fresh random bit that the network cannot see. With probability $p$ they test the pair: each measures their qubit along the same randomly chosen axis ($X$, $Y$ or $Z$), and they record an error if the outcomes show a pattern the ideal state never produces. Otherwise they keep the pair. A standard identity links the three error rates to the fidelity, $1-\Fid=(e_X+e_Y+e_Z)/2$ (Lemma~\ref{lem:F}), so a random test fails with probability $\tfrac23(1-\Fid)$. A pair in a wrong Bell state fails with probability exactly $2/3$, whatever the network did to it. From the running counts of tests and errors one obtains a lower bound on the average fidelity of the kept pairs that holds at all times, against any network behaviour, adaptive or adversarial (Proposition~\ref{prop:C1}). This is a known type of result~\cite{Pallister2018,ZhuHayashi2019,BoumanFehr2010,ZhangGlancyKnill2011}, which we use as a component. It requires trustworthy measurements, and the test decision must stay hidden from the network until the network can no longer change the pair.

\paragraph*{Its price, and a floor.} Even with perfect pairs, certifying an average of $0.9$ takes $1{,}404$, $678$ or $315$ kept pairs when $5\%$, $10\%$ or $20\%$ of pairs are tested [Fig.~\ref{fig:loop}(d)]. Part of this price is unavoidable for this kind of test: any method that uses random Pauli tests and certifies $0.9$ on every record in which all tests pass needs at least $720$, $330$ or $140$ kept pairs (Proposition~\ref{prop:LB}), so ours costs $1.9$--$2.3$ times this floor. Most of the gap is the price of not knowing in advance how many pairs will be kept or how good they will be: a bet planned for the expected number and quality needs $870$, $420$ or $195$ (not pre-registered).

\paragraph*{On real data, the certificate needs trustworthy readout.} The Innsbruck ion--photon data contain the required measurements. From about $7{,}200$ heralded events, the three test settings give a fidelity estimate of $0.914$ (full tomography of all nine settings gives $0.913$) and a certified lower bound of $0.887$ at 95\% confidence (a fixed-sample bound, not an anytime one). These are not test-or-keep data, so they check the identity and the bounds, not the protocol. The Delft data record counts in the three bases without correction for readout errors. With the authors' readout correction, both the post-selected and the delivered states are certified entangled (lower bounds $0.558$ and $0.510$, which rest on that correction). From the uncorrected counts they are not ($0.464$ and $0.399$; both failures were predicted in the pre-registrations). On this link, readout errors cost about $0.07$ in fidelity, so in practice the certificate rests on a separately characterised readout.

\paragraph*{A certificate on randomised test-or-keep data.} The Oxford record lists all $1{,}920{,}000$ heralded pairs of an 8.5-hour run, in time order. After each herald, a quantum random number generator at one node decided whether the pair was a test round or a key round, with probability about one half. This is exactly the fresh choice, made after the pair exists, that our certificate relies on. Its tests are of a different kind (CHSH rounds, a test of entanglement that even a perfect pair fails $14.6\%$ of the time), so we used a bound suited to them (Sec.~\ref{sec:methods-real}). On the $944{,}576$ kept pairs, the certified average fidelity ended at $0.917$ [Fig.~\ref{fig:hwox}(c)] and stayed above $0.9$ at all times after $219{,}185$ kept pairs (point estimate $0.946$). This assumes that the measurement settings were as intended. Without that assumption, a device-independent self-testing bound on the fidelity extractable by local channels~\cite{Kaniewski2016} gives $0.837$. The test failure rate varied between the 96 data chunks of about five minutes each slightly more than chance allows (p-value $0.001$), but by only $1.6\%$ in relative terms, and we found no positive persistence from chunk to chunk (our prediction of persistence failed); other forms or timescales of drift are not excluded.

\begin{figure}[t]
\includegraphics[width=\columnwidth]{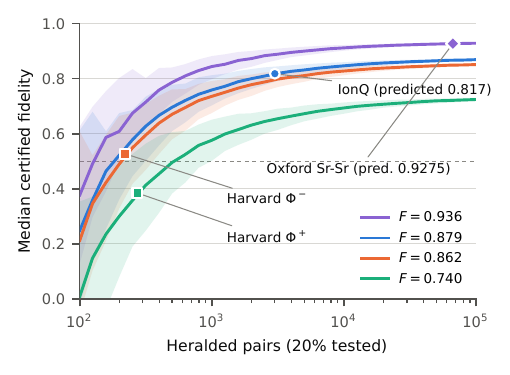}
\caption{How many pairs the certificate needs. Median certificate (Proposition~\ref{prop:C1}, $20\%$ of pairs tested, $\alpha=0.05$) against the number of heralded pairs, for four fidelities (lines: 1{,}000 synthetic draws per point; bands: 10--90\% of draws). Points: medians on real data, Harvard SiV--SiV (S37), IonQ ion--SiV (S38) and Oxford Sr--Sr (S39, fidelity $0.936$ to the fixed target, $66{,}864$ matched-axis pairs); the IonQ and Oxford values were predicted before the analysis was run.}
\label{fig:pairs}
\end{figure}

\paragraph*{On silicon-vacancy links, the number of pairs decides, predictably.} Neither SiV dataset randomises test-or-keep, so we drew the test choice at random after the fact, 1{,}000 times per setting, and compared the certificate with a conservative upper bound computed from the kept pairs' own outcomes. It exceeded that bound in none of $21{,}000$ draws on the IonQ data and in 1 of $8{,}000$ on the Harvard data. Its usefulness was set by the number of pairs [Fig.~\ref{fig:pairs}]. On Harvard's best electron--electron run (about 220 pairs per Bell state; measured fidelity $0.862$), the median certificate with $20\%$ of pairs tested was $0.53$. %
From the published fidelity and the number of pairs alone, we predicted before reading any IonQ data a median of $0.817$ (10--90\% range $0.788$--$0.842$) for the $3{,}000$ Bell-basis pairs of IonQ's main run; the result was $0.818$, above $0.8$ in $82\%$ of draws, from about 3\,s of operation at $1.03$\,kHz. Faster settings of the same link certified less ($0.65$ at $1.35$\,kHz, where the measured fidelity was $0.74$), so the best certified operating point is not the fastest. Both analyses trust the readout (raw thresholded outcomes, no readout correction) and draw the test choice after the fact; the IonQ bases were measured in blocks of 1{,}000, so the anytime property in time order was not exercised.

\paragraph*{With settings drawn per pair, the certificate is nearly tight and holds at all times.} %
In the Oxford Sr--Sr run each node chose $\pm X$, $\pm Y$ or $Z$ at random for every heralded pair, so the $66{,}864$ pairs measured on the same axis at both nodes form a random subset interleaved in time. The entanglement resource is best described by the fidelity to the closest maximally entangled state, $0.960$ in our linear estimate ($\approx0.96$ in the authors' analysis): the relative phase of the pair is stable over days, so in use it is measured once and compensated by one local rotation~\cite{Main2025b}. Our certificate was fixed in advance to $\Psi^+$ in the authors' frame, to which the fidelity is $0.936$ because of that phase offset ($|\phi|\approx0.32$\,rad). With $20\%$ of pairs tested after the fact, the median certificate was $0.927$, within $0.01$ of $0.936$; it exceeded $0.9$ in every draw and $0.8$ already at the first of 20 checkpoints (about $3{,}300$ pairs), and it matched the value predicted before the run ($0.9275$). Because the fidelity to the closest maximally entangled state is never smaller, $0.927$ also bounds it, conservatively. At the 20 checkpoints the certificate exceeded the running conservative reference in 1 of $3{,}000$ draws, the first check of its anytime validity on interleaved Pauli settings from real data. A protocol that applies the phase correction before the random tests would certify against the closest maximally entangled state directly.

\begin{figure*}[t]
\includegraphics[width=\textwidth]{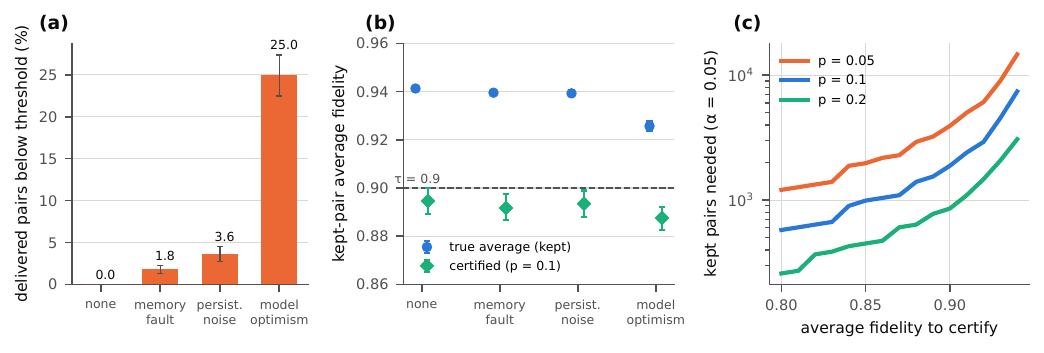}
\caption{Both layers in a network (S27, ion2020, guarded arm, 64 networks per condition: 32 with heavy users and 32 with uniform traffic). (a) Fraction of delivered pairs below threshold under each stressor: the model-based layer's per-pair claim fails. (b) True average fidelity of kept pairs and the model-free certificate when $10\%$ of pairs are tested (each network's pooled value, averaged over networks): the certificate stays valid but certifies only the average, and here stays below $\tau$ in every condition. (c) Kept pairs needed to certify a given average fidelity at 95\% confidence for pairs of fidelity $0.97$ (expected error counts), from the certificate of Proposition~\ref{prop:C1}. Error bars: 95\% intervals from a bootstrap over simulated networks (4{,}000 resamples within hardware type); they have zero width when no network has an event (see Table~\ref{tab:cluster}).}
\label{fig:cert}
\end{figure*}

\paragraph*{In networks, the certificate stays valid but helps few users.} We added the certificate to the simulated networks and compared it with the true average after every kept pair. These checks use ideal measurements and the model's own predictions of the test statistics, so they test the implementation rather than the physics. The certificate stayed valid in every condition, including the 80 cells where the model-based layer was unsafe (wrong for at most $0.34\%$ of eligible users and $1.2\%$ of networks). It also shows the limit of Proposition~\ref{prop:N} at scale [Fig.~\ref{fig:cert}(a,b)]. With an undeclared model offset, the guarded ion2020 networks delivered $126{,}446$ bad pairs out of $506{,}494$ ($25\%$), yet the true average fidelity of the kept pairs was $0.926$, above the threshold, and the certificate (mean $0.888$) stayed below the truth in all 64 networks. No average could have revealed this quarter of bad pairs. The certificate is also expensive in practice. Only $0.8$--$5.3\%$ of users, mostly those who received many pairs, ever reached a certificate above zero, and for users with at least 500 kept pairs it sat $0.07$--$0.16$ below the truth. The low-rate hardware types (NV, SiV) delivered too few pairs to certify a useful level in the simulated 45 minutes.

\subsection{The same tests can check the model, but detection is slow}\label{sec:loop}
The tests do more than certify. Their error rate also shows whether the guard's model is too optimistic. If the model were right, each test would fail with at most a known probability $q_i$, computed from the guard's own prediction plus an allowance for readout errors. A running test of this, the \emph{model check}, raises an alarm when the errors exceed what the model allows, with a false-alarm probability of at most $\beta$ whatever the network does (Propositions~\ref{prop:MC} and~\ref{prop:D}). The check works by betting against the model. A version with fixed bets keeps losing while the model is pessimistic, and this ``debt'' must be repaid before it can alarm, a problem that we met on hardware (Sec.~\ref{sec:hw}). An adaptive version bets only when recent tests look worse than the model, which limits the debt without removing it; our default averages the two (Sec.~\ref{sec:debt}).

Our one new technical step turns the test results into a bound on the model's optimism. For each tested pair we form a simple number $y_i$: the model's most optimistic prediction for that pair, minus one, plus $1.5$ times the test's error indicator (1 for an error, 0 otherwise). If readout errors are symmetric and the link bounds hold, $y_i$ is on average at least the true offset (Lemma~\ref{lem:onesided}, Definition~\ref{def:rule}), so standard betting confidence sequences applied to $y_i$ give a running upper bound on the offset that holds at all times. Used as the guard's margin after an alarm, this bound keeps the chance that a user receives a bad pair from then on below a set level (10\% in our simulations), provided the offset has a declared form (for example, constant, or proportional to storage time) and stays below a declared maximum (Theorem~\ref{thm:post}). Used from the start, it does the same for the whole run, if an offset of that form is present from the start (Corollary~\ref{cor:always}). Protection against arbitrary model drift is not established.

We tested this on the simulated ion networks in three pre-registered studies ($16{,}640$ runs). Unlike the earlier network studies, where the offset was added to the model's prediction, these studies lower the true fidelity. This changes which pairs are wrongly delivered, so their bad-pair rates are not comparable with those above (Sec.~\ref{sec:methods-sim}). In the main study, the margin applied from the start protected every user in every condition with a constant offset, or none, present from the start (0 of 208{,}896 users received a bad pair, in none of the 64 networks), at a cost in throughput: $0.76$ (ion2020) and $0.70$ (ionproj) of the oracle's with no offset, and $0.11$ and $0.61$ at an offset of $0.044$, where ion2020's margins are thin. When the offset began late in the run, breaking the always-on rule's assumption of an offset present from the start, it failed as predicted ($34\%$ of users received a bad pair). Triggered by an alarm, it stopped almost all further harm. In the pre-registered in-class cells, 230 of 89{,}585 users received a bad pair after the alarm, all in 2 of 2{,}048 runs, both with an offset that began late in the run. In one, the check raised a false alarm before the offset began; this case lies outside Theorem~\ref{thm:post}, and Corollary~\ref{cor:O}, which covers it, gives no useful bound for our pooled check. In the other, the offset bound missed the true offset at its first look and, as a running minimum, never recovered, a failure the theorem allows with probability at most $\alpha'$. Counting networks, the independent unit, 2 of 64 had a harmed user after an alarm (conservative 95\% bound $9.5\%$, within the guaranteed 10\% but not far inside it; for ionproj alone, 2 of 32, bound $18.4\%$, so the data cannot confirm the 10\% level for each hardware type); without the late-onset cell, none had (bound $4.6\%$; Table~\ref{tab:cluster}). When the offset broke the declared form, the guarantee failed, as predicted: after the alarm, $14\%$ of users received a bad pair when offsets differed between links, up to $92\%$ when the offset had the wrong shape, and $60\%$ after a false alarm before the offset began.

\begin{figure*}[tp]
\includegraphics[width=\textwidth]{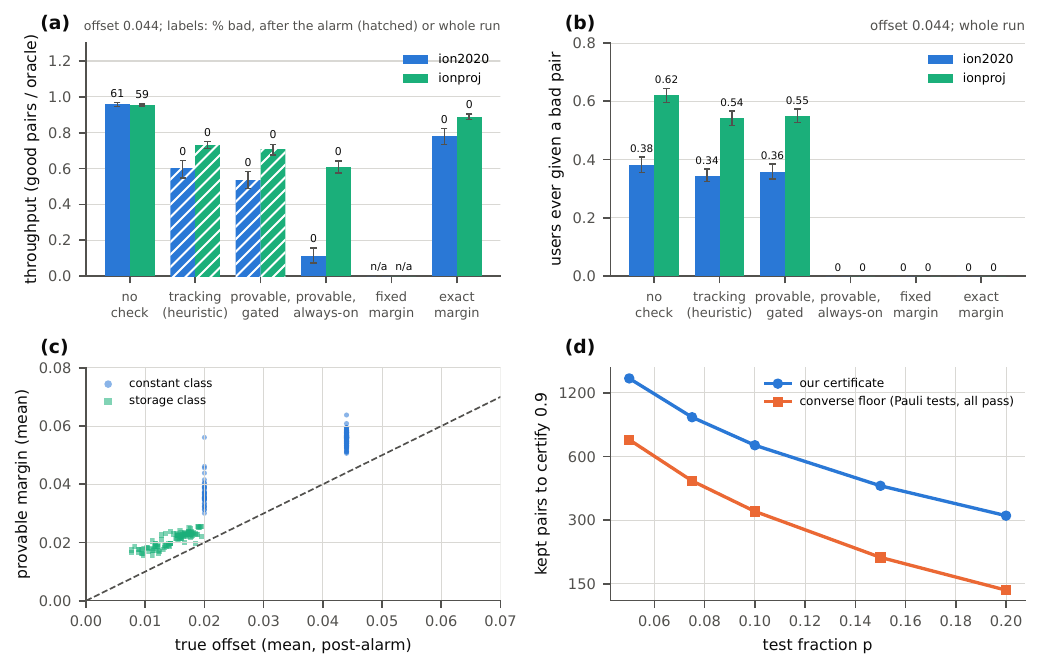}
\caption{Closing the loop with the offset in the true state (S30; both test fractions pooled). (a) Throughput under a constant offset of $0.044$, for ion2020 and ionproj side by side: good delivered pairs divided by the oracle's good pairs. Labels: percentage of delivered pairs that were bad, counted after the first alarm for the two alarm-gated rules (hatched bars) and over the whole run for the others (solid bars); the two windows are not comparable. Rules: no check, the heuristic tracking rule, the provable margin of Theorem~\ref{thm:post} behind the hybrid check (both gated), the always-on margin of Corollary~\ref{cor:always}, a fixed margin of $0.10$ (which delivers nothing) and the exact margin. Without a check $61\%$ (ion2020) and $59\%$ (ionproj) of deliveries were bad; the gated rules delivered none after the alarm, although over the whole run $52\%$ and $29\%$ (tracking) and $57\%$ and $32\%$ (provable, gated) of their deliveries were bad, almost all before the alarm. (b) Share of users who ever received a bad pair over the whole run, for the same rules and hardware types: most harm happens before the first alarm. (c) Mean post-alarm provable margin against the mean true offset of the pairs delivered after the alarm (for the storage class, the margin $\bar\theta\psi_i$ and offset $\Delta_i$ averaged over those pairs), one point per run, for the provable margin behind the hybrid check (constant shape; offsets $0.02$ and $0.044$) and with the storage shape (storage-proportional offset $0.044$); the line marks equality. (d) Kept pairs needed to certify $0.9$ from perfect pairs: the converse floor (Proposition~\ref{prop:LB}) and the anytime certificate (Proposition~\ref{prop:C1}); see Sec.~\ref{sec:cost}. Error bars: 95\% intervals from a bootstrap over simulated networks (4{,}000 resamples within hardware type); they have zero width when no network has an event (see Table~\ref{tab:cluster}).}
\label{fig:loop}
\end{figure*}

\begin{figure*}[t]
\includegraphics[width=\textwidth]{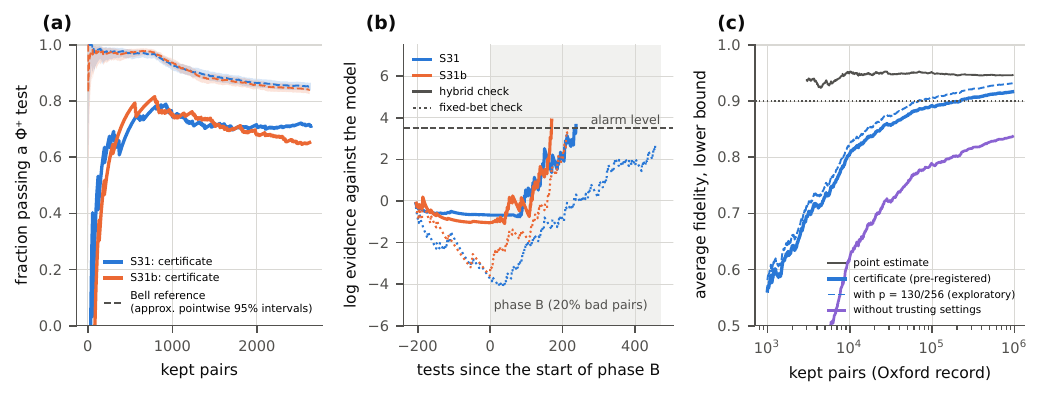}
\caption{Hardware and the Oxford record. (a) IBM runs S31 and S31b: the anytime certificate of Proposition~\ref{prop:C1} (solid) and the Bell-measured pass fraction of the kept pairs (dashed), a noisy reference rather than a ground truth, with pointwise 95\% Clopper--Pearson intervals (shaded; exact intervals for a success rate) that treat the kept pairs as exchangeable, that is, as if their order did not matter; they are approximate, because phases A and B prepare different mixtures, and they are not simultaneous over the run. (b) Log evidence of the model checks against the declared null, from the start of phase B (negative values: tests of phase A). Fixed-bet mixture (dotted) and hybrid check (solid); S31's hybrid curve is a post hoc replay of the recorded outcomes, S31b's was pre-registered. Both checks reset after an alarm; curves stop at the value that crossed the alarm level. (c) Oxford characterisation run (S36): trusted-settings certificate with the pre-registered test probability $1/2$ (solid) and with the confirmed $130/256$ (dashed, exploratory), the self-testing bound that does not trust the settings (purple) and the point estimate $\sqrt2(1-2\hat q)$ (grey); dotted line: $0.9$. Curves are shown from $1{,}000$ kept pairs (the point estimate from $3{,}000$).}
\label{fig:hwox}
\end{figure*}

The limit is detection. In the main study the check detected an offset of $0.044$ in $81$--$100\%$ of runs, but needed tens of the 90 calibration periods to do so (median first alarm in period 27--55 on ion2020 and 15--30 on ionproj). It detected an offset of $0.02$ in only $6$--$56\%$ of runs. Most harm came before the alarm: over the whole run, $47$--$49\%$ of users still received a bad pair at offset $0.044$, against $53\%$ without a check [Fig.~\ref{fig:loop}(b)]. On square grids, detection depended on how many tests the network produced, not on their size. On a $30\times30$ grid it was as fast as on a $5\times5$ grid, and three times the traffic made it $2.3$--$2.5$ times faster. On a cube of $5^3$ nodes it was slower for one hardware type, for reasons we have not established (Sec.~\ref{sec:methods-sim}). A variant that lets the bound rise again after an unlucky early stretch (Remark~\ref{rem:reopen}) recovered in all 128 runs in which we forced such a stretch (median 502 tests), against none for the original.

\subsection{Test-or-keep on quantum processors}\label{sec:hw}
Apart from the Oxford record, whose tests are CHSH rounds, no dataset randomises test-or-keep decisions with Pauli tests pair by pair. We therefore ran the protocol on an IBM superconducting processor (\texttt{ibm\_fez}; two jobs of 3{,}900 shots each on two qubits). Each shot was one ``delivered'' pair, prepared either as the ideal state (good) or as a wrong Bell state, $\Psi^-$ (bad). Because both qubits sit on one chip, every kept pair could afterwards be measured in the Bell basis. This gives a noisy reference estimate of the kept pairs' quality, not a ground truth: gate and readout errors in that measurement can turn good pairs into apparent failures and bad pairs into apparent passes. After 600 calibration pairs came two phases in which $20\%$ of the pairs were tested. In phase A all pairs were good. In phase B, one pair in five was secretly replaced by a bad one, an unannounced degradation that plays the role of model error.

The certificate never rose above this reference at any kept pair in either run and ended far below it [Fig.~\ref{fig:hwox}(a); final values $0.710$ against $0.852$ (pointwise 95\% interval $0.838$--$0.865$), and $0.653$ against $0.840$ ($0.826$--$0.854$)]. These intervals are fixed-sample Clopper--Pearson intervals that treat the kept pairs as exchangeable; the two phases prepare different mixtures, so they are approximate and are not a band valid over the whole run. These runs show that the protocol and its statistics behave as predicted on hardware; they do not validate the certificate's coverage, which would need many runs and a characterised measurement. The model check gave no false alarm in phase A. In the first run, the fixed-bet check missed the degradation: its evidence peaked at $2.71$, below the alarm level of $3.49$ [Fig.~\ref{fig:hwox}(b)]. During phase A the good pairs had erred less often ($5.8\%$) than during calibration ($7.2\%$), so the check built up a debt that the excess errors of phase B could not repay. This failure led us to the adaptive check. Its bet rule was written after we had seen the first run's error rates, so we pre-registered a rerun. In the rerun the adaptive (hybrid) check alarmed at the 170th of 472 phase-B tests. The fixed-bet check, helped by a higher phase-B error rate in this run ($19.5\%$ against $15.5\%$) and a lower null bound ($q=0.077$ against $0.091$), recovered from its debt and alarmed at the 214th; one run shows earlier detection, not a proven difference in power. After the alarm, the offset bound ended at $0.312$, above the offset estimated from the reference ($0.099$): consistent, but loose at this sample size. IBM readout errors are asymmetric, and the injected bad pairs violate the fidelity condition of Lemma~\ref{lem:onesided}, so this is a consistency check rather than a test of Theorem~\ref{thm:post}. The experiment tests the protocol and the statistics, not a network: the pairs were not distributed over any distance.

\paragraph*{On a 36-qubit trapped-ion processor, with every setting fixed in advance.} The IBM runs used two qubits and a reference without ground truth. We therefore ran 50 pre-registered jobs on IonQ Forte~1 (36 trapped-ion qubits, accessed through the Qollab platform). Each job is one circuit on all 36 qubits, measured 1{,}000 times: 18 Bell pairs, or 12 three-qubit or 9 four-qubit GHZ states. The qubit grouping and every state's test setting came from a seeded design table pushed to a public repository before the first job; the submitted circuit was recorded with each job and checked gate by gate; IonQ's error mitigation was off. Because the service runs one circuit per job, a state keeps its setting for all 1{,}000 shots, so settings are random per job and position, not per state: validity under this clustering is an empirical result, not a consequence of Proposition~\ref{prop:C1}. As for the ion--SiV link and the Oxford Sr--Sr run, every state was measured and the test-or-keep choice was drawn afterwards (1{,}000 draws per test fraction).

\begin{figure}[t]
\includegraphics[width=\columnwidth]{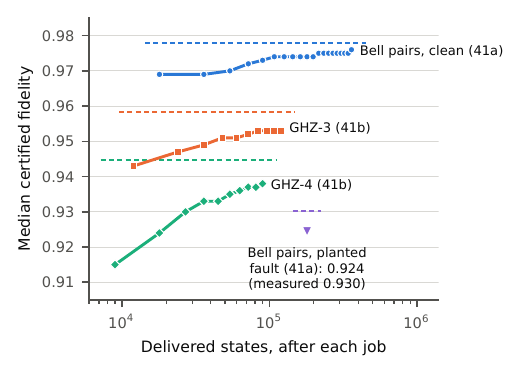}
\caption{The certificate on IonQ Forte~1 with settings fixed in advance (S41a, S41b). Median certificate ($20\%$ of states tested after the fact, $\alpha=0.05$, 1{,}000 draws) after each job, in submission order, against the number of delivered states; dashed lines: measured fidelity of each block. Triangle: the 10 jobs in which a third of the Bell pairs carried a hidden phase error (final value; dashed: that batch's measured fidelity). In $3{,}000$ draws per block the certificate never exceeded a conservative reference at any checkpoint.}
\label{fig:ionq}
\end{figure}

\paragraph*{The certificate is nearly tight and falls when the pairs degrade.} In 20 jobs of clean Bell pairs ($360{,}000$ pair outcomes; measured fidelity $0.978$), the median certificate with $20\%$ of pairs tested was $0.976$ [Fig.~\ref{fig:ionq}], and it never exceeded the conservative reference in $3{,}000$ draws, at the end or at any of the 20 job checkpoints in time order. In 10 further jobs, interleaved with these, 6 of the 18 pairs per job received a hidden phase error of $\pm\pi/4$, a known damage that raises the $X$ and $Y$ error rates by about 14 percentage points and leaves $Z$ unchanged. From the clean jobs we predicted a drop of the certificate by $0.047$; it dropped by $0.051$ (ratio $1.08$; pre-registered expectation $1.05$--$1.1$), and at $0.924$ it stayed below the measured fidelity of the damaged batch, $0.930$ (predicted $0.931$). All five pre-registered hypotheses held.

\paragraph*{The same test certifies GHZ states.} A Greenberger--Horne--Zeilinger (GHZ) state is the many-qubit version of an entangled pair: $n$ qubits whose $Z$ measurements always agree and whose entanglement is shared by all of them at once. For such states, a test that measures all qubits in $Z$ with probability $1/3$ and otherwise one of the $2^{n-1}$ products of $X$ and $Y$ measurements whose outcome the ideal state fixes (its stabilisers) fails with probability exactly $\tfrac23(1-\Fid)$ for every state (Lemma~\ref{lem:ghz}), so the certificate applies unchanged. On 10 jobs each of three- and four-qubit GHZ states ($120{,}000$ and $90{,}000$ states), the median certificates were $0.953$ and $0.938$ against measured fidelities of $0.958$ and $0.945$, never above the reference at any checkpoint; a single job (about $10^4$ states) sufficed to certify that the kept ensemble is genuinely multipartite entangled, that is, entangled across all its qubits rather than only within smaller groups. All four pre-registered hypotheses held. On this device the infidelity grew by about $0.02$ per entangling gate (the two-qubit operation that creates the entanglement) for Bell pairs and both GHZ sizes. Two departures from the idealised assumptions were measurable: errors of states measured in the same shot were mildly correlated (variance of the number of errors per shot $1.1$--$1.2$ times the independent value), and the variance of error rates between jobs and positions reached up to six times the binomial value. The certificate held under both.

\section{Discussion}\label{sec:disc}
\paragraph*{What a network can honestly promise.} Table~\ref{tab:summary} puts the guarantees and the evidence side by side. A promise about every pair needs a model, because tests on a small fraction of the pairs cannot see rare bad pairs (Proposition~\ref{prop:N}). That model has to carry safety margins declared from calibration data, and those margins get expensive quickly as calibration becomes less precise [Fig.~\ref{fig:control}(b)]. It also has to budget for noise beyond chance. Noise that persisted across calibration periods defeated every model-based rule we tried [Fig.~\ref{fig:control}(a)], and no data-driven rule can stop exposure to a sudden fault before data from after the fault arrive (Proposition~\ref{prop:fault}, Appendix~\ref{app:faults}). The model-free certificate of the average covers all of these failures. But it cannot see individual bad pairs, and it costs hundreds to thousands of kept pairs per user, about twice the minimum for random Pauli tests when no test fails. The same tests can also bound the model's optimism. Applied from the start, this protected every simulated user inside its declared class of offsets, at a cost in throughput. Triggered by an alarm, it limits the damage but cannot undo what happened before the alarm, and with realistic offsets the alarm comes late. The tests miss rare or shallow failures, small offsets within short runs and harm before the alarm, and they have no power on low-rate hardware. So a per-user guarantee needs a large enough margin in force from the start, whether it comes from calibration or from the tests. Our advice is simple: a network that promises per-pair fidelity should say which of these layers each promise rests on, and what calibration and test budget it costs.

\begin{table*}[t]
\caption{\label{tab:summary}What each layer guarantees, under which assumptions, and what the evidence showed. Bounds are one-sided 95\% bounds on the share of simulated networks in which any user received a bad pair, conservative when pooled over hardware types (Table~\ref{tab:cluster}).}
\begin{ruledtabular}
\begin{tabular}{p{2.5cm}p{4.9cm}p{5.3cm}p{3.4cm}}
Statement & Guarantee and its conditions & Evidence & Outcome\\
\hline
No per-pair certificate from tests (Prop.~\ref{prop:N}) & Any test, when most pairs are kept (shown for test fractions $p\leq0.37$, or $\leq0.46$ for Pauli tests, at $\alpha=0.05$); concerns realised states & Proof; in simulation, $25\%$ bad pairs behind a true average of $0.926$ (S27) & Illustrated at scale\\
Per-user guard (Thm.~\ref{thm:guard}) & Chance that a user ever gets a bad pair $\leq\alpha$, if the model is not optimistic (or a sufficient margin is declared) and calibration counts are binomial & No bad pair in 128 networks ($\leq2.3\%$); undeclared offset $0.044$: $51\%$ of users harmed; persistent noise: every model-based arm failed (S26) & Held where its assumptions held; failed as predicted otherwise\\
Average certificate (Prop.~\ref{prop:C1}) & Anytime lower bound on the kept pairs' average; hidden test choice, trusted and symmetric measurements & Valid in all simulated conditions (S27); below a noisy reference in two IBM runs (S31, S31b); Oxford $0.917$ with trusted settings (S36); Delft raw counts do not certify (S35, S35B); on two SiV links, 1 of $29{,}000$ post-hoc draws above a conservative bound, usefulness predicted in advance (S37, S38); Oxford Sr--Sr with per-pair random settings, $0.927$ against $0.936$, valid at 20 checkpoints (S39); IonQ Forte, settings fixed in advance: $0.976$ against $0.978$ at every checkpoint, drop under a planted fault as predicted (S41a), GHZ-3 and GHZ-4 at $0.953$ and $0.938$ (S41b) & Held; costly; rests on characterised readout\\
Post-alarm margin (Thm.~\ref{thm:post}) & Chance of a bad pair after the alarm $\leq\alpha+\alpha_{\rm hi}+\alpha'$; declared offset class with $\theta\psi_i\leq\tau-\tfrac34$, symmetric readout, alarm after onset & Harm in 2 of 64 networks ($\leq9.5\%$), in late-onset runs, one of them outside the theorem; outside the class $14$--$92\%$ of users (S30) & Held in class; detection slow, harm before the alarm\\
Always-on margin (Cor.~\ref{cor:always}) & Same bound for the whole run; constant offset present from the start & No harmed user in any network ($\leq4.6\%$ per study); late onset: $34\%$ (S30, S32, S34) & Held in class; costs throughput\\
\end{tabular}
\end{ruledtabular}
\end{table*}

\paragraph*{Limitations.} All network results are simulations with simplified physics, mostly on square grids of side 5 to 37, with safety checks on cubes and on a grid with diagonal links. The real-data analyses re-use eight datasets; model offsets were measured on three of them, and excess calibration noise on three. On the SiV links and the Oxford Sr--Sr run the test choice was drawn after the fact; only the Oxford Sr--Sr run draws the Pauli settings per pair. Only the Oxford record randomises test-or-keep per pair between separate nodes; those nodes are about 2\,m apart, its tests are CHSH rounds, its certificate trusts the nominal measurement settings and has no ground truth for the kept pairs, and it is a single 8.5\,h run. Our IBM runs randomise Pauli tests, but on local pairs, in two jobs on one device, and are checked against a noisy reference rather than a ground truth. Our IonQ runs (50 jobs) also use local qubits of one device, fix each state's setting per job rather than per state, draw test-or-keep after the fact, and have a ground truth only for the planted fault. The certificate assumes trustworthy measurements, which was binding on the Delft data. The guarantee after an alarm needs an offset of a declared form and covers no pair delivered before the first alarm. The full list is in Sec.~\ref{sec:limits}.

\paragraph*{Outlook.} Where could this go next? The Oxford record shows that experiments already randomise test-or-keep per pair between separate nodes and record the decisions. If the same protocol used the Pauli tests of Proposition~\ref{prop:C1}, the certificate would become a working guarantee for delivered entanglement, with far fewer kept pairs, because a CHSH test fails even on perfect pairs. By our estimate (exploratory; expected error counts), Pauli tests at the Oxford test fraction would certify $0.9$ after about 770 kept pairs at the Oxford point estimate of $0.946$, against about $2.2\times10^5$ here. At kilohertz rates the same certificate reaches the SiV threshold of $0.8$ in seconds. Our analysis gives the pair budgets such a protocol would need. And the GHZ runs show that the same certificate, with the same constant, carries over to the multipartite states that network applications beyond pairs will need.

\section{Related work}\label{sec:related}
\paragraph*{Certification and verification.} Measuring a random subset of copies and inferring the quality of the rest without a model is the basis of quantum system certification and state verification~\cite{Kliesch2021,Eisert2020,Pallister2018,Takeuchi2018,ZhuHayashi2019,Gocanin2022,BoumanFehr2010,Bhavsar2025}; fidelity estimation from few Pauli measurements is standard~\cite{FlammiaLiu2011,daSilva2011}. The sample complexities of state verification~\cite{Pallister2018,ZhuHayashi2019,Takeuchi2018} are classical, including adaptive local strategies~\cite{WangHayashi2019,LiHanZhu2019,YuShangGuhne2019} and optimal Pauli strategies for GHZ states~\cite{LiHanZhu2020}; Proposition~\ref{prop:LB} is their kept-pair counterpart for a test-or-keep protocol.

\paragraph*{Anytime-valid statistics.} These methods keep a running score of the evidence against a hypothesis that may be checked at any time without inflating the error rate. Ville's inequality, mixture confidence sequences and random-sampling certification are classical~\cite{Ville1939,Robbins1970,Howard2021,Ramdas2023,BoumanFehr2010}, as are e-processes for composite nulls~\cite{Ramdas2023}. Test supermartingales have been used to analyse quantum experiments~\cite{ZhangGlancyKnill2011,Knill2020} and for anytime-valid tomography~\cite{Cumitini2026}. Sampling-without-replacement confidence sequences~\cite{WaudbySmith2020} are a natural alternative construction for the certificate, and betting-based bounds~\cite{WaudbySmith2024} may be tighter.

\paragraph*{QKD parameter estimation.} Test-or-keep sampling with a virtual Bell-basis reduction~\cite{LoChau1999} is also the core of finite-key parameter estimation in quantum key distribution~\cite{BoumanFehr2010,Tomamichel2012}. A device-independent key distribution experiment randomises test and key rounds per pair~\cite{Nadlinger2022}, so its record lets the certificate be run under its own assumptions. Our certificate is of this known type; what is specific here is the adaptive delivery of unmeasured copies, the per-user scope and the coupling to a model.

\paragraph*{Fidelity estimation and link selection.} Fixed-confidence and per-estimate fidelity methods~\cite{Seshadri2024,RuanDirkse2024,Oka2026,AditiBecker2026,LinkSelFiE2024,BeQuP2025}, for example conformal prediction intervals for fidelity whose coverage is averaged over a prior of states~\cite{AditiBecker2026}, answer a different question from the one a user of delivered pairs needs answered (Remark~\ref{rem:baserate}). Sequential tests that decide whether a link or path currently supports entanglement above a time-dependent fidelity threshold have also been proposed as a network diagnostic primitive~\cite{QPing2025}; they consume the tested pairs and make a decision about the link, not a statement about the pairs that are later delivered.

\paragraph*{Fidelity-guaranteed routing.} Routing and purification protocols promise a threshold for every delivered pair, computed from a model at estimated parameters~\cite{EFiRAP2022,QPATH2022,QGUARD2026}; memory cut-offs~\cite{Rozpedek2018} bound the storage time instead. Our baselines are class representatives of these designs, re-implemented in our model.

\paragraph*{Change detection.} The model check is a change-detection problem~\cite{Page1954,Lorden1971}. A one-shot e-process controls the probability of any false alarm under an adaptive, composite null without a change-point model, but it pays for slack before the change (Secs.~\ref{sec:debt} and~\ref{sec:hw}). E-detectors~\cite{ShinRamdasRinaldo2024} avoid this debt and control the average run length instead; we chose the probability of any false alarm because it composes with Corollary~\ref{cor:O}, and we have not compared the two empirically.

\paragraph*{What is new.} The statistical components are standard: union bounds (adding up the failure probabilities of many statements), mixture and betting confidence sequences, and test supermartingales (running scores of evidence of the kind just described). The new element is the modelling reduction of Theorem~\ref{thm:post}: the proxy $y_i=\Fid_{\rm M}(\bm\lambda^{\rm hi}_{i-1};\bm w_i)-1+1.5E_i$ has conditional mean at least the model offset, which turns the users' test outcomes into a time-uniform upper confidence sequence on the model's optimism; together with the map of which guarantees hold or fail at measured hardware levels.

\paragraph*{Complementary work.} Calibration scheduling~\cite{Kumar2026} and rate--fidelity control~\cite{RateFidelity2026} are complementary; combining such schemes with the guarantees and certificates here is open.

\section{Methods}\label{sec:methods}
This section states the setting and the results formally and gives the details of the analyses. Proofs are in Appendix~\ref{app:proofs}.

\subsection{Setting and notation}\label{sec:setting}
A network has a link set $\mathcal{E}$. Link $e$ produces heralded Werner-like pairs of level $\lambda_e\in[0,1]$ (fidelity $(3\lambda_e+1)/4$), possibly varying in time. A fraction of heralded link pairs is used for calibration: a parity check returns a bit with success probability $\mu_e=(1+\eta\lambda_e)/2$, where $\eta$ is a known readout factor. Pairs are stored in memories with finite coherence, swapped along a path, and delivered to a \emph{user}, an ordered pair of end nodes. A user's request can be served on any path in a candidate set $\mathcal{P}_u$, chosen by any data-dependent rule. The network's declared physics model gives the delivered fidelity $\Fid_P(\bm\lambda_P;\bm w)$ as a function of the link levels and the storage times $\bm w$. The true fidelity $\Fid_{\rm true}$ of a delivered pair may differ; we call $\Delta=\Fid_P-\Fid_{\rm true}$ the \emph{model offset} ($\Delta>0$: the model is optimistic). A pair is \emph{bad} if its fidelity to $|\Phi^+\rangle$ is below a threshold $\tau$.

\paragraph*{Which state a fidelity refers to.} A pair's \emph{realised} state is its state given every classical record of the experiment, including the network's internal ones (which state a source emitted, whether a fault occurred); per-pair promises, and the word ``bad'', refer to realised states. Averaging over records the users cannot see gives a coarser description: in the second network of Proposition~\ref{prop:N}, a user who sees only test outcomes assigns every pair fidelity $1-c/N$, above $\tau$ for large $N$, although its realised state is $\Psi^-$ with probability $c/N$. Proposition~\ref{prop:C1} is stated for each pair's conditional fidelity given everything before it. Because the test choice is fresh and hidden, its supermartingale argument also holds when ``everything before'' includes the network's records, and its count $B_t$ is a count of realised virtual labels. Averages over realised states can therefore be certified; Proposition~\ref{prop:N} shows that statements about each realised pair cannot.

We use three risk notions. The \emph{delivered-pair risk} $r=\Pr(\text{bad}\mid\text{delivered})$ is the fraction of delivered pairs that are bad. The \emph{per-user risk} is $\Pr(\text{user }u\text{ ever receives a bad pair})$. The \emph{certified average} is a lower bound, valid at all times, on the average conditional fidelity of the pairs a user has kept (made precise in Proposition~\ref{prop:C1}).

Users may test a delivered pair instead of keeping it; $p$ always denotes this test fraction. Table~\ref{tab:notation} lists the main symbols.

\begin{table}[!t]
\caption{\label{tab:notation}Main symbols.}
\begin{ruledtabular}
\begin{tabular}{p{2.1cm}p{6.0cm}}
$\tau$ & fidelity threshold; a pair is bad if $\Fid<\tau$\\
$\lambda_e$; $\lambda^{\rm lo}_e,\lambda^{\rm hi}_e$ & Werner level of link $e$; lower and upper ends of its confidence sequence\\
$\mu_e$, $\eta$ & calibration success probability $(1+\eta\lambda_e)/2$; known readout factor\\
$\Lambda_t(\mu)$ & mixture likelihood ratio of a link's confidence sequence\\
$\alpha$, $a_e$ & per-user error level of the guard; per-link level\\
$U_u$, $\Umax$, $\Lmax$ & links user $u$ could be routed over; $\max_u|U_u|$; maximum path length\\
$r$ & delivered-pair risk\\
$d$, $\phi$ & declared model margin; declared dispersion\\
$\Delta$ & model offset $\Fid_{\rm model}-\Fid_{\rm true}$\\
$p$ & test fraction\\
$E_i$ & error indicator of test $i$\\
$L_t$ & certified lower bound on the kept pairs' average\\
$\kappa$ & minimum failure probability of a test on states orthogonal to $\Phi^+$\\
$q_i$ & null bound on the error probability of test $i$\\
$s$, $s_{\max}$, $\varepsilon_{\max}$ & parity-flip probability, its bound, per-qubit readout-error bound\\
$\nu$ & bets of a test supermartingale\\
$M_t$, $\beta$ & model-check e-process; its false-alarm level\\
$\theta$, $\psi_i$, $\bar\theta_t$ & offset scale, declared offset shape, post-alarm upper confidence bound on $\theta$\\
\end{tabular}
\end{ruledtabular}
\end{table}

\subsection{What can be certified without a model}\label{sec:certifiable}
\subsubsection{No per-pair certificate}\label{sec:nogo}
\begin{proposition}[No model-free per-pair certificate]\label{prop:N}
Let each of $N$ delivered pairs be tested independently with probability $p<1$, by any measurement (possibly chosen adaptively from earlier outcomes), and kept unmeasured otherwise. Fix any rule that sees only the test outcomes and declares ``no kept pair has realised fidelity below $\tau$''. Compare a network that always delivers $|\Phi^+\rangle$ with one that independently replaces each pair by $|\Psi^-\rangle$ with probability $c/N$ and records the replacement only internally. Let $\kappa_T\leq1$ bound the total-variation distance between the outcome distributions on $\Phi^+$ and on $\Psi^-$ of every measurement that may be used as the test ($\kappa_T=\tfrac23$ when the test is the random Pauli test of Sec.~\ref{sec:cert}, with a uniformly random basis that the rule does not choose; $\kappa_T=1$ in general, for example for a fixed $X$ basis or a Bell measurement). The test-outcome distributions of the two networks differ in total variation by at most $\kappa_Tpc$, whereas the second network delivers at least one kept pair with realised state $\Psi^-$ with probability $1-(1-(1-p)c/N)^N\to1-e^{-(1-p)c}$.
\end{proposition}
For $p=0.1$ and $c=1$ with random Pauli tests: a rule that certifies the good network with probability $\geq1-\alpha$ also certifies the bad one with probability $\geq0.93-\alpha$ ($\geq0.90-\alpha$ for any test), although with probability $\geq1-e^{-(1-p)c}=0.59$ (for every $N$) a bad pair is kept, so the rule falsely certifies the bad network with probability at least $1-\alpha-\kappa_Tpc-e^{-(1-p)c}$, about $0.48$ for Pauli tests and $0.44$ for any test (proof in Appendix~\ref{app:cert}). The statement is about sampling: it concerns protocols that leave a fraction of the pairs unmeasured, and it uses nothing about the test except $\kappa_T\leq1$. It describes a regime, not every $p<1$. Maximising the bound over $c$, it rules out such certification at $\alpha=0.05$ for every $p\leq0.46$ with random Pauli tests and every $p\leq0.37$ for any test. For larger $p$ it is silent, and when the untested fraction $1-p$ is at most $\alpha$, the rule that certifies whenever no random Pauli test fails succeeds against these replacement networks (its false-certification probability is at most $1-p$, for any number of replaced pairs). The regime of interest, in which most pairs are kept ($p=0.05$--$0.2$ in our simulations), lies well inside the first range; the Oxford record, with a test fraction near $1/2$, is used only for an average certificate. In that regime, per-pair threshold statements about realised states cannot be certified in this way without a model; aggregate quantities (averages, counts of failed checks) can. Verification protocols for the adversarial scenario~\cite{ZhuHayashi2019,ZhuHayashi2019PRA} are consistent with this: they certify the fidelity of a randomly chosen untested copy, that is, an average over the untested copies, not a bound on each of them.

\begin{remark}[Averages bound the bad fraction, weakly]\label{rem:markov}
By Markov's inequality, at most $(1-\bar\Fid)/(1-\tau)$ of pairs with average fidelity $\bar\Fid$ lie below $\tau$, and the bound is tight. With $\tau=0.9$, certifying a bad-pair fraction of $1\%$ needs a certified average of $0.999$. So the delivered-pair risk $r$ can be bounded without a model, but only weakly; what needs a model is the per-user (``ever'') and zero-defect statement.
\end{remark}
In the limited-testing regime covered by Proposition~\ref{prop:N}, random testing alone cannot support a zero-defect guarantee; our framework obtains that guarantee from explicit model assumptions, and it is only as good as the model and its declared margins. Proposition~\ref{prop:N} is elementary; its value is the consequence, which Sec.~\ref{sec:sim} shows at scale: a true average above $\tau$ with a quarter of the pairs below it, which no average certificate could rule out.

\subsubsection{An anytime certificate on kept pairs}\label{sec:cert}
After a pair is delivered and its Pauli-frame correction has arrived, the two users draw fresh shared randomness hidden from the network. With probability $p$ they test the pair in a basis $B$ chosen uniformly from $\{X,Y,Z\}$, recording an error if the outcomes show a pattern $|\Phi^+\rangle$ never produces; otherwise they keep it.
\begin{lemma}[Pauli identity; standard~\cite{FlammiaLiu2011,daSilva2011,Pallister2018}]\label{lem:F}
For any two-qubit state, $1-\Fid(\rho,\Phi^+)=(e_X+e_Y+e_Z)/2$ with $e_X=(1-\langle XX\rangle)/2$, $e_Y=(1+\langle YY\rangle)/2$, $e_Z=(1-\langle ZZ\rangle)/2$. A random-basis test errs with probability $\tfrac23(1-\Fid)$.
\end{lemma}
Because $XX$, $YY$, $ZZ$ and $|\Phi^+\rangle\langle\Phi^+|$ commute, one may imagine each pair measured in the Bell basis first. A pair whose virtual label is not $\Phi^+$ errs on a random-basis test with probability exactly $2/3$, whatever the network did. This yields the following known-type certificate~\cite{Pallister2018,ZhuHayashi2019,BoumanFehr2010,ZhangGlancyKnill2011}.

\begin{proposition}[Anytime-valid kept-pair certificate]\label{prop:C1}
Assume the test choice is fresh, kept private from the network (for example drawn from pre-shared randomness or sent over an encrypted channel) and revealed only after the pair's frame correction; tests with no outcome are not discarded in an outcome-dependent way; the network has no access to a pair after the keep decision; and the measurements are trusted. For $\nu\in(0,\nu_{\max})$, $\nu_{\max}=\ln[(1-p/3)/(1-p)]$, let $k(\nu)$ solve $(1-p)e^{\nu}+p[1-\tfrac23(1-e^{-1.5\nu k})]=1$. Let $B_t$ be the number of kept pairs with a non-$\Phi^+$ virtual label and $\mathrm{Err}_t$ the number of test errors. Then, for any network strategy, quantum or classical, adaptive or adversarial, and any finite mixture over $\nu$ with weights $w_g$,
\begin{equation}
\Pr\Big(\exists t:\ \sum_g w_g\, e^{\nu_g[B_t-1.5k(\nu_g)\mathrm{Err}_t]}\ge\frac1\alpha\Big)\le\alpha .
\label{eq:C1}
\end{equation}
Inverting in $B_t$ gives an upper bound $B^{\rm ub}(\mathrm{Err}_t)$ with $\Pr(\exists t: B_t>B^{\rm ub}(\mathrm{Err}_t))\leq\alpha$. So $L_t=1-B^{\rm ub}/K_t$ lower-bounds, at all times simultaneously, the fraction of the $K_t$ kept pairs that would pass a $\Phi^+$ projective test. The same bound holds for the sum of the kept pairs' conditional infidelities $1-\Fid_i$ (each pair's fidelity given everything before it), so $L_t$ also lower-bounds their average conditional fidelity. When delivered pairs are not entangled with each other or with systems the network retains, this is the kept pairs' average fidelity.
\end{proposition}
Each factor has conditional mean at most $1$, so the product is a nonnegative supermartingale. For the fidelity version, the conditional mean of a factor is $g(x)=(1-p)e^{\nu x}+p[1-\tfrac23x(1-e^{-1.5\nu k})]$ with $x=1-\Fid_i$, which is convex with $g(0)=g(1)=1$, hence at most $1$ (Appendix~\ref{app:cert}). The measurements must be trusted. Readout errors are assumed symmetric (parity flips independent of the outcome); they can \emph{lower} the error rate of states with $\Fid<1/4$, and a known bound on such flips is handled by rescaling the $\tfrac23$. As a consistency check of the implementation, Monte Carlo runs with adaptive adversaries and non-Bell-diagonal states gave a worst violation rate of $4.6\%$ at $\alpha=5\%$ (95\% upper bound $5.3\%$). The certificate concerns pairs at delivery, not after further storage.

\paragraph*{GHZ states.} The same constant holds for a multipartite test.
\begin{lemma}[GHZ test; the Pauli strategy of Ref.~\cite{LiHanZhu2020}]\label{lem:ghz}
Let $|{\rm GHZ}_n\rangle=(|0\cdots0\rangle+|1\cdots1\rangle)/\sqrt2$. Test with probability $1/3$ by measuring every qubit in $Z$ (an error unless all outcomes agree), and otherwise by measuring a string $s\in\{X,Y\}^n$ with an even number of $Y$, chosen uniformly among the $2^{n-1}$ such strings (an error unless the product of the outcomes equals $\sigma_s=(-1)^{\#Y/2}$). For any $n$-qubit state the test errs with probability exactly $\tfrac23(1-\Fid(\rho,{\rm GHZ}_n))$, and $\Fid=1-e_Z/2-2^{1-n}\sum_s e_s$.
\end{lemma}
The averaged pass operator is $\Omega=\tfrac13P_Z+\tfrac23\,2^{1-n}\sum_s\tfrac12(I+\sigma_s s)$ with $P_Z=|0\cdots0\rangle\langle0\cdots0|+|1\cdots1\rangle\langle1\cdots1|$. Since $2^{1-n}\sum_s\sigma_s s=|0\cdots0\rangle\langle1\cdots1|+\mathrm{h.c.}$, $\Omega=\tfrac13 I+\tfrac23|{\rm GHZ}_n\rangle\langle{\rm GHZ}_n|$; for $n=2$ this is Lemma~\ref{lem:F}. A state orthogonal to ${\rm GHZ}_n$ therefore fails with probability exactly $2/3$, and Proposition~\ref{prop:C1} holds verbatim with $\Phi^+$ replaced by ${\rm GHZ}_n$. Because biseparable states have GHZ fidelity at most $1/2$~\cite{GuhneToth2009}, a certified average above $1/2$ shows that the kept ensemble is genuinely multipartite entangled (not all kept states are biseparable), assuming trusted measurements.

\subsubsection{Cost and a converse}\label{sec:cost}
The certificate is anytime-valid against arbitrary strategies, and this makes it conservative. Even with perfect pairs, certifying $0.9$ requires $1{,}404$, $678$ or $315$ kept pairs at $p=0.05$, $0.1$ or $0.2$ [Fig.~\ref{fig:loop}(d)]; for pairs of fidelity $0.97$ it takes about $850$--$3{,}900$ ($p=0.2$ to $0.05$) [Fig.~\ref{fig:cert}(c)]. Part of this cost is unavoidable:
\begin{proposition}[Converse]\label{prop:LB}
Let a rule map the observed test record to a lower bound on the fraction of kept pairs that would pass a $\Phi^+$ test, valid at level $\alpha$ for every source, and let it certify $\tau$ on every all-pass record with $K$ kept pairs. Then
\begin{equation}
\max_{f>1-\tau}\Big(\frac{1-p}{1-p+p\kappa f}\Big)^{K}\,\Pr\big(\mathrm{Bin}(K,f)>(1-\tau)K\big)\leq\alpha ,
\end{equation}
with $\kappa=2/3$. For small $p(1-\tau)$ and $\alpha$ this gives approximately $K\approx(1-p)\ln(1/2\alpha)/[p\kappa(1-\tau)]$, so the floor scales as $1/[p(1-\tau)]$.
\end{proposition}
The proof mixes $\Phi^+$ with states orthogonal to it; each such state fails a random-basis test with probability exactly $2/3$ by Lemma~\ref{lem:F} (Appendix~\ref{app:loop}). To certify $0.9$ the floor is $720$, $330$ and $140$ kept pairs at $p=0.05$, $0.1$, $0.2$ [Fig.~\ref{fig:loop}(d)], so at zero observed errors Proposition~\ref{prop:C1} needs $1.9$--$2.3$ times the converse floor. The floor is attained by a fixed-sample rule against i.i.d.\ sources. A computation separates the two parts (not pre-registered; Appendix~\ref{app:prov}, key S33). A single bet chosen in advance for the planned number of kept pairs and the expected fidelity is just as valid as the mixture; with perfect pairs it needs $870$, $420$ and $195$ kept pairs at $p=0.05$, $0.1$, $0.2$, that is $1.2$--$1.4$ times the floor, and Proposition~\ref{prop:C1} needs about $1.6$ times as many as this planned bet, also at fidelities $0.97$ and $0.95$ ($1.5$--$1.75$). Most of the gap to the floor is therefore the price of not planning $K$ or the fidelity in advance (the eight-bet mixture); how much of the rest is unavoidable against adaptive sources is open. Non-adaptive local projective test schemes have $\kappa\leq2/3$ for Bell states~\cite{Pallister2018}; adaptive local strategies~\cite{WangHayashi2019,LiHanZhu2019,YuShangGuhne2019} are not covered by this statement, and a Bell measurement ($\kappa=1$) would lower the floor by about a third.

\subsection{The model-based guard}\label{sec:guard}
Each link maintains a confidence sequence (CS) for $\mu_e$: the set of $\mu$ for which the mixture likelihood ratio
\begin{equation}
\Lambda_t(\mu)=\frac{\int u^{K_t}(1-u)^{N_t-K_t}\,\pi(du)}{\mu^{K_t}(1-\mu)^{N_t-K_t}}
\end{equation}
stays below $1/a_e$, where $K_t$ of the link's $N_t$ calibration bits succeeded and $\pi=\mathrm{Beta}(\tfrac12,\tfrac12)$~\cite{Robbins1970,Howard2021}. Under any predictable sampling, $\Lambda_t(\mu_e)$ is a test martingale, so by Ville's inequality~\cite{Ville1939} the CS covers $\mu_e$ at all times with probability $\geq1-a_e$. The guard maps the CS lower end to a lower level $\lambda^{\rm lo}_e(t)$ and delivers an attempt only if the model evaluated at these lower levels clears $\tau$. Any other controller (a risk-controlled threshold, a learned policy) may only remove deliveries.

\begin{theorem}[Per-user anytime guard]\label{thm:guard}
Assume (A0) the delivered fidelity is $\Fid_P(\bm\lambda_P;\bm w)$ for the declared physics model (no model error, or a declared margin that covers it); (A1) calibration bits on link $e$ are conditionally Bernoulli$(\mu_e)$ given the past, with arbitrary predictable sampling, and whether a pair is calibrated or delivered does not depend on its own outcome; (A2) $\Fid_P$ is nondecreasing in every link level; and (A3) levels are constant. Let $U_u=\bigcup_{P\in\mathcal P_u}P$ be the links user $u$ could ever be routed over (the candidate set $\mathcal P_u$ is fixed in advance; the choice within it may depend on data), and run every link at $a_e=\alpha/\Umax$ with $\Umax=\max_u|U_u|$. Then for every user, under any data-dependent routing,
\begin{equation}
\Pr(\text{user }u\text{ ever receives a pair with }\Fid<\tau)\leq\alpha .
\end{equation}
\end{theorem}
\noindent The proof is a union bound over $U_u$ (Appendix~\ref{app:guardproof}). Running the links at $\alpha/\Lmax$ instead, as is natural for a single fixed path, gives only $|U_u|\alpha/\Lmax$ per user: data-driven routing preferentially selects links whose estimates are optimistic. On a $5\times5$ grid with $\Lmax=2$ ($3$), $\Umax=4$ ($7$). The guarantee is per user; it is not simultaneous over all users of a network. For that, a larger union suffices:
\begin{corollary}[Network-wide guard]\label{cor:NW}
If every link's sequence runs at level $\alpha/|\mathcal E|$ instead of $\alpha/\Umax$, then under the assumptions of Theorem~\ref{thm:guard}, $\Pr(\text{any user ever receives a bad pair})\leq\alpha$.
\end{corollary}
\noindent The proof is the same union argument over all links (Appendix~\ref{app:loop}).

\paragraph*{What the guard needs.} The guarantee holds only if the physics model is not optimistic and the calibration bits are binomial. Real hardware can violate both (Sec.~\ref{sec:real}). Hence the guard must \emph{declare} a model margin $d$ (it requires $\Fid_{\rm model}-d\geq\tau$, so that (A0) holds when $\Delta\leq d$) and a dispersion $\phi$ (it deflates counts by $\phi$). We estimate both from calibration data with one-sided 95\% bounds (level $\alpha_d=5\%$): $d$ from $M$ tomography points (fidelity measurements, each of the precision of one USTC data point, against which the model is compared); $\phi$ from a 30-epoch campaign, either fleet-wide (a $t$ bound over independent epochs) or per link (a $\chi^2$ bound). With a margin declared from data at level $\alpha_d$, the per-user guarantee becomes $\alpha+\alpha_d$. Deflating counts by $\phi$ is a quasi-likelihood correction, not a theorem. The mixing prior $\pi$ may also be chosen from other data; this affects only efficiency when the prior is independent of the link's own stream (Appendix~\ref{app:priors}).

\paragraph*{Extensions} (Appendix~\ref{app:drift}). The guard extends to bounded drift between re-alignments (restarting the CS each cycle at level $6\alpha/(\pi^2c^2\Umax)$ per link), to drift known up to a set of rates, and to a declared bound on the change at each re-alignment. With a declared bound the guarantee becomes $\alpha$ plus the probability that the declared bound is violated.

\begin{remark}[Per-estimate intervals do not control delivered pairs]\label{rem:baserate}
Let a rule deliver a link's pairs when an estimate of the link's fidelity margin exceeds a fixed multiple of its standard error. If the rule delivers, with probability $\alpha_\xi>0$, from links whose true margin is $-\xi<0$, then for every $r^\ast<1$ there is a population of links (a law of true margins) on which the delivered-pair risk is at least $r^\ast$. The proof (Appendix~\ref{app:faults}) is the base-rate identity $r=\pi_0\alpha_\xi/(\pi_0\alpha_\xi+(1-\pi_0)\gamma)$, with $\pi_0$ the share of marginal links and $\gamma$ the delivery probability of good links~\cite{Storey2003}.
\end{remark}
A confidence statement about each estimate says nothing about the fraction of delivered pairs that are bad, because that fraction depends on how many marginal links are offered. The guard's statement is of a different kind: with probability at least $1-\alpha$, no bad pair reaches the user at all.

\subsection{The model check and the provable margin}\label{sec:loop-methods}
\subsubsection{The model check}\label{sec:mc}
The guard approves pair $i$ only if its conservative fidelity $\Fid_g(i)$ (lower confidence bounds on the links, the declared model and the declared margin $d$) is at least $\tau$. Its guarantee rests on one condition: $\Fid_{\rm true}(i)\geq\Fid_g(i)$ for every delivered pair. By Lemma~\ref{lem:F}, with readout acting in basis $b$ as a symmetric parity flip of probability $s_b\le s_{\max}$, a tested pair errs with probability $\tfrac13\sum_b[s_b+(1-2s_b)e_b]$, which is at most $s_{\max}+(1-2s_{\max})\tfrac23(1-\Fid_{\rm true})$ when every $e_b\leq\tfrac12$ (implied by $\Fid_{\rm true}\geq\tfrac34$). So the test outcomes the users already collect can check that condition.
\begin{proposition}[Model check]\label{prop:MC}
Let $\tau\geq\tfrac34$. For tested pairs let $q_i=s_{\max}+(1-2s_{\max})\tfrac23(1-\Fid_g(i))$, where $s_{\max}=2\varepsilon_{\max}(1-\varepsilon_{\max})$ bounds the parity-flip probability in every basis for a per-qubit readout error $\varepsilon_{\max}$ (readout acts as a symmetric, state-independent parity flip whose probability may depend on the basis), and let $E_i$ be the error indicator. For fixed bets $\nu_g\geq0$ with weights $w_g$, capped as $\nu_{g,i}=\min(\nu_g,0.9/q_i)$,
\begin{equation}
M_t=\sum_g w_g\prod_{i\leq t}\big[1+\nu_{g,i}(E_i-q_i)\big]
\end{equation}
is a nonnegative supermartingale until the first delivered pair with $\Fid_{\rm true}<\Fid_g$. Hence the probability of an alarm ($M_t\geq1/\beta_1$, $\beta_1=6\beta/\pi^2$) before any such pair is at most $\beta_1$, whatever the network does. Restarting at level $6\beta/(\pi^2(k+1)^2)$ after the $k$-th restart (an alarm, or a decrease of the margin) keeps the total false-alarm probability below $\beta$.
\end{proposition}
The proof (Appendix~\ref{app:loop}) uses mainly that $q_i$ is fixed before the test outcome, which holds because the test choice is fresh and hidden. Detection needs a mean shift: if every tested pair has error probability at least $q_i+\delta$, the expected number of tests to an alarm is at most about $\ln(1/(\beta_1 w))/h(\delta)$, with $h$ the growth rate of the best bet and $w$ its weight. Failures that are sparse (few bad pairs) or shallow (many pairs just below $\tau$) give almost no shift in the mean; Proposition~\ref{prop:N} says the sparse ones cannot be seen.

\subsubsection{A check with limited debt}\label{sec:debt}
A fixed-bet mixture keeps betting when the guard is conservative, i.e.\ when the true error rate $\bar e$ lies below $q$. It then loses log-wealth at a rate of about $\sum_g w_g\nu_g(q-\bar e)$ per test, and a later excess must first repay this debt. On hardware this debt prevented an alarm (Sec.~\ref{sec:hw}). Bets may instead be chosen from the data:
\begin{proposition}[Model check with predictable bets]\label{prop:D}
Let $H_0$: $\Pr(E_i=1\mid\text{past})\leq q_i$ for every tested pair, with $q_i$ as in Proposition~\ref{prop:MC}. For any bets $\nu_i\in[0,1/q_i)$ fixed before the outcome of test $i$, $M_t=\prod_{i\le t}[1+\nu_i(E_i-q_i)]$ is a nonnegative supermartingale under $H_0$, and so is any average of such processes. With the alarm and restart levels of Proposition~\ref{prop:MC} (segment $k=1,2,\dots$ at level $6\beta/(\pi^2k^2)$), the probability of any false alarm is at most $\beta$ under $H_0$. Under (A0), let $\mathcal E_c$ be the links whose pairs feed the check. The probability of an alarm before the first tested pair with $\Fid_{\rm true}<\Fid_g$ is at most $\beta$, and such a pair requires a lower-end failure on some link in $\mathcal E_c$; so the probability of a false alarm is at most $\beta+\sum_{e\in\mathcal E_c}a_e$. This is at most $\beta+\alpha$ for a check fed only by user $u$'s pairs, or with the links at $\alpha/|\mathcal E|$ (Corollary~\ref{cor:NW}).
\end{proposition}
In our simulations the check is pooled per network: it uses the tested pairs of all users, so $\mathcal E_c=\mathcal E$; with links at $\alpha/\Umax$ the $\alpha_c$ term of Corollary~\ref{cor:O} is then not small ($|\mathcal E|\alpha/\Umax$), and the simulations exercise only its $\beta$ term (the simulated model is exact before an onset). Our plug-in bet, of the aGRAPA type~\cite{WaudbySmith2024}, is $\nu_i=\mathrm{clip}(\hat\delta_i/(\hat v_i+\hat\delta_i^2),0,0.5/q_i)$, with $\hat\delta_i$ and $\hat v_i$ the mean and variance of $E-q$ over the last 100 tested pairs ($\nu_i=0$ until there are 10). It bets nothing while the tests look better than the null, so its debt is limited, though not zero: when the guard is nearly exact, noise in $\hat\delta_i$ triggers positive bets. Against a change at $t=0$ it is somewhat slower than the best fixed bet. The hybrid $M_t=(M^{\rm mix}_t+M^{\rm plug}_t)/2$ is an e-process for the same null and costs $\ln2$ nats against the better component; it is our default. Debt is the textbook weakness of a one-shot e-process used for change detection; CUSUM- and Shiryaev--Roberts-type e-detectors~\cite{ShinRamdasRinaldo2024} avoid it by restarting, and control the average run length to a false alarm rather than the probability of any false alarm. We control the latter because it composes with the post-alarm guarantee: Corollary~\ref{cor:O} pays exactly $\beta$ for a false alarm before an onset. An e-detector would plausibly detect faster after a conservative stretch, but would give no finite $\beta$ term there.

\subsubsection{Responding to an alarm}\label{sec:response}
\paragraph*{A heuristic predecessor.} Before Theorem~\ref{thm:post}, we used a \emph{tracking} rule, tested in simulation (Sec.~\ref{sec:methods-sim}): after an alarm, the margin was reset every 10 epochs to a Clopper--Pearson upper bound on the model offset, with margin-blocked pairs probe-tested (Appendix~\ref{app:loopsim}). It is a heuristic: Theorem~\ref{thm:guard} covers the guard only while the margin in force is at least the true offset, and no margin set from test data, rising or falling, is guaranteed to be. The rest of this section replaces the rule by one with a proof.

\paragraph*{A provable margin.} Pair $i$ has features $\bm w_i$ (path, link waits, storage time), known when it is made; $\Fid_{\rm M}(\bm\lambda;\bm w_i)$ is the declared model's fidelity for it (that is, $\Fid_P(\bm\lambda_P;\bm w)$ for its path and waits) and $\Fid_i=\Fid_{\rm M}(\bm\lambda;\bm w_i)-\Delta_i$ its true fidelity. With $e_{i,b}$ the error probability of pair $i$ in basis $b$ and $s_{i,b}\in[0,\tfrac12)$ an unknown parity-flip probability in that basis, a tested pair errs with probability $\Pr(E_i=1\mid\text{past})=\tfrac13\sum_b[s_{i,b}+(1-2s_{i,b})e_{i,b}]$.
\begin{lemma}[One-sided test of fidelity]\label{lem:onesided}
If every basis error probability satisfies $e_{i,b}\leq\tfrac12$, which is implied by $\Fid_i\geq\tfrac34$, and readout in each basis acts as a symmetric (state-independent) parity flip with probability $s_{i,b}\in[0,\tfrac12)$, possibly different for each basis, then $1-\Fid_i\leq1.5\,\Pr(E_i=1\mid\text{past})$.
\end{lemma}
\noindent\emph{Proof.} By Lemma~\ref{lem:F}, $\tfrac13\sum_be_{i,b}=\tfrac23(1-\Fid_i)$, so $\Pr(E_i=1\mid\text{past})-\tfrac23(1-\Fid_i)=\tfrac13\sum_bs_{i,b}(1-2e_{i,b})\geq0$; and $e_{i,b}\leq\sum_{b'}e_{i,b'}=2(1-\Fid_i)\leq\tfrac12$ when $\Fid_i\geq\tfrac34$. $\square$

\noindent We assume:
(B1) link levels are constant, and upper ends $\lambda^{\rm hi}_e$ from the same mixture CS at level $\alpha_{\rm hi}/|\mathcal E|$ hold for all links at all times with probability $\geq1-\alpha_{\rm hi}$;
(B2) for every pair generated, tested or delivered after the \emph{first} alarm, $\Delta_i=\theta\psi_i$, with one deterministic $\theta\in[0,\theta_{\max}]$ and a declared shape $\psi_i\in[0,\psi_{\max}]$ known when the pair is made (for example $\psi_i=1$, or $\psi_i$ proportional to storage time, capped); pairs before the first alarm are unrestricted, but the first alarm must come after any onset of the offset;
(B3) the test basis is uniform, independent of everything the network holds and revealed only after the frame correction, while test selection may follow any predictable rule (probe tests are selected by the guard); and the readout and fidelity conditions of Lemma~\ref{lem:onesided} hold for tested and probe-tested pairs. On the coverage events below, such a pair has $\Fid_i\geq\tau-\theta\psi_i$, so the fidelity condition is guaranteed if $\theta\psi_i\leq\tau-\tfrac34$ for every pair ($0.15$ at $\tau=0.9$; in our ion simulations $\theta\psi_i\leq0.08$ for the constant class and $\leq0.132$ for the storage class);
(B4) a declared maximum offset $\theta_{\max}$ per unit $\psi$, for example $0.10$, the one-sided bound on the measured offset (Sec.~\ref{sec:real}). Without a physical bound, a data-free choice is valid only in special cases. For the constant shape ($\psi_i=1$), $\theta_{\max}=1$ is valid, because an offset cannot exceed 1; for a general shape, $\theta\psi_i\leq1$ gives $\theta_{\max}=1/\psi_{\min}$ if every pair has $\psi_i\geq\psi_{\min}>0$, and no data-free bound otherwise. With such a fallback the first post-alarm pairs are all probe-tested until the bound falls. A large $\theta_{\max}$ does not by itself secure (B3) for these probe pairs: on the coverage events a probe-tested pair has $\Fid_i\geq\tau-\theta\psi_i$, so the fidelity condition of Lemma~\ref{lem:onesided} ($\Fid_i\geq\tfrac34$) is guaranteed only under the condition $\theta\psi_i\leq\tau-\tfrac34$ of (B3), an assumption on the true offset. Since that condition already bounds $\theta$, the data-free fallback matters only for how long the first post-alarm pairs are probe-tested.

\begin{definition}[Post-alarm margin rule]\label{def:rule}
Fix bets $\nu_1,\dots,\nu_J>0$ and a level $\alpha'$.
(i) Before the first alarm, run the guard of Theorem~\ref{thm:guard} with margin $0$ and a model check (Proposition~\ref{prop:MC} or~\ref{prop:D}).
(ii) At an alarm, open window $k$ (the $k$-th window) at level $\alpha'_k=6\alpha'/(\pi^2k^2)$ and set $\bar\theta=\theta_{\max}$.
(iii) Each tested or probe-tested pair $i$ in the window contributes $y_i=\Fid_{\rm M}(\bm\lambda^{\rm hi}_{i-1};\bm w_i)-1+1.5E_i$, which satisfies $y_i\leq y^{\max}_i:=\Fid_{\rm M}(\bm\lambda^{\rm hi}_{i-1};\bm w_i)+\tfrac12$; here $\bm\lambda^{\rm hi}_{i-1}$ are the upper ends before pair $i$.
(iv) With $\nu_{ij}=\min(\nu_j,0.9/y^{\max}_i)$ and the product over the window's pairs, let
\begin{equation}
W_t(\vartheta)=\frac1J\sum_{j=1}^J\prod_{i\leq t}\big[1+\nu_{ij}(\vartheta\psi_i-y_i)\big],
\end{equation}
and $\bar\theta_t=\min\big(\bar\theta_{t-1},\sup\{\vartheta\in[0,\theta_{\max}]:W_t(\vartheta)<1/\alpha'_k\}\big)$, with $\bar\theta_t=0$ if even $\vartheta=0$ is excluded.
(v) Deliver pair $j$ only if $\Fid_{\rm M}(\bm\lambda^{\rm lo};\bm w_j)-\bar\theta_t\psi_j\geq\tau$. A pair blocked only by the margin is probe-tested instead and joins the window. The model check keeps running against the new guard; a later alarm returns to (ii).
\end{definition}
\noindent $W_t(0)\geq1/\alpha'_k$ means that the tests look better than the model at its upper link bounds (a pessimistic model); we report it as a diagnostic.

\begin{theorem}[Post-alarm guarantee]\label{thm:post}
Under (A1)--(A3) for user $u$'s links and (B1)--(B4), the rule of Definition~\ref{def:rule} gives, for every user $u$,
\begin{equation}
\Pr(\text{bad pair to }u\text{ after the first alarm})\leq\alpha+\alpha_{\rm hi}+\alpha' ,
\end{equation}
where a bad pair has $\Fid<\tau$. With the links at $\alpha/|\mathcal E|$ (Corollary~\ref{cor:NW}), the same bound holds for ``any user''.
\end{theorem}
\noindent\emph{Proof sketch} (details in Appendix~\ref{app:post}). Let $G_1$ be the event that the lower ends hold for $u$'s links at all times ($\Pr(G_1^c)\leq\alpha$, Theorem~\ref{thm:guard}), $G_2$ the event that all upper ends hold ($\Pr(G_2^c)\leq\alpha_{\rm hi}$, B1) and $G_3$ the event that $\theta\leq\bar\theta_t$ at all times. While the upper ends hold, (A2) and Lemma~\ref{lem:onesided} give $\E[y_i\mid\text{past}]\geq\Fid_{\rm M}(\bm\lambda;\bm w_i)-\Fid_i=\Delta_i=\theta\psi_i$: each $y_i$ over-estimates the offset on average. Every factor of $W_t$ is at least $0.1$, hence positive, and at $\vartheta=\theta$ its conditional mean is at most $1$ because the bets are fixed in advance; so $W_t(\theta)$ is a nonnegative supermartingale in each window, and Ville's inequality with the spending $\alpha'_k$ gives $\Pr(G_2\cap G_3^c)\leq\alpha'$, because $W_t$ is nondecreasing in $\vartheta$. On $G_1\cap G_2\cap G_3$, every pair delivered after the alarm has $\Fid_j=\Fid_{\rm M}(\bm\lambda;\bm w_j)-\theta\psi_j\geq\Fid_{\rm M}(\bm\lambda^{\rm lo};\bm w_j)-\bar\theta_t\psi_j\geq\tau$. $\square$

\begin{remark}[A bound that can recover]\label{rem:reopen}
The running minimum in Definition~\ref{def:rule} is not needed. Because $W_t(\vartheta)$ is nondecreasing in $\vartheta$, the current-look bound $\bar\theta'_t=\sup\{\vartheta\in[0,\theta_{\max}]:W_t(\vartheta)<1/\alpha'_k\}$ falls below $\theta$ only on the event $\{\exists t: W_t(\theta)\geq1/\alpha'_k\}$ that Ville's inequality already bounds, so Theorem~\ref{thm:post} holds with $\bar\theta'_t$ in place of $\bar\theta_t$ (as is standard for confidence sequences~\cite{Howard2021}). Unlike the running minimum, $\bar\theta'_t$ rises again when later tests contradict an early unlucky stretch. Alternatively, a new window may be opened at any stopping time; it restarts at $\theta_{\max}$ and spends $6\alpha'/(\pi^2k^2)$. A pre-registered test of $\bar\theta'_t$ is in Sec.~\ref{sec:loop}.
\end{remark}

Theorem~\ref{thm:post} does not need (A0) after the alarm: (B2) and (B4) replace it. It does need the first alarm to come after the offset appears. A false alarm before an onset breaks (B2); the next statement bounds that case.
\begin{corollary}[Offset with an onset]\label{cor:O}
Suppose instead that $\Delta_i=\theta\psi_i\,\mathbf 1\{t_i\geq t_0\}$ with an unknown onset $t_0$ (including $t_0=0$ and $t_0=\infty$) that is deterministic or a stopping time. Let $\mathcal E_c$ be the links whose pairs feed the model check and $\alpha_c=\sum_{e\in U_u\cup\mathcal E_c}a_e$. Then $\Pr(u\text{ receives a bad pair after the first alarm})\leq\alpha_c+\alpha_{\rm hi}+\alpha'+\beta$; $\alpha_c\leq\alpha$ when the check is fed only by $u$'s pairs or the links run at $\alpha/|\mathcal E|$.
\end{corollary}
\noindent\emph{Proof sketch} (full proof in Appendix~\ref{app:post}). Before $t_0$ the model is exact, so on the event that the links in $U_u\cup\mathcal E_c$ are covered, no tested pair violates the check's null, and a first alarm before $t_0$ has probability at most $\beta$ (Propositions~\ref{prop:MC} and~\ref{prop:D}). If the first alarm comes at or after $t_0$, every pair in a window has $t_i\geq t_0$, and the proof of Theorem~\ref{thm:post} applies with the predictable indicator $\mathbf 1\{t_i\geq t_0\}$ in each factor. $\square$

The always-on variant opens the window at $t=0$ instead of at an alarm. It replaces (A0) by (B2) and (B4) for every pair:
\begin{corollary}[Always-on margin]\label{cor:always}
Under the other assumptions of Theorem~\ref{thm:post}, if (B2) holds from $t=0$ (with a fixed $\theta$ and no onset), running Definition~\ref{def:rule} with its first window opened at $t=0$ instead of at an alarm (a later alarm opens a further window as in step (v)) gives, for every user $u$, $\Pr(u\text{ ever receives a bad pair})\leq\alpha+\alpha_{\rm hi}+\alpha'$.
\end{corollary}
\noindent The proof is that of Theorem~\ref{thm:post} with the first window opened at the stopping time $t=0$ (Appendix~\ref{app:post}). The price is throughput: the margin starts at $\theta_{\max}$ and every pair is guarded by it until the tests bring it down.

\paragraph*{What the theorem does not cover.} Pairs delivered before the first alarm are not covered; by Proposition~\ref{prop:N} and the fault-exposure bound (Proposition~\ref{prop:fault} in Appendix~\ref{app:faults}), no data-driven rule can protect all of them. The always-on margin (Corollary~\ref{cor:always}) covers every pair only if (B2) holds from $t=0$ with a fixed $\theta$ (no onset); under a late onset its running minimum sits near the pre-onset slack until the check alarms and opens a new window. A fixed margin $\theta_{\max}$ from $t=0$ is equally provable for the constant class and needs no data. (B2) needs exact proportionality: with heterogeneous offsets the CS estimates a weighted average over the tested and probed pairs, and pairs with above-average offset are not protected. A change after an alarm that breaks (B2) is not covered either; the check is still watching, and protection resumes after the next alarm. Finally, the margin is not tight: even with many tests it is expected to exceed $\theta$ by the slack of the upper link bounds and a readout term (Appendix~\ref{app:post}), the price of a one-sided, model-free reading of the tests.

\subsection{Real-data analyses}\label{sec:methods-real}
All eight analyses were pre-registered, with the analysis scripts frozen before the reported statistic was computed. For the Innsbruck certificate the counts had been seen in an earlier stage, so only the method was fixed in advance. Each Delft dataset, the Oxford record and the Harvard, IonQ and Oxford Sr--Sr data were used only after their pre-registrations had been committed and pushed. Details, hashes and the order of events are in Appendix~\ref{app:real} and the ledger~\cite{Repo2026}.

\paragraph*{Excess noise.} The ICFO dispersion ratios (observed over binomial variance of the block-to-block echo probability) range from $0.73$ to $5.6$; the USTC ratio is $3.57$. The relative fluctuation beyond chance quoted in the text is the relative standard deviation beyond binomial noise.

\paragraph*{USTC offset.} The memory model uses only the single-ion coherence curves. Because the four fidelity gaps share the $t=0$ reference, we fitted a common offset by generalised least squares ($0.029\pm0.044$).

\paragraph*{Innsbruck certificate.} The data contain heralded outcomes in all nine Pauli settings, each measured twice with swapped detectors. We used only the three correlated settings $XX$, $YY$, $ZZ$ and Lemma~\ref{lem:F}, with $e_B$ the error rate in basis $B$. The bound $\Fid\geq0.887$ (95\%) holds simultaneously over all eight candidate target states and detector assignments; this union accounts for the target state and detector mapping having been chosen after an earlier analysis. It uses fixed-sample Chernoff--KL bounds that remain valid when error probabilities vary adaptively. The detector-swapped halves are consistent in every basis (p-value $\geq0.056$). The certificate is for the herald-conditioned state and assumes trusted measurements, fair sampling and stationarity across the six scans.

\paragraph*{Delft (S35, S35B).} The data contain outcome counts in the $XX$, $YY$ and $ZZ$ bases without readout correction (they differ from the authors' corrected probabilities by up to $0.05$), together with the authors' model predictions, summed per basis without time order, so only fixed-sample analyses were possible. Stage 1 used the post-selected entanglement of their Fig.~3 (10\,ns herald window, $1{,}415$--$1{,}535$ events per basis); stage 2 the fully heralded, delivered $\Psi^-$ states of their Fig.~4 (15\,ns window, about $450$ events per basis). The certificate uses Lemma~\ref{lem:F} with Clopper--Pearson bounds at level $\alpha/3$ per basis and no readout model; the readout-corrected bounds are normal approximations on the authors' corrected probabilities. The model-offset tests passed if the offset was within twice its uncertainty plus $0.02$ and its one-sided lower bound was at most zero. Descriptively, the offset of the post-selected states grows to $+0.016$ at a 20\,ns window.

\paragraph*{Oxford (S36).} The record contains $1{,}920{,}000$ heralded pairs; according to the authors, every herald produced a row. Proposition~\ref{prop:C1} relies on a fresh test choice made after the pair exists, which holds here; its other conditions hold by assumption (trusted measurements), by the authors' statement that every herald produced a row, and because each pair stays in the nodes' own ions. A CHSH test cannot use the zero-error property of Lemma~\ref{lem:F}. We therefore used the linear bound $\Fid\geq\sqrt2(1-2q)$, valid for any two-qubit state measured at the nominal settings, with $q$ the conditional probability of losing the CHSH game and the target the maximally entangled state of the measurement frame (for each herald pattern), together with a Hoeffding test supermartingale for the kept pairs' summed loss probability, over a grid of 20 bets, each at level $\alpha/20$ (Appendix~\ref{app:real}). The bound is exact only at the nominal settings; a setting error $\delta$ can make it exceed the true fidelity by up to $|\sin\delta|$ times the pair's weight outside the target's two-state subspace. As a sanity check, the certificate never exceeded a loose upper bound set by the kept pairs' own key-basis errors. The sign convention of the CHSH game was fixed by reproducing the published CHSH value; a union over all eight conventions would lower the final bound to $0.914$. The pre-registered test probability $1/2$ was slightly below the realised fraction $0.5080$ ($130/256$, drawn from 8 random bits, as the authors confirmed), which makes the bound conservative; with $p=130/256$ it reaches $0.932$ (exploratory, not pre-registered). The self-testing bound on the fidelity extractable by local channels~\cite{Kaniewski2016} is affine in the CHSH value and so applies to the average. For dispersion, the CHSH loss rate across the 96 acquisition chunks gave $p=0.001$ in a $\chi^2$ test, with a relative excess of $1.6\%$ and a lag-1 autocorrelation of $-0.17$. At this chunk scale (about 5\,min) we therefore found no period-to-period persistence of the kind that defeated every model-based arm on ion2020 in S26 (pre-run power $0.35$ at $1\%$ and $0.87$ at $2\%$ relative fluctuation), and the excess is far below the USTC level used there; this is one 8.5\,h run. The chunks are a software convenience, the authors expect slow drifts over the whole run, and calibrations ran about hourly (D.~P.~Nadlinger, private communication); we did not detect positive lag-1 persistence at this chunk scale, which does not exclude other forms or timescales of drift.

\paragraph*{Harvard (S37).} %
The authors' own analysis functions were executed from their notebooks, and gate G0 required our per-herald records to reproduce their Bell-count tables exactly. The test choice was drawn after the fact (probability $p$, 1{,}000 seeded draws); the certificate of Proposition~\ref{prop:C1} used the tested pairs, and a draw counted as a violation if it exceeded $1-\tfrac12\sum_B\mathrm{CP}^{\rm lo}(e_B)$ on the kept pairs, each Clopper--Pearson bound at level $0.025/3$. Details in Appendix~\ref{app:real}.

\paragraph*{Oxford Sr--Sr (S39).} %
Eigenvalues follow from bright/dark with the sign reversed for $-X$, $-Y$, under the single convention that reproduces the authors' stored fidelities (gate G2). The certificate of Proposition~\ref{prop:C1} uses the matched-axis pairs; the reference and the draws are those of S37, with the reference also evaluated at 20 checkpoints in time order (each per-basis bound at level $0.025/3/20$). An independent code review on synthetic data, made after the pre-registered run, corrected only the sign of a descriptive phase; all verdicts were identical (Appendix~\ref{app:real}).

\paragraph*{IonQ (S38).} The public HDF5 records (commit 8a08f69 of Ref.~\cite{Hartung2026}) give, for each heralded pair, the setting, both outcomes and timestamps; we used the $XX$, $YY$ and $ZZ$ settings (1{,}000 pairs each per run) with the same draws and reference as S37. The H5 prediction band was simulated from the published fidelity at the run's number of pairs before the data were read.

\subsection{Simulations}\label{sec:methods-sim}
\paragraph*{Setup.} A discrete-event simulator runs a $5\times5$ grid with greedy best-path routing, link levels drawn from a spatially correlated fleet law, calibration at 30\% of the heralded rate, and memory decoherence. It uses the hardware parameters of trapped ions (two parameter sets, ion2020 and ionproj), NV centres and SiV centres (Table~\ref{tab:hw} in Appendix~\ref{app:sim}). Delivered fidelities use a Pauli-transfer model of swapping, storage and one DEJMPS purification round~\cite{DEJMPS1996}, validated against an exact density-matrix model cross-checked with NetSquid~\cite{Coopmans2021}. Each cell has 32 networks, paired across arms. The safety bar is a one-sided 95\% cluster-bootstrap bound of $\leq5\%$ on per-user risk and $\leq1\%$ on delivered-pair risk. Test outcomes are drawn from the true correlators of each delivered pair. Throughput is the number of good delivered pairs divided by the oracle's good pairs (Appendix~\ref{app:sim}).

\paragraph*{Rates, units and bounds.} The simulations report four quantities that answer different questions: the share of all users ever given a bad pair (per-user risk, including users never served), the same over served users, the share of delivered pairs that are bad (delivered-pair risk) and, for the closed loop, the same quantities after the first alarm. Theorems~\ref{thm:guard} and~\ref{thm:post} bound yet another quantity: for each user, the probability of ever (or, after an alarm, of later) receiving a bad pair. Users and pairs within one simulated network are correlated, and so are the runs of one network under different rules and cells, which share its links and seeds; the independent replication unit is the network. The pre-registered bounds are one-sided 95\% bootstraps over networks. When no network has any bad pair, such a bootstrap returns 0, which carries no information. Table~\ref{tab:cluster} therefore re-reports the headline claims with Clopper--Pearson bounds on the share of distinct networks in which any user received a bad pair in any of the pooled runs, treating all runs of one network as one unit, as the pre-registered bootstrap does. The sampling model is as follows. Within each claim, every network was run under the same collection of arms and cells (except S34, whose $30\times30$ grid has more load variants), and the networks of a stratum (hardware type, and grid size in S34) are independent draws from the same design, with 32 per stratum. Network seeds are derived from the study, hardware type, grid and index, so networks of different strata are also independent. The target is the probability that a network drawn from the design has a harmed user in some pooled run. Within a stratum the networks are independent draws with a common probability, so the Clopper--Pearson bound is exact. The pooled bound treats the target as an equal-weight average over strata with possibly different probabilities; for that average it is conservative rather than exact, under the conditions stated in Appendix~\ref{app:pooled}~\cite{Hoeffding1956}. The last column gives the largest per-stratum bound, which holds whatever weight each stratum is given. In any run, the chance that a given user is harmed is at most the chance that anyone is, which is at most the network's risk, so the per-stratum bound also bounds per-user risk in that stratum's pooled runs. With 32 networks per stratum, a stratum with no harmed network can only be bounded at $8.9\%$, and the ionproj stratum of S30, with 2 harmed networks (one of them outside Theorem~\ref{thm:post}), at $18.4\%$: the simulations confirm the theorem's $10\%$ level for every stratum without a harmed network, but not for S30 ionproj ($18.4\%$) or S32 ion2020 ($14.0\%$), where they are consistent with it but cannot confirm it. The runs column is shown for reference only: runs of one network are not independent, so its bound is too optimistic. This re-reporting was not pre-registered, and it would not change any pre-registered verdict.

\begin{table*}[t]
\caption{\label{tab:cluster}Headline simulation claims with their replication units. Users: harmed over exposed users, as reported. Runs, networks: runs and networks with any harmed user; worst stratum: the stratum (hardware type, and grid size for S34; 32 networks each) with the largest bound. The one-sided 95\% upper bound on the share of networks is the valid bound (exact within a stratum, conservative when pooled); the run bound, shown for reference, treats correlated runs as independent. Script \texttt{cluster\_bounds.py}; not pre-registered.}
\begin{ruledtabular}
\footnotesize\setlength{\tabcolsep}{3pt}\begin{tabular}{lrrrrrl}
Claim & Users & Runs & Bound & Networks & Bound & Worst stratum\\
\hline
S26: guard, per-user budget, no stressor & 0\,/\,91{,}392 & 0\,/\,384 & 0.8\% & 0\,/\,128 & 2.3\% & each: 0\,/\,32, 8.9\%\\
S26: finite memory (pairs, not users) & 0\,/\,4{,}340{,}857 & 0\,/\,864 & 0.3\% & 0\,/\,96 & 3.1\% & each: 0\,/\,32, 8.9\%\\
S30: provable margin after alarm, in class & 230\,/\,89{,}585 & 2\,/\,2{,}048 & 0.3\% & 2\,/\,64 & 9.5\% & ionproj: 2\,/\,32, 18.4\%\\
\quad same, without the late-onset cell & 0\,/\,74{,}370 & 0\,/\,1{,}664 & 0.2\% & 0\,/\,64 & 4.6\% & each: 0\,/\,32, 8.9\%\\
\quad same, hybrid check only & 0\,/\,22{,}945 & 0\,/\,640 & 0.5\% & 0\,/\,64 & 4.6\% & each: 0\,/\,32, 8.9\%\\
S30: always-on margin, whole run & 0\,/\,208{,}896 & 0\,/\,768 & 0.4\% & 0\,/\,64 & 4.6\% & each: 0\,/\,32, 8.9\%\\
S32: current-look bound after alarm, in class & 23\,/\,11{,}800 & 1\,/\,256 & 1.8\% & 1\,/\,64 & 7.2\% & ion2020: 1\,/\,32, 14.0\%\\
\quad same, without the late-onset cell & 0\,/\,6{,}339 & 0\,/\,128 & 2.3\% & 0\,/\,64 & 4.6\% & each: 0\,/\,32, 8.9\%\\
S32: after alarm, $\tau$ and calibration variants & 0\,/\,15{,}606 & 0\,/\,192 & 1.5\% & 0\,/\,64 & 4.6\% & each: 0\,/\,32, 8.9\%\\
S32: always-on margin, variants & 0\,/\,139{,}264 & 0\,/\,512 & 0.6\% & 0\,/\,64 & 4.6\% & each: 0\,/\,32, 8.9\%\\
S34: after alarm, $10\times10$ and $30\times30$ & 0\,/\,940{,}391 & 0\,/\,1{,}280 & 0.2\% & 0\,/\,128 & 2.3\% & each: 0\,/\,32, 8.9\%\\
S34: always-on, larger grids (served users) & 0\,/\,1{,}495{,}612 & 0\,/\,960 & 0.3\% & 0\,/\,128 & 2.3\% & each: 0\,/\,32, 8.9\%\\
\end{tabular}
\end{ruledtabular}
\end{table*}

\paragraph*{Model-based layer (S26, S28, G3).} The cost of the per-user budget on SiV ($2.4$ points) exceeded the pre-registered 2-point bar. In a separate study (S28), the network-wide guard of Corollary~\ref{cor:NW} cost $0.5$ points on ions and $2.4$ on SiV relative to the per-user guard, and no network delivered a bad pair; the per-user guard also had none in those cells, so this checks validity rather than showing a difference. The plug-in rule (our simplified re-implementation of the EFiRAP/Q-PATH/Q-GUARD class~\cite{EFiRAP2022,QPATH2022,QGUARD2026}, not the published protocols) delivers when the model at point estimates clears $\tau$. With a margin tuned on ground truth from separate networks it was safe and competitive on ions, but it carries no guarantee off its tuning data, and on thin hardware it fell well behind the guards with informed priors, with one non-significant exception (Table~\ref{tab:secondary} in Appendix~\ref{app:sim}). A memory cut-off rule~\cite{Rozpedek2018} had no setting in its tuning grid that was safe on the development networks for three of four hardware types; we attribute this to its blindness to link-to-link differences. The $30\times30$ study (G3) ran the guard with links at $\alpha/\Lmax$ (a per-path guarantee), with about 27{,}000 requests per network and 64 networks each run under two priors (128 runs per controller): 0 of 930{,}435 delivering paths received a bad pair, while the same controller without the guard kept its pooled risk below $1\%$ but $1$--$3\%$ of its paths delivered a bad pair. That study used an exact physics model, stationary links and 2--3-hop requests, so it tests scale, not model error. At the USTC noise level the learned-prior guard that ignored the noise failed the safety bar, and the declared-noise cost on SiV ($4.5$ points) exceeded the pre-registered 3-point bar. Under persistent noise, the undeclared and fleet-recipe arms also reached $5.0$--$12.8\%$ per-user risk; the oracle and per-link arms stayed below $5\%$ over all users ($2.7$--$3.6\%$) but not over served users ($5.7$--$7.2\%$). With the declared margin, delivered-pair risk stayed up to $5\%$ at $M=4$, because some networks under-declared the margin (24 of 512 overall). In an earlier simulation set (S25), the guards added no measurable latency on ions; the finite-memory runs had $2.9\times10^6$ deliveries with 2 or 4 qubits per node.

\paragraph*{Certification layer (S27).} The four conditions were no stressor, undeclared model optimism, persistent USTC-level noise and a memory fault (all coherence times drop fivefold mid-run, unseen by the models). The certificate was checked against the true running average after every kept pair. These checks use ideal measurements and the model's own correlators, so they test the implementation rather than the physics. ``Unsafe'' cells have point per-user risk $>5\%$ or $r>1\%$ (Table~\ref{tab:secondary}). The undeclared-optimism networks were 64 (two traffic patterns). When $10\%$ of pairs were tested, the pooled certificate stayed below $\tau$ in every condition, including the one without bad pairs, so it would not have reassured a user here; the point is that the true average itself hides a quarter of bad pairs, which by Remark~\ref{rem:markov} no average certificate could exclude. Faults and persistent noise produced $1.8\%$ and $3.6\%$ bad pairs with similarly unremarkable averages. Eligible users are those whose certificate ever rose above 0. For users with $\geq500$ kept pairs on ion hardware, the certificate sat $0.16$, $0.10$ and $0.07$ below the truth at $p=0.05$, $0.1$, $0.2$ (ion2020). NV network-pooled certificates reached $0.09$--$0.44$ and SiV $\approx0$. Testing costs $5\%$ ($p=0.1$) or $16\%$ ($p=0.2$) of the usable pairs relative to $p=0.05$.

\paragraph*{Closed-loop studies in detail.} The paragraphs below give the full results of the model-check studies, in the order they were run.

\paragraph*{With the offset added to the model (S28, S29), the check detected a large constant offset; a small one late or not at all.} We first added the check of Proposition~\ref{prop:MC} and two heuristic responses to these networks for 90 epochs, on ion and thin hardware, at $p=0.1$ and $0.2$ (Appendix~\ref{app:loopsim}, Table~\ref{tab:superseded}). In both studies (S28 and S29) the offset was added to the model rather than to the truth, which inflated detection power (Sec.~\ref{sec:limits}); with the offset in the truth (S30, below) detection was much slower, so their detection results are superseded. Two results do not rest on detection power. Without model error, false alarms occurred in $3$ of $1{,}024$ runs ($0.29\%$; level $3.04\%$); on persistent noise, memory faults and thin hardware the check raised 2 alarms in 1,792 runs, both false: there the failures are shallow or the tests too few (after the memory fault $8$--$23\%$ of ion pairs fell just below $\tau$, but the mean test error rate barely moved). And a response must not trust a few tests: the first response failed its pre-registered post-alarm bar ($0.80\%$ of post-alarm deliveries bad, upper bound $2.27\%$ against $1\%$), all from one network whose alarm, after only 8 tests, raised the margin by a fixed step of $0.02$ against a true offset of $0.044$. The tracking rule of Sec.~\ref{sec:response}, designed after this failure on the same simulator and pre-registered separately, delivered no bad pair after the alarm in any of 512 runs (S29); with the offset in the truth (S30) it failed in 1 of 128 runs with an alarm forced after 8 tests. In S29, pairs delivered before the first alarm remain delivered (Table~\ref{tab:superseded}), and probe tests need the end nodes to measure pairs that the network declines to deliver.

\paragraph*{The provable response holds; detection is the bottleneck.} We tested Theorem~\ref{thm:post}, Corollary~\ref{cor:always} and Proposition~\ref{prop:D} on the same ion networks, now with the offset placed in the truth (extra depolarizing noise), so that model fidelities never exceed one (pre-registered; 8{,}384 runs; Appendix~\ref{app:s30}) [Fig.~\ref{fig:loop}(a)--(c)]. The levels were $\alpha=0.05$ and $\alpha_{\rm hi}=\alpha'=0.025$, so the per-user bound of Theorem~\ref{thm:post} is $0.10$, and $\beta=0.05$; the declared maximum offset was $\theta_{\max}=0.10$ per unit $\psi$, with $\psi_i=1$ for the constant class and $\psi_i=\min(h_i/\bar H,3)$ for the storage class ($h_i$ the storage time, $\bar H=0.0105$\,s, so $\psi_{\max}=3$). The check is pooled per network over all users' tests while the links run at $\alpha/\Umax$, so the $\alpha_c$ term of Corollary~\ref{cor:O} is not informative for this setup; the simulations test its $\beta$ term (a false alarm before an onset). In the pre-registered in-class cells (constant offsets, including a late onset, and a storage-proportional offset with matching shape), after an alarm, 230 of 89{,}585 users received a bad pair (bootstrap upper bound $0.60\%$; by networks, 2 of 64, conservative bound $9.5\%$, Table~\ref{tab:cluster}; the pre-registered bar compared this pool with the theorem's per-user bound of $0.10$, but the pool includes one run outside Theorem~\ref{thm:post}, see below) and $0.21\%$ of post-alarm deliveries were bad (bound $0.50\%$). All of these came from two of 2{,}048 runs: one false alarm before the onset of a late offset, which is outside (B2) and is the case Corollary~\ref{cor:O} pays $\beta$ for, and one run in which the offset confidence sequence excluded the true offset at its first look and, being a running minimum, never recovered, which is the failure Theorem~\ref{thm:post} allows with probability at most $\alpha'$ (Remark~\ref{rem:reopen} gives a valid way to recover). The hybrid check had none. With an alarm forced after 8 tests, no provable run delivered a bad pair after the alarm (0/384), while the heuristic tracking rule failed in one of 128. At offset $0.044$, the median post-alarm margin sat $0.011$--$0.013$ above the true offset (constant class) and $0.006$--$0.007$ above it (storage-proportional class); at offset $0.02$ the median excess was larger, $0.016$ (maximum $0.036$, recomputed from the run records) [Fig.~\ref{fig:loop}(c)]. At offset $0.044$, provability cost $1.3$--$8.5$ points of throughput against tracking. The new checks raised few false alarms without an offset (plug-in 1 of 128 runs, hybrid none), and the certificate stayed valid under this feedback (violated in $1.8\%$ and $3.1\%$ of runs at $p=0.1$, $0.2$; upper bounds $2.0\%$, $3.4\%$).

\paragraph*{A bound that can recover, and robustness.} A second pre-registered study (S32; 3{,}328 runs) tested the current-look bound of Remark~\ref{rem:reopen} and varied the threshold and the calibration data. Inside the offset class the current-look bound kept the guarantee (23 of 11{,}800 users and 661 of 450{,}432 post-alarm pairs bad, all in one run with a false alarm before the onset of a late offset, the Corollary~\ref{cor:O} case, in which the running minimum delivered 1{,}236) at a throughput cost of $0.3$ points. When the first tests after an alarm were forced to read as passes, so that the bound started below the true offset, the current-look bound came back above it in all 128 runs (median 502 tests), against none for the running minimum; post-alarm bad pairs fell from $46\%$ to $10\%$ and re-alarms from $0.88$ to $0.02$ per run. After a false alarm before an onset it halved the post-alarm bad-pair rate ($0.33\to0.16$), but per-user risk stayed high because (B2) fails. At thresholds $0.85$ and $0.92$ and with a third of the calibration data, no user received a bad pair after an alarm (0 of 15{,}606) and the always-on margin protected every user (0 of 139{,}264); its throughput cost with no model error ranged from $1.5$ to $28$ points. On the thin hardware types the check never alarmed and the always-on margin delivered $1.5\%$ of oracle: the closed loop needs test volume. A third pre-registered study (S34; 4{,}928 runs) repeated the closed loop on grids of $10\times10$ (180 links) and $30\times30$ (1{,}740 links), with the same total request rate, with demand restricted to as many users as on the $5\times5$ grid (matched exposure), and with three times the rate: after an alarm no pair was bad (0 of $4.2\times10^6$), the always-on margin delivered no bad pair (0 of $7.8\times10^6$), and the post-alarm margin (about $0.056$ at offset $0.044$) and throughput were essentially unchanged. Detection depended on the test volume, not on the grid size: at offset $0.044$ the check alarmed in $94\%$ of runs on both the $5\times5$ and the $30\times30$ grid, with a median first alarm at epoch $26$--$28$, and at epoch $10$--$12$ (2.3--2.5 times earlier) with three times the rate; harm before the alarm dominated on every grid.

Without the alarm gate, the always-on margin of Corollary~\ref{cor:always} protected every user over the whole run (0 of 208{,}896) in every cell with a constant offset (or none) from the start. It ran at $0.76$ (ion2020) and $0.70$ (ionproj) of oracle with no stressor, against $0.90$ and $0.92$ without a margin, and at $0.11$ and $0.61$ at offset $0.044$, where ion2020's margins are thin; for comparison, a margin declared from 64 or 256 tomography points reached $0.67$ or $0.81$ of oracle in S26, where the offset was added to the model. A fixed margin of $0.10$ delivered nothing. Outside the class the guarantee failed as predicted: with per-link heterogeneous offsets $0.14$ of post-alarm users received a bad pair, with a wrong offset shape up to $0.92$, after a false alarm before an onset $0.60$, and with the always-on margin under a late onset $0.34$ over the whole run.

The limit is detection. With physical offsets the check needed tens of epochs (median first alarm at epoch $27$--$55$ on ion2020 and $15$--$30$ on ionproj at offset $0.044$), detected an offset of $0.02$ in only $6$--$56\%$ of runs, and under a late onset alarmed in $41$--$100\%$ of runs, so over the whole run $0.47$--$0.49$ of users still received a bad pair at offset $0.044$, against $0.53$ without a check [Fig.~\ref{fig:loop}(b)]. The plug-in check alarmed more often than the mixture under a late onset ($83\%$ against $70\%$), and the hybrid's alarm rate stayed within $0.04$ of the better of the two in each of the four pre-registered offset scenarios (the pre-registered bar required at most $0.05$ below the worse); this, more than the single hardware rerun of Sec.~\ref{sec:hw}, is the evidence for the value of Proposition~\ref{prop:D}. Throughput counts good pairs only: without a check, the guard at offset $0.044$ kept $0.96$ of the oracle's good pairs, but $61\%$ (ion2020) and $59\%$ (ionproj) of its deliveries were bad [Fig.~\ref{fig:loop}(a)]. This is far above the $25\%$ of the S27 guarded arm at the same offset because the scenarios differ: in S26 and S27 the offset was added to the model's fidelity, over 30 epochs, whereas here it lowers the true fidelity, over 90 epochs. With the offset in the model, the wrongly delivered pairs are those whose true fidelity lies less than $\Delta$ below $\tau$; with the offset in the truth, they are all delivered pairs whose model fidelity lies less than $\Delta$ above $\tau$. The two rates are therefore not comparable.

\begin{table*}[p]
\caption{\label{tab:results}Consolidated simulation and hardware results (keys: Appendix~\ref{app:prov}). Per-user risk: users ever given a bad pair over all users; $r$: bad over delivered pairs; bounds: one-sided 95\% cluster-bootstrap, where reported (such a bound is 0 when no network has an event; cluster-level bounds are in Table~\ref{tab:cluster}); the closed-loop and hardware rows continue in Table~\ref{tab:results2}; safe: passed the safety bar; throughput: good delivered pairs over the oracle's good pairs (Appendix~\ref{app:sim}); pt: percentage points; fleet, learned: guard priors (Appendix~\ref{app:priors}). S26 and S27 added the offset to the model's fidelity, S30, S32 and S34 to the true state; their rows are not directly comparable. S28 and S29 (offset also added to the model, which inflated detection power; Sec.~\ref{sec:limits}) are in Table~\ref{tab:superseded}; secondary rows of S26--S28 are in Table~\ref{tab:secondary}.}
\scriptsize
\setlength{\tabcolsep}{2pt}\renewcommand{\arraystretch}{0.9}
\begin{ruledtabular}
\begin{tabular}{p{2.5cm}p{1.9cm}p{2.8cm}p{1.8cm}p{2.9cm}p{4.2cm}p{0.7cm}}
Setting & Arm & Per-user risk & Delivered-pair risk $r$ & Throughput or cost & Certificate or model check & Key\\
\hline
\multicolumn{7}{l}{\emph{Model-based layer, no stressor}}\\
All hardware types & guard, links at $\alpha/\Umax$ & 0/91{,}392 users & 0 in $1.01\times10^6$ & cost vs links at $\alpha/\Lmax$: $\leq0.75$\,pt (ions), up to $2.4$\,pt (SiV) & -- & S26\\
Ions & plug-in & 17--19\% & -- & -- & -- & S26\\
\multicolumn{7}{l}{\emph{Over-dispersion at measured levels} [ion2020 unless stated; Fig.~\ref{fig:control}(a)]}\\
ICFO levels & every guard & safe & safe & -- & -- & S26\\
USTC level & $\phi$ ignored & 1.1--5.5\% & -- & -- & -- & S26\\
USTC level & $\phi$ declared & safe & $\leq67$ bad in $10^5$ & cost $\leq4.5$\,pt (SiV) & -- & S26\\
USTC, persistent & undeclared; fleet recipe & 5.0--12.8\% & 1.4--7.3\% (all model-based arms) & -- & -- & S26\\
USTC, persistent & oracle; per-link & 2.7--3.6\% (served users 5.7--7.2\%) & (as above) & -- & -- & S26\\
\multicolumn{7}{l}{\emph{Model optimism, undeclared offset $\Delta=0.044$} [Fig.~\ref{fig:control}(b)]}\\
ion2020 & guard & 51\% & -- & -- & -- & S26\\
Ions & margin from $M$ tomography points; exact margin & $\leq0.9\%$ (bounds $\leq2.7\%$) & up to 5\% ($M=4$) & 16\% ($M=4$), 67\% ($M=64$), 81\% ($M=256$); exact 90\% & -- & S26\\
Ions, offset $0.044$ in the truth & always-on margin (Cor.~\ref{cor:always}) & 0 & 0 & 0.11 (ion2020), 0.61 (ionproj); exact margin 0.78, 0.89 & -- & S30\\
\multicolumn{7}{l}{\emph{Both layers} [ion2020, guarded arm, 64 networks per condition; Fig.~\ref{fig:cert}]}\\
Undeclared optimism & guard, certificate & -- & 25\% & -- & true average $0.926$; certificate $0.888$, below the truth in 64/64 networks & S27\\
Fault; persistent noise & guard, certificate & -- & 1.8\%; 3.6\% & -- & valid & S27\\
All cells, $p=0.05$, $0.1$, $0.2$ & certificate & -- & -- & testing costs 5\% ($p=0.1$), 16\% ($p=0.2$) of usable pairs & wrong for 0/3{,}283, 12/3{,}584, 11/21{,}358 users; in 0.9\%, 1.4\%, 0.3\% of networks & S27\\
\end{tabular}
\end{ruledtabular}
\end{table*}

\begin{table*}[p]
\caption{\label{tab:results2}Consolidated results, continued: the closed loop with the offset in the true state, and hardware. Columns and conventions as in Table~\ref{tab:results}; cluster-level bounds for the zero-event rows are in Table~\ref{tab:cluster}.}
\scriptsize
\setlength{\tabcolsep}{2pt}\renewcommand{\arraystretch}{0.9}
\begin{ruledtabular}
\begin{tabular}{p{2.5cm}p{1.9cm}p{2.8cm}p{1.8cm}p{2.9cm}p{4.2cm}p{0.7cm}}
Setting & Arm & Per-user risk & Delivered-pair risk $r$ & Throughput or cost & Certificate or model check & Key\\
\hline
\multicolumn{7}{l}{\emph{Provable response, offset in the true state} [S30, S32, S34; ion2020, ionproj; Fig.~\ref{fig:loop}]}\\
Offsets in the truth: constant 0.02--0.08, late onset, storage-proportional (in class) & provable margin (Thm.~\ref{thm:post}), alarm-gated & after alarm 230/89{,}585 users (bound $0.60\%$; 2 of 64 networks, one run outside Thm.~\ref{thm:post}); whole run $0.47$--$0.49$ at $0.044$ (no check $0.53$) & after alarm $0.21\%$ (bound $0.50\%$) & $0.54$/$0.71$ at $0.044$ (ion2020/ionproj) & forced early alarm 0/384 runs; margin $+0.011$--$0.013$ at $0.044$ & S30\\
Constant offsets (0.02--0.08) or none, no alarm gate & always-on margin (Cor.~\ref{cor:always}) & 0/208{,}896 users (whole run) & 0 & $0.76$/$0.70$ no stressor; $0.11$/$0.61$ at $0.044$ & late onset: $0.34$ (predicted failure); fixed margin 0.10 delivers 0 & S30\\
Outside (B2) & provable margin & after alarm $0.14$ (heterogeneous)--$0.92$ (wrong shape); $0.60$ (false alarm before onset) & -- & -- & predicted & S30\\
Offset $0.044$, forced miscoverage after the alarm & current-look bound vs running minimum & after alarm $0.37$--$0.56$ vs $0.64$--$0.68$ & after alarm $10\%$ vs $46\%$ & -- & recovered 128/128 (median 502 tests) vs 0/128 & S32\\
Threshold $0.85$, $0.92$; less calibration data & provable margin; always-on & after alarm 0/15{,}606; always-on 0/139{,}264 & 0 & always-on $0.64$--$0.99$ (no offset), $0.05$--$0.74$ at $0.044$ & thin hardware: no alarms & S32\\
Grids $10\times10$, $30\times30$ (same total rate; matched exposure; $3\times$ rate on $30\times30$) & provable margin; always-on & after alarm 0/940{,}391; always-on 0/1{,}495{,}612 (served) & 0 & gated $0.52$--$0.60$, always-on $0.37$--$0.50$ at $0.044$ & alarm rate $0.94$ on $5\times5$ and $30\times30$ & S34\\
Cubes $5^3$, $10^3$; $32\times32$ grid with diagonals; square grids of side 11--37; $\Umax\leq104$ & provable margin after an alarm; always-on & no harmed user in any of 36 cells of 32 networks & 0 in $8.8\times10^6$ & -- & alarm rate on the $5^3$ cube (ion2020) $0.69$, against $0.97$ on $11\times11$ & S3D\\
No offset; all runs & checks; certificate & -- & -- & -- & false alarms in 1/128 (plug-in), 0/128 (hybrid) runs; certificate violated in $1.8\%$, $3.1\%$ of 8{,}384 runs ($p=0.1$, $0.2$) & S30\\
\multicolumn{7}{l}{\emph{Hardware} (IBM \texttt{ibm\_fez}, local pairs, $p=0.2$; Sec.~\ref{sec:hw})}\\
Phases A, B; rerun & certificate & -- & -- & -- & valid at every kept pair; final $0.710$ vs $0.852$ (S31), $0.653$ vs $0.840$ (S31b) & S31, S31b\\
Phase A; phase B ($20\%$ $\Psi^-$) & fixed-bet check & -- & -- & -- & no alarm in 206; 457 tests (peak log-wealth $2.71<3.49$) & S31\\
Phase A; phase B (rerun) & hybrid check; fixed-bet; offset bound & -- & -- & -- & no alarm in 205 tests; alarm at test 170 (hybrid), 214 (fixed-bet) of 472; after it $\bar\theta=0.312\geq0.099$ (302 tests) & S31b\\
\end{tabular}
\end{ruledtabular}
\end{table*}

\paragraph*{Other topologies.} Three further pre-registered studies (S3D-1 to S3D-3; 2{,}976 runs; Appendix~\ref{app:s3d}) repeated the provable margin after an alarm and the always-on margin on lattices that are not square grids: a cube of $5^3$ nodes with the ion presets and near-neighbour demand, and, with projected links (link fidelity $0.97$--$0.99$) and requests between uniformly random nodes within a service radius of 5--8 hops, a cube of $10^3$ nodes and a $32\times32$ grid with diagonal links, alongside square grids of side 11, 13, 32 and 37. The routing sets reached $\Umax=104$ links, against at most 7 in the studies above. In each of the 36 lattice--hardware--arm cells (32 networks each), no network had a harmed user (bound $8.9\%$ per cell), and none of $8.8\times10^6$ guarded deliveries was bad. Detection was the exception: on the $5^3$ cube with ion2020 hardware, the check alarmed in 22 of 32 networks at offset $0.044$, against 31 of 32 on an $11\times11$ grid of similar size; with ionproj hardware the rates were 31 and 32 of 32. These studies also compared how much demand each lattice serves; those results concern network design rather than the guarantees and are reported with the studies~\cite{Repo3D2026}.

\subsection{Hardware experiment}\label{sec:methods-hw}
Kept pairs were measured in the Bell basis; this pass fraction is a noisy reference, not a ground truth. For these states, a single gate or readout error turns a $\Phi^+$ pair into an apparent failure, whereas a $\Psi^-$ pair reads as $\Phi^+$ only after an error that flips both bits (two readout flips, or one $Y$-type gate error before the Bell circuit), so the reference may be biased low; we did not characterise the Bell measurement, so the direction and size of its bias are unknown, and the reference also has sampling error (pointwise intervals in Sec.~\ref{sec:hw}). A pre-drawn, seeded schedule had three phases: 600 calibration pairs, all tested in a random basis, which fix the declared fidelity $\Fid_{\rm decl}$ (a one-sided 95\% Clopper--Pearson bound via Lemma~\ref{lem:F}) and the null bound $q=\tfrac23(1-\Fid_{\rm decl})$ of the check; phase A, with test fraction $p=0.2$ and no injection; and phase B, with $p=0.2$ and $20\%$ of the pairs replaced by $\Psi^-$. There are no links here: the declared model is the constant $\Fid_{\rm M}=\Fid_{\rm decl}$, with no link parameters, so $\bm\lambda^{\rm hi}$ is not used. The certificate of Proposition~\ref{prop:C1} ($\alpha=0.05$, declared parity-flip bound $0.05$, not verified for the device) was updated after every kept pair, and a model check ($\beta=0.05$) ran on the tests of phases A and B. Each run pre-registered its binary outcomes (Appendix~\ref{app:hw}).

\paragraph*{First run (S31).} The fixed-bet check of Proposition~\ref{prop:MC} raised no false alarm in the 206 tests of phase A; its peak log-wealth in phase B was $2.71$, below the threshold $\ln(1/\beta_1)=3.49$, so the pre-registered detection outcome failed. An exploratory analysis identifies the mechanism. Calibration gave $\hat\Fid=0.8925$ (43 errors in 600 tests), a declared fidelity of $0.8628$ and a null bound $q=0.0914$. In phase A the good pairs erred less often ($5.8\%$) than in calibration ($7.2\%$), so the null was conservative and the fixed-bet mixture lost log-wealth, reaching $-3.75$ by the end of phase A. Phase B raised the test error rate to $15.5\%$ (the injected $\Psi^-$ pairs erred in $97\%$ of $X$ and $Z$ tests and $0\%$ of $Y$ tests, as Lemma~\ref{lem:F} predicts), an excess of about $+0.064$ over 457 tests, which lifted the log-wealth by about $+6.4$: not enough to repay the debt. A local noisy simulation of a different device, run first as a dry run, did alarm in phase B (peak log-wealth $4.98$). This is the debt mechanism of Sec.~\ref{sec:debt}. A post hoc replay of the recorded outcomes with the plug-in and hybrid checks of Proposition~\ref{prop:D} (whose bet rule was written after this run's summary rates had been seen) alarmed in phase B, so we pre-registered a rerun.

\paragraph*{Rerun (S31b).} The rerun used the same device and design with a fresh schedule seed, the hybrid check as the primary check and the fixed-bet mixture as a paired comparison. The certificate ended at $0.6526$ against $0.8403$; the hybrid check raised no alarm in the 205 tests of phase A. Calibration gave $\Fid_{\rm decl}=0.8851$ ($q=0.0766$); the test error rates were $5.8\%$, $4.9\%$ and $19.5\%$ in the three phases. The fixed-bet mixture again fell into deficit in phase A (log-wealth $-3.37$), but the larger phase-B excess of this run let it recover. Before the run, the hybrid's power at the first run's error rates was estimated at $79\%$ (mixture $45\%$). After the alarm, the offset confidence sequence of Theorem~\ref{thm:post} ran on the post-alarm tests (level $0.025$, constant shape, no declared maximum, i.e.\ $\theta_{\max}=1$ so that $\bar\theta$ starts at 1, with $\Fid_{\rm decl}$ in place of the model, $y_i=\Fid_{\rm decl}-1+1.5E_i$), for 302 tests; its final value is about three times the Bell-measured offset of the kept pairs after the alarm. IBM readout is asymmetric (decay-biased), which Lemma~\ref{lem:onesided} does not cover; the checks' null $q$ also omits the readout term of Proposition~\ref{prop:MC}, because $\Fid_{\rm decl}$ is estimated from tests that include readout errors. The experiment is two jobs on one device, the pairs are not distributed over distance, and the Bell-measured reference has its own, uncharacterised error.

\paragraph*{IonQ Forte (S41a, S41b).} Each job was submitted from the Qollab browser environment (Qiskit 2.6.2, qiskit-ionq 1.0.2) with debiasing switched off; the job script, the design table derived from fixed seeds, the analysis and the certificate code were frozen and pushed before the first job (S41a: 30 jobs, 18 Bell pairs $\Phi^+$ per job, axes stratified as 6 $X$, 6 $Y$, 6 $Z$ per job; 10 jobs, chosen by seed, with $R_z(\pm\pi/4)$ on one qubit of 6 pairs, balanced in sign within each axis. S41b: 10 jobs of 12 GHZ-3 and 10 of 9 GHZ-4 states in a seeded order, prepared by a CNOT chain; tests stratified to the marginals of Lemma~\ref{lem:ghz}). Gates required the design and the submitted OpenQASM circuit to match the frozen table, IonQ's job record to show hardware without a noise model and without mitigation, and every job to show each test error below $0.25$ and a linear fidelity above $0.5$. Test-or-keep was drawn after the fact with probability $p\in\{0.05,0.1,0.2\}$, 1{,}000 seeded draws each, as binomial counts per job, position and outcome. The reference was $1-\tfrac12\sum_B\mathrm{CP}^{\rm lo}(e_B)$ for Bell pairs and $1-\tfrac12\mathrm{CP}^{\rm lo}(e_Z)-2^{1-n}\sum_s\mathrm{CP}^{\rm lo}(e_s)$ for GHZ states, on the kept states, with the level split across terms and across the job checkpoints. The planted-fault prediction was $\Delta=\tfrac12f\sin^2(\pi/8)[(1-2e_X)+(1-2e_Y)]$ with $f=1/3$ and $e$ from the clean jobs; with balanced signs it is exact for Bell-diagonal noise and robust to a shared coherent phase. Independent AI code reviews before each freeze found two critical issues in S41a (a circuit-size fallback and output loss after submission), fixed before any hardware job.

\subsection{Detailed limitations}\label{sec:limits}
\begin{itemize}
\item \emph{Simulation scope.} All network results are simulations with simplified (Werner/Pauli-transfer) physics, on a $5\times5$ grid except for the scaling checks on $10\times10$ and $30\times30$ grids (S34, G3) and the topology checks on cubes, a grid with diagonal links and square grids up to $37\times37$ (S3D-1 to S3D-3, the last two with projected link parameters); the recent studies have no purification beyond one DEJMPS round. Requests are served one at a time (S27--S34, S3D-1 to S3D-3, and S26 except its finite-memory cells, which served disjoint paths in parallel), so contention between users who need the same link or memory at the same time is not tested beyond those cells. The guard's prediction uses each qubit's actual storage time, so contention should cost throughput rather than validity, but this is untested. The baselines are class representatives re-implemented in our model, not the original codes.
\item \emph{Real data.} The real-data analyses re-use published data. The ICFO primary visibility estimator failed because phase settings were not recorded. The Innsbruck ion--ion event counts did not match the publication. Model offsets were measured on three experiments: optimistic but not significantly so at USTC, no optimism detected at Delft (upper bounds $0.009$ for the post-selected and $0.018$ for the delivered states; the latter test had power $0.63$ at an offset of $0.03$) or at Harvard (pooled $-0.077$, upper bound $0.050$); over-dispersion rests on three (ICFO, USTC, Harvard; the IonQ runs, about 9\,s each, showed none). The Delft data are summed per basis, without time order, so only fixed-sample analyses were possible.
\item \emph{Hardware.} Only one network dataset (Oxford) randomises test-or-keep per pair, between two nodes about 2\,m apart; its tests are CHSH rounds, its certificate assumes trusted nominal measurement settings and has no ground truth for the kept pairs, and it is a single 8.5\,h run. Our IBM runs randomise Pauli tests, but only for local pairs, in two jobs on one device, checked against a Bell-measured reference that has its own error and was not characterised. The IonQ runs (S41a, S41b) cover 50 jobs on one device over about four days; each state's test setting is fixed per job and position (one circuit per job), so validity under this clustering is empirical; errors of simultaneous states were mildly correlated; test-or-keep was drawn after the fact; the reference is conservative and the only ground truth is the planted phase fault; one Bell job's output was lost to a server error and re-run, as pre-registered. The fixed-bet check missed the injected degradation in the first run; in a pre-registered rerun the hybrid check of Proposition~\ref{prop:D} detected it, but so did the fixed-bet check, 44 tests later, so a single run shows earlier detection, not a power difference (pre-run power at the first run's rates: $79\%$ against $45\%$). IBM readout is asymmetric, which Lemma~\ref{lem:onesided} does not cover, so the post-alarm bound on hardware is a consistency check. The SiV analyses (S37, S38) and S39 draw the test choice after the fact; the IonQ settings were measured in blocks, and the Harvard herald rule (counting an odd number of detector clicks) follows the authors' code.
\item \emph{Certificate.} The certificate assumes trusted measurements; on the Delft data this assumption was binding, since raw counts did not certify entanglement without the authors' readout correction. It is loose by $0.05$--$0.16$ at practical sizes; its pair cost is within a factor $1.9$--$2.3$ of the converse at zero observed errors. Tighter anytime constructions (e.g., betting-based~\cite{WaudbySmith2024}) are left open.
\item \emph{Pre-alarm exposure.} The model check limits harm rather than preventing it. Pairs delivered before the first alarm stay delivered; Theorem~\ref{thm:post} covers only pairs after it, and by Proposition~\ref{prop:N} (when most pairs are kept) and Proposition~\ref{prop:fault} no data-driven rule can protect all earlier pairs. The check did not restore the per-user guarantee: with the offset in the truth at $0.044$, $47$--$49\%$ of users (pooled over ion2020 and ionproj) still received a bad pair, against $53\%$ without a check (S30).
\item \emph{Offset class of Theorem~\ref{thm:post}.} (B2) needs exact proportionality to the declared shape; with heterogeneous offsets, pairs with above-average offset are not protected. Offsets above $\theta_{\max}$ violate (B4) and are not covered. A false alarm before the onset of an offset breaks (B2): the first window starts with no offset, its bound falls toward the slack, and after the onset the true offset exceeds it until the check alarms again. Corollary~\ref{cor:O} pays $\beta$ for this case, but protection resumes only after the later alarm. The guarantee also assumes symmetric, outcome-independent readout (Lemma~\ref{lem:onesided}). In simulation all three failure modes appeared as predicted (post-alarm per-user risk $0.14$--$0.92$), and the running minimum let one early miscoverage persist for the rest of a run (S30); the current-look bound of Remark~\ref{rem:reopen} recovered from a forced miscoverage in all 128 runs of S32 (median 502 tests), while re-opening a window at a stopping time has not been tested.
\item \emph{Topology.} All guarantees are stated per link and per user and hold on any graph. The closed loop was simulated on square grids of side 5 to 37, on cubes of $5^3$ and $10^3$ nodes and on a $32\times32$ grid with diagonal links (S34, S3D-1 to S3D-3), with routing sets of up to 104 links, and the guard alone also at $30\times30$ (G3); irregular and long-haul topologies were not simulated, and the diagonal links were given the same quality as the others. On square grids, detection delay depended mainly on the number of tests the network generates, not on the grid size; on the $5^3$ cube with ion2020 hardware it was slower than on a square grid of similar size (alarm in 22 against 31 of 32 networks at offset $0.044$), for reasons we have not established.
\item \emph{Detection power.} In the model-check simulations S28 and S29 the offset was added to the model's fidelity rather than to the truth, so model fidelities could exceed 1 for some attempts; for those pairs the null bound was clipped to a near-zero floor, which inflated the check's detection power, and a central prediction above 1 may also have inflated the tracking rule's offset estimate. Scoring was correct (bad pairs, false alarms and certificate validity were judged against the true fidelity). With the offset in the truth (S30, Sec.~\ref{sec:loop}) detection was much slower: a constant offset of $0.044$ was first detected after a median of $15$--$55$ epochs, an offset of $0.02$ in only $6$--$56\%$ of runs, and most harm to users happened before the first alarm.
\item \emph{Scope of the closed-loop evidence.} The responses to an alarm were tested in simulation on ion hardware only: in S30--S34 on square grids of side 5, 10 and 30, for constant, storage-proportional, per-link heterogeneous and late-onset offsets; in S3D-1 to S3D-3 on cubes, a grid with diagonal links and square grids of side 11--37, for a constant offset of $0.044$ only (with projected ion links in S3D-2 and S3D-3); thresholds $0.85$--$0.92$ and reduced calibration data were also tested (S32). The check itself also ran on NV and SiV, where it raised no true alarm (too few tests; S28, S32). Offsets that change after an alarm and offsets above $\theta_{\max}$ are not covered. The tracking response is a heuristic, and probe tests require cooperating end nodes. It was designed after the first detection study failed, on the same simulator; fresh seeds do not remove that dependence.
\item \emph{Metric.} Our per-user risk metric averages over all users, including users never served; it does not report the worst user, which Theorem~\ref{thm:guard} bounds.
\item \emph{Stressors.} The noise and fault models are stylised: ``persistent noise'' is an AR(1) common-mode shift of the calibration parity probability, and the simulated memory fault also affects storage that began before it.
\end{itemize}

\begin{acknowledgments}
We thank the authors of Refs.~\cite{LagoRivera2021,Krutyanskiy2023,USTC2026,Stolk2024} for making their data public, D.~P.~Nadlinger for providing the raw data of Ref.~\cite{Nadlinger2022} and answering our questions about them, Y.-C.~Wei and M.~D.~Lukin for providing the data and analysis notebooks of Ref.~\cite{Knaut2024}, the authors of Ref.~\cite{Hartung2026} for publishing their data and code, and G.~Araneda for providing the Sr--Sr data and explaining the protocol and the fidelity conventions.
This work used Bridges-2 at the Pittsburgh Supercomputing Center through allocation CIS261606
from the Advanced Cyberinfrastructure Coordination Ecosystem: Services \& Support (ACCESS) program~\cite{Boerner2023}, which is supported by U.S. National Science Foundation grants \#2138259, \#2138286, \#2138307, \#2137603, and \#2138296.
We thank Qollab and IonQ for the hardware credits awarded in the Qollab $\times$ IonQ hackathon, which funded the IonQ Forte~1 runs. We acknowledge the use of IBM Quantum services for this work. The views expressed are those of the authors, and do not reflect the official policy or position of IBM or the IBM Quantum team.
\end{acknowledgments}

\paragraph*{Author contributions.} T.S.R.\ started the project, set its direction, made the design decisions, ran all simulations, hardware experiments and analyses, arranged the data with the experimental groups, and checked the code, the results and the manuscript. M.M.\ supervised the work, reviewed the analyses and the manuscript, and suggested the latency analysis.

\paragraph*{Use of AI.} We used an AI assistant (Claude, by Anthropic) a lot in this work, and we want to be clear about how. It wrote most of the simulation and analysis code, drafted the pre-registration documents and most of the manuscript text, and independent instances of it reviewed the proofs and code as adversarial reviewers (Methods and Appendix~\ref{app:prov}). The Abstract, Introduction, Box~1 and Discussion were then revised to follow T.S.R.'s own explanations and writing style. T.S.R.\ decided what to study and what to keep, ran every simulation and hardware job himself, and checked the outputs against the frozen code and the raw data. Every confirmatory analysis was fixed and made public before it was run, so the results do not depend on who wrote the code. The authors take full responsibility for the content of this paper.

\paragraph*{Data and code availability.} Code, pre-registrations, run outputs, analysis scripts and the evidence ledger are at Ref.~\cite{Repo2026} (release v1.0.0, which matches this manuscript, DOI 10.5281/zenodo.24447997; concept DOI for all releases 10.5281/zenodo.23287414). The topology studies S3D-1 to S3D-3 (simulator fork, pre-registrations, run outputs and analyses) are in a separate repository~\cite{Repo3D2026} (release v1.0.1, DOI 10.5281/zenodo.23468140).

Third-party data are available from their original deposits (for Ref.~\cite{Hartung2026}, its public repository at commit 8a08f695fca8), or for Refs.~\cite{Nadlinger2022,Knaut2024} and the Oxford Sr--Sr run from their authors, and are not redistributed.

\appendix
\input{appendix}

\FloatBarrier
\end{document}

%% file: appendix.tex
\section{Statements and proofs}\label{app:proofs}
\subsection{Setting and standing assumptions}
Links $e\in\mathcal E$ have levels $\lambda_e\in[0,1]$. A calibration bit on link $e$ is $1$ with probability $\mu_e=(1+\eta\lambda_e)/2$, with $\eta\in(0,1]$ known. The filtration $(\mathcal F_t)$ is generated by all calibration bits, waits, herald records, routing and delivery decisions, and the controller's independent randomisation.
\begin{itemize}
\item[(A1)] For every link $e$ and calibration bit $i$, $\Pr(X_{e,i}=1\mid\mathcal F_{e,i-1})=\mu_e$. How many bits are measured, when, and on which links may depend arbitrarily on the past. Whether a given pair is calibrated or delivered must not depend on that pair's own outcome.
\item[(A0)] A pair delivered on path $P$ with waits $\bm w$ has fidelity $\Fid_P(\bm\lambda_P;\bm w)$ (or the declared margin covers the difference).
\item[(A2)] $\Fid_P$ is nondecreasing in each link level. This holds for Werner compositions under depolarising swaps, and was checked numerically for the Pauli-transfer model used here.
\item[(A3)] Levels are constant in time. Appendix~\ref{app:drift} relaxes this.
\end{itemize}
The guard keeps, for each link, a confidence sequence $\mathrm{CS}_e(t)$ for $\mu_e$ at level $a_e$, with lower end $\ell_e(t)$, and sets $\lambda^{\rm lo}_e(t)=\mathrm{clip}((2\ell_e(t)-1)/\eta,0,1)$. An attempt on $P$ with waits $\bm w$ may be delivered only if $\Fid_P(\bm\lambda^{\rm lo}_P(t);\bm w)\geq\tau$.

\paragraph*{Classical tools.}
\begin{itemize}
\item \emph{Ville's inequality}~\cite{Ville1939}: a nonnegative supermartingale with initial value at most $1$ reaches $1/a$ at some time with probability at most $a$.
\item \emph{Method of mixtures}~\cite{Robbins1970,Howard2021,Ramdas2023}: for a prior $\pi$ fixed in advance, $\Lambda_t(\mu)=\int u^{K_t}(1-u)^{N_t-K_t}\pi(du)/[\mu^{K_t}(1-\mu)^{N_t-K_t}]$ is a test martingale under (A1) and any predictable sampling. Hence $\mathrm{CS}(t)=\{\mu:\Lambda_t(\mu)<1/a\}$ covers $\mu$ at all times with probability $\geq1-a$.
\item \emph{Weighted Bonferroni}: $\sum_{c\geq1}6/(\pi^2c^2)=1$.
\end{itemize}

\subsection{Lemmas on the link confidence sequence}
\begin{lemma}[Interval]\label{lem:interval}
$\mathrm{CS}(t)$ is an interval containing $\hat\mu_t=K_t/N_t$, since $\log \Lambda_t$ is quasi-convex with minimum at $\hat\mu_t$ and $\Lambda_t(\hat\mu_t)\leq1$.
\end{lemma}
\begin{lemma}[Informed priors]\label{lem:prior}
Let $\mathcal G$ be a $\sigma$-field with $\Pr(X_{e,i}=1\mid\mathcal G\vee\mathcal F^{\rm own}_{e,i-1})=\mu_e$. Then for any $\mathcal G$-measurable prior $\pi$, $\Lambda_t(\mu_e)$ is a test martingale conditionally on $\mathcal G$, and coverage holds unconditionally.
\end{lemma}
\noindent\emph{Proof.} Apply the method of mixtures conditionally on $\mathcal G$, then take expectations. $\square$

A sufficient condition is that $\pi$ is a function of data that are conditionally independent of the link's own calibration stream, given the realised parameters. Priors fitted on the link's own data violate this: a prior concentrated at the own maximum-likelihood estimate gave a 14\% miss rate at nominal 5\% after 2{,}000 looks. So do priors from neighbours that share \emph{per-trial} noise with the link.

\begin{lemma}[Monotone lower end]\label{lem:mono}
For the $\mathrm{Beta}(\tfrac12,\tfrac12)$ mixture, $\ell(N,K+1)\geq\ell(N,K)$.
\end{lemma}
\noindent\emph{Proof.} $\Lambda_{K+1}(\mu)/\Lambda_K(\mu)=[(K+\tfrac12)/(N-K-\tfrac12)](1-\mu)/\mu$, which is $\geq1$ for $\mu\leq(K+\tfrac12)/N$. Every $\mu<\ell(N,K)$ satisfies this, by Lemma~\ref{lem:interval}. $\square$

\begin{lemma}[Heterogeneous bits within an epoch]\label{lem:het}
If an epoch's $n$ bits are independent with mean probability $\bar\mu$, then $\E[(\bar u/\bar\mu)^K((1-\bar u)/(1-\bar\mu))^{n-K}]\leq1$, by AM--GM. A likelihood that treats epoch counts as binomial therefore yields a supermartingale.
\end{lemma}
\begin{lemma}[Restarts]\label{lem:restart}
A CS built only from bits after a stopping time $\sigma$ has coverage $\geq1-a$ conditionally on $\mathcal F_\sigma$. This covers alarm-triggered restarts.
\end{lemma}

\subsection{The guard}\label{app:guardproof}
\begin{lemma}[Per-path guard]\label{thm:path}
Under (A0)--(A3), with each link at level $a_e$, for every fixed path $P$: $\Pr(\text{a pair delivered on }P\text{ ever has }\Fid<\tau)\leq\sum_{e\in P}a_e$.
\end{lemma}
\noindent\emph{Proof.}
\begin{enumerate}
\item Let $G_e=\{\forall t:\mu_e\in\mathrm{CS}_e(t)\}$, so $\Pr(G_e^c)\leq a_e$.
\item On $G_e$, $\lambda^{\rm lo}_e(t)\leq\lambda_e$ for all $t$.
\item On $\bigcap_{e\in P}G_e$, every delivery has true fidelity $\Fid_P(\bm\lambda_P;\bm w)\geq\Fid_P(\bm\lambda^{\rm lo}_P;\bm w)\geq\tau$, by (A0), (A2) and the guard.
\item A union bound completes the proof. $\square$
\end{enumerate}

\noindent\emph{Proof of Theorem~\ref{thm:guard} (per user).} User $u$ only ever uses links in $U_u$, whatever the routing. Repeat the argument with $\bigcap_{e\in U_u}G_e$, which gives $|U_u|\alpha/\Umax\leq\alpha$. $\square$

\noindent\emph{Tightness of the factor $|U_u|/\Lmax$.} Running links at $\alpha/\Lmax$ gives $|U_u|\alpha/\Lmax$ per user. With $k$ marginal candidate links and a near-tight CS, Monte Carlo gave a per-user miss of $0.21$ at $k=40$ against nominal $0.05$. The Beta mixture is conservative enough to hide this in practice. In a $d_x\times d_y$ grid box, $|U_u|=2d_xd_y+d_x+d_y$.

\noindent\emph{Scope.} The bound concerns the event ``ever''. It does not imply a bound on the fraction of delivering paths that contain a bad pair, which is a selected quantity; that fraction appeared as a secondary bar in our earlier studies.

\subsection{Drift, re-alignment and restarts}\label{app:drift}
\paragraph*{Drift model.} At re-alignment times $t_{e,1}<t_{e,2}<\dots$ (known, or stopping times), the level in cycle $c$ satisfies $\theta_{e,c}-g(t'-t_{e,c})\leq\lambda_e(t')\leq\theta_{e,c}$ at times $t'$ in the cycle, with a known rate bound $g$.

\begin{corollary}[Bounded drift]\label{cor:drift}
Restart each link's CS at each re-alignment, at level $6\alpha/(\pi^2c^2\Umax)$, and set $\lambda^{\rm lo}_e(t)=\mathrm{clip}((2\ell_{e,c}(t)-1)/\eta,0,1)-g(t-t_{e,c})$. Then Theorem~\ref{thm:guard} holds unchanged.
\end{corollary}
\noindent\emph{Proof.} Couple the drifting bits with bits of constant probability $\mu(\theta_{e,c})$, so that $X_i\leq Y_i$. Lemma~\ref{lem:mono} gives $\ell(N,K^X)\leq\ell(N,K^Y)$. Coverage for $Y$, weighted Bonferroni over cycles, and the drift bound complete the proof. A pair stored across a re-alignment must be checked against the bound of the cycle in which it was generated. $\square$

\begin{corollary}[Known drift]
If the drift rate $g$ is known exactly (or lies in a declared set), a mixture over $\theta$ with epoch-averaged probabilities is a supermartingale by Lemma~\ref{lem:het}. Taking the minimum over the declared set costs no extra $\alpha$.
\end{corollary}

\begin{corollary}[Declared re-alignment bound]
Let $\mathrm{comb}_{e,c}=\max(\theta^{\rm lo}_{e,c},\mathrm{comb}^-_{e,c-1}-B)$, and let $V_e$ be the event that some jump $\theta_{e,c-1}-\theta_{e,c}$ exceeds $B$. Then $\Pr(\exists c,t:\mathrm{comb}_{e,c}(t)>\theta_{e,c})\leq\alpha/\Umax+\Pr(V_e)$.
\end{corollary}
\noindent After violations, the overshoot accumulates as the sum of the excesses along the carry chain, not their maximum.

\subsection{Delivered-pair risk and faults}\label{app:faults}
\noindent\emph{Proof of Remark~\ref{rem:baserate}.} Take a two-point law of true margins: $\pi_0$ at $-\xi$ and $1-\pi_0$ at a large positive value. Bayes' identity gives $r=\pi_0\alpha_\xi/(\pi_0\alpha_\xi+(1-\pi_0)\gamma)$ with $\gamma\leq1$ the delivery probability of good links, so $r\geq r^\ast$ once $\pi_0\geq r^\ast/(r^\ast+\alpha_\xi(1-r^\ast))$. $\square$

\begin{proposition}[Fault exposure]\label{prop:fault}
Couple a faulted and an unfaulted world on the same randomness. At an unannounced time $t_f$, independent of the no-fault process and of the controller's randomisation, link $e$ starts producing bad pairs, and nothing observable differs between the two worlds before $t_f+\delta_f$. Then
$\Pr(\text{faulted world delivers a bad pair})\geq\int\Pr(D(t))\,\Pr_{t_f}(dt)$,
where $D(t)$ is the event that the unfaulted world delivers a pair generated on $e$ in $[t,t+\delta_f)$. A per-user bound $\alpha$ that holds for every fault time therefore requires $\sup_t\Pr(D(t))\leq\alpha$.
\end{proposition}
\noindent\emph{Proof.} Decisions coincide pathwise before $t_f+\delta_f$. $\square$

\subsection{Model-free certification}\label{app:cert}
\noindent\emph{Proof of Lemma~\ref{lem:F}.} $|\Phi^+\rangle\langle\Phi^+|=(I+XX-YY+ZZ)/4$. $\square$

\noindent\emph{Proof of Proposition~\ref{prop:N}.}
\begin{enumerate}
\item Couple the two networks so that they deliver the same pair whenever no replacement occurs, and use the same test decisions and the same (possibly adaptive) choice of measurement. Pair $i$ can change the joint distribution of the test outcomes only if it is replaced and tested, which happens with probability $p\,c/N$; given that, its own outcome distribution changes by at most $\kappa_T$ in total variation ($\kappa_T=\tfrac23$ for a uniformly random Pauli test, since $\Psi^-$ and $\Phi^+$ give identical $Y$ outcomes and opposite $X$ and $Z$ parities). Later adaptive choices are functions of earlier outcomes and add nothing.
\item Summing over the $N$ pairs bounds the total variation by $\kappa_Tpc$.
\item The probability of at least one replaced kept pair is $1-(1-(1-p)c/N)^N$. $\square$
\end{enumerate}

\noindent\emph{Proof of Proposition~\ref{prop:C1}.}
\begin{enumerate}
\item XX, YY, ZZ and $|\Phi^+\rangle\langle\Phi^+|$ are simultaneously diagonal in the Bell basis. So the joint statistics of the users' tests are unchanged if each pair is first measured in the Bell basis (as in Lo--Chau~\cite{LoChau1999}).
\item Let $b_i$ indicate a non-$\Phi^+$ virtual label and $T_i$ a test. Conditionally on the past, with test choice and basis independent of everything the network holds:
\begin{itemize}
\item if $b_i=0$, then $E_i=0$ and $\E[e^{\nu((1-T_i)b_i-1.5kT_iE_i)}]=1$;
\item if $b_i=1$, the expectation is $(1-p)e^\nu+p[1-\tfrac23(1-e^{-1.5\nu k})]=1$ by the choice of $k$.
\end{itemize}
\item The product is a nonnegative supermartingale, and so is a mixture over $\nu$. Ville's inequality and monotone inversion complete the proof. $\square$
\end{enumerate}

\noindent\emph{Average conditional fidelity.} Let $x_i=1-\Fid_i$, the infidelity of pair $i$'s state conditional on the past. Replace the label indicator by $x_i$ in the kept-pair exponent. Given the past, a tested pair errs with probability $\tfrac23x_i$ (Lemma~\ref{lem:F}), so the factor's conditional mean is $g(x_i)=(1-p)e^{\nu x_i}+p[1-\tfrac23x_i(1-e^{-1.5\nu k})]$. $g$ is convex, and $g(0)=1$ and $g(1)=1$ by the choice of $k$, so $g\leq1$ on $[0,1]$. The resulting process is a nonnegative supermartingale, and the same inversion bounds $\sum_{\rm kept}(1-\Fid_i)$. $\square$

\noindent\emph{Readout errors.} If each basis's parity-flip probability is at most $s_{\max}$, the error probability of a non-$\Phi^+$ label is at least $\tfrac23(1-2s_{\max})$. Replacing $\tfrac23$ by this value restores validity. Without such a bound, readout noise can lower error rates for states with $\Fid<1/4$: a Monte Carlo of an all-$\Psi^-$ network with 1--2\% readout flips gave violation rates of 5.8--11.8\% at nominal 5\%.

\noindent\emph{Implementation.} We use eight values of $\nu$, geometrically spaced in $[\nu_{\max}/16,0.98\nu_{\max}]$, with $\nu_{\max}=\ln[(1-p/3)/(1-p)]$ and uniform weights. This grid was chosen on synthetic Werner streams before any study data.

\noindent\emph{Monte Carlo.} We ran 1{,}000 runs per scenario, checked after every kept pair, at $p\in\{0.05,0.2\}$, over iid, drift, fault, rare-bad, adaptive and test-timing adversaries. The maximum violation rate was $2.5\%$. A second check (3{,}000 runs $\times$ $10^4$ pairs, adaptive adversaries, non-Bell-diagonal states) gave at most $4.6\%$.

\subsection{Converse, model check and network-wide guard}\label{app:loop}
\paragraph*{Proposition~\ref{prop:LB} (converse).} Take the source that emits $\Phi^+$ with probability $1-f$ and some state orthogonal to $\Phi^+$ with probability $f$, independently. For an orthogonal state $e_X+e_Y+e_Z=2$ (Lemma~\ref{lem:F} with $\Fid=0$), so a test in a uniformly random basis fails with probability exactly $\kappa=2/3$, whatever the per-basis rates. Kept and tested pairs are disjoint and independent, and the test/keep pattern does not depend on the states. Summing over patterns with $n_K$ tests before the $K$-th kept pair,
$\Pr_f(\text{all tests pass},\,B_K>(1-\tau)K)=\E[(1-\kappa f)^{n_K}]\,\Pr(\mathrm{Bin}(K,f)>(1-\tau)K)$,
where $B_K$ counts orthogonal kept pairs. On this event the rule certifies $\tau$ (the record is all-pass) while the kept failed fraction exceeds $1-\tau$. Since $n_K$ is negative binomial, $\E[x^{n_K}]=[(1-p)/(1-px)]^K$; maximising over $f$ gives the statement. $\square$

\noindent\emph{Remarks.} (i) The hypothesis ``on every all-pass record'' is needed: a rule that certifies only on the all-pass records with the most tests escapes part of the bound (at $\tau=0.9$, $p=0.1$ the floor is $330$, $300$ and $205$ when it certifies on all, $90\%$ and $50\%$ of them; at $10\%$ the bound is vacuous). The certificate of Proposition~\ref{prop:C1} is deterministic at zero errors. (ii) The rule ``after exactly $K$ kept pairs, output $\tau$ if every test passed'' is valid against every i.i.d.\ source exactly when the condition holds, so the converse is tight for such sources. (iii) For another test, the same argument with the state orthogonal to $\Phi^+$ that the test rejects with the smallest probability $\kappa_{\min}$ gives the bound with $\kappa=\kappa_{\min}$. The direction matters: a bound with a larger $\kappa$ is weaker, so a test that rejects some orthogonal states more often than others is limited by the state it detects worst. For random Pauli tests every orthogonal state is rejected with probability exactly $2/3$, so $\kappa_{\min}=2/3$.

\paragraph*{Proposition~\ref{prop:MC} (model check).} Let $\mathcal G_i$ contain everything up to and including the generation of pair $i$, including the unobservable $\Fid_{\rm true}(i)$. Then $q_i$ and the capped bets are $\mathcal G_i$-measurable, and the test indicator, basis and readout flip are independent of $\mathcal G_i$. Let readout in basis $b$ act as a symmetric parity flip with probability $s_b\leq s_{\max}$, and let $e_b$ be the pair's error probability in basis $b$. If $\Fid_{\rm true}(i)\geq\Fid_g(i)$, then $\Fid_{\rm true}(i)\geq\tau\geq\tfrac34$, so every $e_b\leq2(1-\Fid_{\rm true}(i))\leq\tfrac12$ (Lemma~\ref{lem:F}). Each term $s_b+(1-2s_b)e_b$ then increases in $s_b$, so $\Pr(E_i=1\mid\mathcal G_i,\text{tested})=\tfrac13\sum_b[s_b+(1-2s_b)e_b]\leq s_{\max}+(1-2s_{\max})\tfrac23(1-\Fid_{\rm true}(i))\leq q_i$, the last step because the expression decreases in $\Fid$. Every factor $1+\nu_{g,i}(E_i-q_i)$ is then nonnegative, at least $0.1$, and has conditional mean at most $1$. The process stopped just before the first violating pair is therefore a nonnegative supermartingale with initial value 1, and Ville's inequality bounds the probability that it ever reaches $1/\beta_1$ by $\beta_1$. The same argument applies to each segment after a restart (restart times are stopping times; cf.\ Lemma~\ref{lem:restart}); the first level is $\beta_1=6\beta/\pi^2$ and the $k$-th restart uses $6\beta/(\pi^2(k+1)^2)$, so all levels sum to $\beta$. The statement also holds with ``first delivered violating pair'' replaced by ``first tested violating pair''. $\square$

\noindent\emph{Detection.} The expected log-increment of bet $\nu$ is linear in $\Pr(E_i=1)$. If $\Pr(E_i=1\mid\mathcal G_i)\geq q+\delta$ for every tested pair, it is at least $h=(q+\delta)\ln(1+\nu(1-q))+(1-q-\delta)\ln(1-\nu q)$. Wald's identity then bounds the expected number of tests to an alarm by $[\ln(1/(\beta_1w))+\ln(1+\nu(1-q))]/h$. In Monte Carlo runs with constant $q$, the mean delays (135--5{,}035 tests) were below this bound. Before a change point the log-wealth drifts downward, so detection after a change takes longer than the bound suggests.

\noindent\emph{Numerical checks.} With errors drawn exactly at the null bound, the first-alarm rate was $2.09$--$2.26\%$ (20,000 runs of $10^4$ tests each; level $3.04\%$). With restarts, the rate was $0.021$ alarms per run.

\paragraph*{Corollary~\ref{cor:NW} (network-wide guard).} If every link's sequence runs at level $\alpha/|\mathcal E|$, then with probability at least $1-\alpha$ every link's sequence covers its level at all times. On that event every delivered pair has $\Fid_{\rm true}\geq\Fid(\bm\lambda^{\rm lo})-\Delta\geq\tau+d-\Delta\geq\tau$ whenever the declared margin covers the offset ($\Delta\leq d$, assumption (A0)). Hence $\Pr(\text{any user ever receives a bad pair})\leq\alpha$ under Theorem~\ref{thm:guard}'s assumptions. $\square$

\subsection{The closed loop with a guarantee}\label{app:post}
\noindent\emph{Lemma~\ref{lem:onesided}: refinements.} With a calibrated lower bound $s_{\rm lo}\leq s_{i,b}$ for every basis, the sharper bound $1-\Fid_i\leq1.5(P_i-s_{\rm lo})/(1-2s_{\rm lo})$ holds, where $P_i=\Pr(E_i=1\mid\text{past})$: under the conditions of the lemma each term $s_{i,b}+(1-2s_{i,b})e_{i,b}$ increases in $s_{i,b}$, so $P_i\geq s_{\rm lo}+(1-2s_{\rm lo})\tfrac23(1-\Fid_i)$. With asymmetric readout, where the flip probability depends on the outcome, the lemma needs an additional condition that relates the error rate to the two flip probabilities; we do not use that case.

\noindent\emph{Proof of Proposition~\ref{prop:D}.} Each factor satisfies $1+\nu_i(E_i-q_i)\geq1-\nu_iq_i>0$, and under $H_0$ its conditional mean is $1+\nu_i(\Pr(E_i=1\mid\text{past})-q_i)\leq1$ because $\nu_i$ is fixed before the outcome. A product of such factors is a nonnegative supermartingale, and an average of supermartingales is one. Ville's inequality and the spending over segments give the false-alarm bound. For the last statement, let $V_i$ indicate that tested pair $i$ has $\Fid_{\rm true}(i)<\Fid_g(i)$; $V_i$ is fixed when the pair is generated, before its test, as in the proof of Proposition~\ref{prop:MC}. Replace each factor by $1$ from the first $i$ with $V_i=1$ on; the stopped process is a nonnegative supermartingale whatever the network does, because every remaining factor satisfies the null bound (by the argument of Proposition~\ref{prop:MC}, with $\tau\geq\tfrac34$). So the probability of an alarm before the first violating tested pair is at most $\beta$. Under (A0), a violating tested pair requires a lower-end failure on a link that carried it, i.e.\ on some link in $\mathcal E_c$, which has probability at most $\sum_{e\in\mathcal E_c}a_e$; this is at most $\alpha$ when $\mathcal E_c\subseteq U_u$ or the links run at $\alpha/|\mathcal E|$. (After an alarm the same device applies to the guard in force, with the coverage events $G_2\cap G_3$ below.) $\square$

\noindent\emph{Debt.} When the guard is clearly conservative, the plug-in mean $\hat\delta_i$ is negative and $\nu_i=0$, so wealth stays put; the fixed-bet mixture of Proposition~\ref{prop:MC} keeps betting and loses log-wealth at a rate of about $\sum_g w_g\nu_g(q-\bar e)$. When the guard is nearly exact, noise in $\hat\delta_i$ triggers positive bets and the plug-in also loses wealth. The plug-in needs some tens of tests to notice an excess. After an alarm, the provable rule restarts the check (with spending) whenever $\bar\theta_t$ has fallen by $0.01$ since the last restart, so the check watches the current guard with fresh wealth (at most $\theta_{\max}/0.01$ restarts per window).

\noindent\emph{Proof of Theorem~\ref{thm:post}.}
\begin{enumerate}
\item \emph{Good events.} $G_1$: the lower ends hold for $u$'s links at all times; $\Pr(G_1^c)\leq\alpha$ (Theorem~\ref{thm:guard}). $G_2$: the upper ends hold for every link at all times; $\Pr(G_2^c)\leq\alpha_{\rm hi}$ (B1). $G_3$: $\theta\leq\bar\theta_t$ at all times after the first alarm.
\item \emph{Each $y_i$ over-estimates the offset on average.} For a pair $i$ in a window, let $c_i=\mathbf 1\{\bm\lambda\leq\bm\lambda^{\rm hi}_{i-1}\}$. Because $\bm\lambda$ is constant, $c_i$ is determined by the past. On $c_i=1$, by (A2) and Lemma~\ref{lem:onesided},
$\E[y_i\mid\text{past}]\geq\Fid_{\rm M}(\bm\lambda;\bm w_i)-1+(1-\Fid_i)=\Delta_i=\theta\psi_i$, the last step by (B2).
\item \emph{A supermartingale at the true $\theta$.} Let $\tilde W_t(\vartheta)=J^{-1}\sum_j\prod_i[1+c_i\nu_{ij}(\vartheta\psi_i-y_i)]$. Every factor is at least $0.1$, because $\nu_{ij}\leq0.9/y^{\max}_i$, $y_i\leq y^{\max}_i$ and $\vartheta\psi_i\geq0$. At $\vartheta=\theta$ each factor has conditional mean at most $1$, by step 2 and because $\nu_{ij}$ is fixed in advance. Positivity gives nonnegativity, and the mean bound gives the supermartingale property: $\tilde W(\theta)$ is a nonnegative supermartingale that starts at $1$ in each window; a window starts at an alarm, which is a stopping time.
\item \emph{Ville.} $\Pr(\exists t:\tilde W_t(\theta)\geq1/\alpha'_k)\leq\alpha'_k$. On $G_2$, $\tilde W=W$. Each $\nu_{ij}\geq0$ and $\psi_i\geq0$, so $W_t(\vartheta)$ is nondecreasing in $\vartheta$; hence $W_t(\theta)<1/\alpha'_k$ implies $\bar\theta_t\geq\theta$, given $\theta\leq\theta_{\max}$ (B2, B4). Summing over windows, $\Pr(G_2\cap G_3^c)\leq\sum_k\alpha'_k\leq\alpha'$.
\item \emph{No bad delivery on the good events.} On $G_1\cap G_2\cap G_3$, every pair delivered to $u$ after the first alarm satisfies $\Fid_j=\Fid_{\rm M}(\bm\lambda;\bm w_j)-\theta\psi_j\geq\Fid_{\rm M}(\bm\lambda^{\rm lo};\bm w_j)-\bar\theta_t\psi_j\geq\tau$. So the probability of a bad delivery to $u$ after the alarm is at most $\Pr(G_1^c)+\Pr(G_2^c)+\Pr(G_2\cap G_3^c)$. $\square$
\end{enumerate}

\noindent\emph{Remarks.} (i) The union bound is conservative: both link ends come from the same mixture martingale, so per link the failure of either end is bounded by the larger of the two levels. (ii) Heuristically, as the number $n$ of window tests grows, $\bar\theta_t-\theta$ approaches the average of $[\Fid_{\rm M}(\bm\lambda^{\rm hi})-\Fid_{\rm M}(\bm\lambda)]+1.5\bar s(1-\tfrac43(1-\bar\Fid))$, plus a stochastic term of order $\sqrt{2\,\mathrm{Var}(y)\ln(J/\alpha')/n}$. The first two terms are the price of a one-sided, model-free reading of the tests; the $s_{\rm lo}$ version of Lemma~\ref{lem:onesided} removes the readout part. (iii) The simulations of S30 used the bet grid $\nu_j=0.01\cdot2^j$, $j=0,\dots,6$, the levels $\alpha=0.05$, $\alpha_{\rm hi}=\alpha'=0.025$, $\beta=0.05$, and $\theta_{\max}=0.10$.

\noindent\emph{Proof of Corollary~\ref{cor:O}.} Let $\mathcal E_c$ be the links whose pairs feed the check, $G'_1$ the event that the lower ends of all links in $U_u\cup\mathcal E_c$ hold at all times ($\Pr(G_1'^c)\leq\alpha_c$), and $T_1$ the time of the first alarm. Define $\tilde W$ as in step 3 of the proof of Theorem~\ref{thm:post}, with each factor also multiplied inside by $\mathbf 1\{t_i\geq t_0\}$, i.e.\ $1+c_i\mathbf 1\{t_i\geq t_0\}\nu_{ij}(\vartheta\psi_i-y_i)$. Because $t_0$ is deterministic or a stopping time, the indicator is fixed before the outcome of test $i$, and on $t_i\geq t_0$ we have $\Delta_i=\theta\psi_i$, so steps 2--4 give $\Pr(\exists k,t:\tilde W_t(\theta)\geq1/\alpha'_k)\leq\alpha'$. On $\{T_1\geq t_0\}$ every pair in a window has $t_i\geq T_1\geq t_0$, so $\tilde W=W$ on $G_2\cap\{T_1\geq t_0\}$, and step 5 applies to every pair delivered after $T_1$ on $G_1'\cap G_2\cap G_3$. On $\{T_1<t_0\}\cap G_1'$, the model is exact before $t_0$, so every tested pair before $t_0$ satisfies the check's null, and $T_1<t_0$ is an alarm before the first violating tested pair, which has probability at most $\beta$ (Propositions~\ref{prop:MC} and~\ref{prop:D}). Hence
$\Pr(\text{bad pair to }u\text{ after }T_1)\leq\Pr(G_1'^c)+\Pr(G_2^c)+\alpha'+\Pr(T_1<t_0,G_1')\leq\alpha_c+\alpha_{\rm hi}+\alpha'+\beta$. $\square$

\noindent\emph{Proof of Corollary~\ref{cor:always}.} Open window $k=1$ at the stopping time $t=0$, with $\bar\theta_0=\theta_{\max}$ (B4). Because (B2) holds for every pair from $t=0$, steps 1--4 of the proof of Theorem~\ref{thm:post} apply verbatim to this window and to any later window opened at an alarm, so $\Pr(G_2\cap G_3^c)\leq\sum_k\alpha'_k\leq\alpha'$, where $G_3$ is now the event $\theta\leq\bar\theta_t$ at all $t\geq0$. Step 5 then holds for every pair delivered to $u$, not only those after an alarm, which gives $\Pr(u\text{ ever receives a bad pair})\leq\alpha+\alpha_{\rm hi}+\alpha'$. The same argument covers windows re-opened at other stopping times (Remark~\ref{rem:reopen}). $\square$

\subsection{Further remarks}
\paragraph*{Occam identity.} $\mathrm{CS}_t(\pi)=\{\mu:\ell_t(\hat\mu)-\ell_t(\mu)<\log(1/a)+O_t(\pi)\}$, with $\ell_t$ the log-likelihood and $O_t(\pi)\geq0$. A prior affects efficiency only through $O_t$; its validity is governed by Lemma~\ref{lem:prior}.

\paragraph*{Identifiability index.} When the fleet law is estimated from $N$ windows on \emph{distinct} links, the relative standard error of the spread estimate is about $[\sqrt{N/2}/(1+(\sigma_w^2+A^2/12)/s^2)]^{-1}$. Windows that repeat links count roughly as the number of distinct links.

\section{Simulation methods}\label{app:sim}
\paragraph*{Network and service.}
\begin{itemize}
\item A $5\times5$ grid (40 links); users are all ordered node pairs within $\Lmax$ hops (204 users for $\Lmax=2$, 340 for $\Lmax=3$); candidates are all shortest paths.
\item Requests arrive as one Poisson process for the whole network, at the total rate $\Gamma$ of Table~\ref{tab:hw}, and each request is for a user drawn uniformly at random. Each is served by greedy best-path routing on the controller's current estimates, with up to four attempts. Requests are served one at a time in order of arrival (except in the finite-memory cells of S26, where requests on disjoint paths are served in parallel, limited by the memory qubits per node), and each delivered pair is recorded for the user of its request.
\item Each attempt generates two chains of link pairs with exponential waits; the first chain is stored while the second is generated, then one DEJMPS round is applied.
\item 30 epochs of 100\,s; the first 3 are warm-up.
\end{itemize}

\paragraph*{Links and calibration.}
\begin{itemize}
\item Link levels follow a fleet law with a spatial field (length scale 1.5 grid units, correlation 0.7).
\item Calibration uses 30\% of heralded pairs, pre-drawn per epoch.
\item Optional node common-mode noise and model offsets are applied as described in Sec.~\ref{sec:methods-sim}.
\end{itemize}

\paragraph*{Physics.} The Pauli-transfer model (memory dephasing, depolarising gates, readout) agrees with an exact density-matrix model to $<10^{-15}$, which had been cross-checked against NetSquid~\cite{Coopmans2021} in earlier studies. Correlators of delivered pairs come from the same bilinear forms; the check $1-\Fid=(e_X+e_Y+e_Z)/2$ holds to $2\times10^{-16}$.

\begin{table*}[t]
\caption{\label{tab:hw}Hardware parameter sets, anchored to demonstrated links: ion2020 to Ref.~\cite{Stephenson2020}, NV to Ref.~\cite{Pompili2021}, SiV to Ref.~\cite{Knaut2024}; ionproj is a projected ion link; fut97--fut99 (S3D-2, S3D-3 only) are projected links with the rate, memory and gates of a future ion link (ionfut) and fleet link fidelity $0.97$--$0.99$, for which $\Lmax$ is the service radius and $\Gamma$ counts only requests within it. $R$: heralded link-pair rate; $T_2$: memory coherence; $\varepsilon_{\rm g}$: gate error; $\varepsilon_{\rm ro}$: per-qubit readout error; $\Fid_0$, sd: fleet mean and spread of link fidelity; $\tau$: threshold; $\Lmax$: maximum path length; $\Gamma$: total request rate of the network (each request is for a user drawn at random).}
\begin{ruledtabular}
\begin{tabular}{lccccccccc}
 & $R$ (s$^{-1}$) & $T_2$ (s) & $\varepsilon_{\rm g}$ & $\varepsilon_{\rm ro}$ & $\Fid_0$ & sd & $\tau$ & $\Lmax$ & $\Gamma$ (s$^{-1}$) \\
ion2020 & 182 & 1.0 & 0.005 & 0.001 & 0.94 & 0.012 & 0.9 & 2 & 3.0 \\
ionproj & 200 & 1.0 & 0.003 & 0.001 & 0.97 & 0.006 & 0.9 & 3 & 3.0 \\
NV & 9 & 1.0 & 0.005 & 0.03 & 0.82 & 0.036 & 0.8 & 2 & 0.3 \\
SiV & 1 & 2.0 & 0.005 & 0.01 & 0.86 & 0.028 & 0.8 & 2 & 0.05 \\
fut97, fut98, fut99 & 1000 & 1.0 & 0.003 & 0.001 & 0.97, 0.98, 0.99 & 0.006, 0.004, 0.002 & 0.9 & 5, 6, 8 & 3.0 \\
\end{tabular}
\end{ruledtabular}
\end{table*}

\paragraph*{Controllers.}
\begin{itemize}
\item The risk controller chooses a delivery threshold on the posterior probability that an attempt is bad, from a sliding window, targeting $r\leq1\%$.
\item Guards:
  \begin{itemize}
  \item \texttt{bg}: Beta$(\tfrac12,\tfrac12)$ CS.
  \item \texttt{kg}: mixture CS with the fleet law as the prior.
  \item \texttt{ag2t}: a hedged mixture of a learned neighbour-based prior frozen after epoch 2, the fleet law and Jeffreys, combined with own data after the freeze only (Lemma~\ref{lem:prior}).
  \end{itemize}
\item Baselines (no controller, no guard): \texttt{plug} delivers if the model at point estimates clears $\tau$; \texttt{plugM} adds a margin $m$; \texttt{cut} delivers if the storage wait is below $x/((1-\rho)R)$.
\item $m$ and $x$ were tuned per hardware on 16 dev networks by a pre-registered rule: $m\in\{0.01,0.01,0.06,0.10\}$ and $x\in\{0.01,0.1,0.01,0.01\}$ for ion2020, ionproj, NV and SiV.
\end{itemize}

\subsection{Informed and learned priors}\label{app:priors}
The mixing prior $\pi$ may be chosen from other data, for example by a model that predicts a link's quality from its neighbours. This affects efficiency only through the ``Occam'' term of the CS. It preserves validity if the prior is conditionally independent of the link's own calibration stream (Lemma~\ref{lem:prior}). A prior frozen from neighbours' data and combined only with the link's later data satisfies this when noise does not persist across the freeze. A prior fitted on the link's own data does not: it raised the miss rate to 14\% at nominal 5\% after 2{,}000 looks in our checks. In our simulations informed priors raised throughput on data-starved hardware. On SiV, either the known fleet law or a learned prior took it from $4\%$ to $20$--$21\%$ of oracle. On NV, only the learned prior helped (from $21\%$ to $39\%$). On ions the fleet law added 1--2 points and the learned prior cost 3--4. We treat the learned prior as an optional component.

\subsection{Stressors and metrics}
\paragraph*{Stylised stressors.} ``Persistent noise'' is an AR(1) common-mode shift of each node's calibration parity probability across epochs (correlation $0.9$). The simulated memory fault (S27) applies the faulted $T_2$ to all storage in an attempt, including storage that began before the fault time. Measurements and test outcomes are ideal and drawn from the model's own correlators.

\paragraph*{Metrics.} Per-user risk is the number of users ever served a bad pair divided by all users, including users never served; over served users only, it is higher (for example $5.7$--$7.2\%$ instead of $2.7$--$3.6\%$ for the persistent-noise oracle and per-link arms). Delivered-pair risk is bad pairs divided by delivered pairs. Both have one-sided 95\% cluster-bootstrap bounds over networks. Throughput is the number of good (above-threshold) delivered pairs divided by the number of good pairs an oracle obtains on the same requests: the oracle knows the true link levels and the true fidelity of each attempt, routes on the true best path, and counts a request as served if one of its attempts is good. Bad deliveries therefore do not count towards throughput.

\paragraph*{Secondary results.} Table~\ref{tab:secondary} lists the results of S26--S28 that Sec.~\ref{sec:sim} discusses but Table~\ref{tab:results} omits.

\begin{table*}[t]
\caption{\label{tab:secondary}Secondary simulation results (S26--S28), moved from Table~\ref{tab:results}. Columns as in Table~\ref{tab:results}; n.s.: not significant.}
\footnotesize
\setlength{\tabcolsep}{2.5pt}
\begin{ruledtabular}
\begin{tabular}{p{2.35cm}p{2.1cm}p{2.75cm}p{2.2cm}p{3.0cm}p{3.45cm}l}
Setting & Arm & Per-user risk & Delivered-pair risk $r$ & Throughput or cost & Certificate or model check & Key\\
\hline
\multicolumn{7}{l}{\emph{Model-based layer, no stressor}}\\
Ions, SiV & network-wide guard & 0/128 networks (per-user guard also 0/128) & -- & cost vs per-user guard: $0.5$\,pt (ions), $2.4$\,pt (SiV) & -- & S28\\
Ions & tuned margin & safe & safe & guard $+2.4$ to $+4.2$\,pt (fleet), $-3.0$ to $+0.4$\,pt (learned) & -- & S26\\
NV, SiV & tuned margin & -- & SiV: 4 pairs (0.25\% of oracle) & learned: guard $+17$ (NV), $+19$ (SiV)\,pt; fleet: $+20.5$ (SiV), $-1.6$ (NV, n.s.) & -- & S26\\
Finite memory, $4\times$ load & 2, 4, $\infty$ qubits & -- & 0 in $4.3\times10^6$ & latency $+20$--$60\%$ (2 qubits; NV, SiV) & -- & S26\\
\multicolumn{7}{l}{\emph{Both layers} [ion2020, guarded arm, 64 networks per condition]}\\
80 unsafe cells & certificate & -- & -- & -- & wrong for $\leq0.34\%$ of users, $\leq1.2\%$ of networks & S27\\
\end{tabular}
\end{ruledtabular}
\end{table*}

\subsection{Model-check simulations (S28 and S29)}\label{app:loopsim}
\begin{itemize}
\item \emph{Horizon and tests.} The horizon is 90 epochs. Every delivered pair carries two independent test streams ($p=0.1$ and $0.2$); an arm with feedback uses the stream of its own $p$. The per-qubit readout bound $\varepsilon_{\max}$ is each hardware type's specified readout error, and $s_{\max}=2\varepsilon_{\max}(1-\varepsilon_{\max})$.
\item \emph{Model offset.} The offset was added to the model's fidelity, not removed from the truth, so model fidelities could exceed 1; for those pairs the null bound $q_i$ was clipped to a near-zero floor. Scoring used the true fidelity (Sec.~\ref{sec:limits}); this inflated the check's detection power, and S30 places the offset in the true state instead (Appendix~\ref{app:s30}). The model-optimism cells of S26 and S27 also added the offset to the model's fidelity.
\item \emph{Alarm response in S28.} On an alarm the margin rises by $1.5$ times an empirical-Bernstein upper estimate of the mean of $E_i-q_i$ over the segment, clipped to $[0.02,0.10]$. Segments with fewer than 50 tests use a fixed step of $0.02$.
\item \emph{Tracking in S29.} After an alarm the margin is $0.10$ until 50 tests have arrived. Every 10 epochs it is then set to $1.5[\mathrm{CP}_{\rm ub}(n_E,n;\beta/9)-\bar q_c]+\bar d$, clipped to $[0,0.10]$. Here $n_E$ and $n$ count errors and tests since the alarm, including probe tests of margin-blocked pairs, and the Clopper--Pearson bound is recomputed at most 9 times per run (hence $\beta/9$). The quantity $q_c=\tfrac23(1-\Fid_c)$ uses the guard's central prediction $\Fid_c$ at the point estimates, net of the margin in force, with zero readout error, and $\bar d$ is the mean margin in force.
\item \emph{Scenarios.} S28: offset $0.044$, persistent noise, a memory fault at epoch 45, and readout at the declared bound, each on all four hardware types. S29 (ion hardware): offsets $0.02$, $0.044$ and $0.08$, and an alarm forced at the 8th test (with no stressor, and with offset $0.044$).
\item \emph{Baselines.} Margins from 16 tomography points and the exact margin. In the S29 dev pilot (not the confirmatory run), a replay of S28's failing network on the same seeds reproduced its 6,232 bad pairs under S28's rule; tracking delivered none.
\end{itemize}

\paragraph*{The tracking rule as a heuristic.} The tracking margin of S29 is held at $0.10$ until $50$ tests have arrived after the alarm; from then on, every 10 epochs it is set to a Clopper--Pearson upper bound (level $\beta/9$) on the model offset, estimated from the model's central prediction. Pairs blocked only by the margin are probe-tested rather than discarded, so a high margin cannot starve the estimate. The margin may fall, and each decrease restarts the e-process. The rule is a heuristic: the Clopper--Pearson bound assumes i.i.d.\ trials, whereas the tested pairs have different $q_i$ and are selected through the margin itself; the level $\beta/9$ is a split over the looks of a fixed horizon; and the bound concerns the average offset over tested pairs.

\paragraph*{Results of S28 and S29 (superseded detection power).} Table~\ref{tab:superseded} lists the results. Because the offset was added to the model, the detection results (alarm rates and epochs) and the pre-alarm harm are superseded by S30, where the offset lowers the true fidelity: there detection was much slower (Sec.~\ref{sec:loop}), whole-run harm with a gated check at offset $0.044$ was $34$--$36\%$ of users on ion2020 and $54$--$56\%$ on ionproj, reversing the hardware ranking of S29, and the tracking rule failed in 1 of 128 runs with a forced early alarm. In S28, on ion hardware with an undeclared offset of $0.044$, the check alarmed in all 128 runs (median epoch 8--23) and cut bad delivered pairs by $70$--$92\%$. The first response raised the margin by $1.5$ times an empirical-Bernstein estimate, or by a fixed step of $0.02$ when fewer than 50 tests were available; all 6,232 post-alarm bad pairs came from one network whose alarm came after only 8 tests, and its post-alarm margins were about twice the true offset (throughput about half of oracle). In S29 the tracking rule delivered no bad pair after the alarm in any of 512 runs ($5{,}765{,}233$ deliveries); with alarms forced after 8 tests it was clean in $128/128$ runs, against $1/128$ for the first rule, which also leaked in 4 and 13 of 128 fresh runs at offsets $0.044$ and $0.08$. The tracking margin sat $0.006$--$0.011$ above the true offset, and throughput rose by $24$--$34$ points to about $0.78$ of oracle at offset $0.044$; a 16-point tomography campaign reached $0.35$--$0.53$, and the exact margin $0.91$--$0.94$. After a forced false alarm the margin returned to about $0.01$; a rule that jumps to the maximum margin stayed there and delivered almost nothing. Without the upper bound, 102 of 512 runs leaked bad pairs after the alarm; without the alarm gate (tracking from the start), throughput fell by $14$--$16$ points with no model error. The tracking margin fell below the true offset in 2 of 512 runs (8,119 deliveries, none bad), where Theorem~\ref{thm:guard} no longer covered the guard. At offset $0.02$ the check alarmed in only 56 of 128 runs within the horizon, and runs without an alarm kept delivering bad pairs ($19\%$). In S29, pairs delivered before the first alarm remain delivered: at offset $0.044$, $45$--$48\%$ of ion2020 users and $19$--$20\%$ of ionproj users had received a bad pair (against $53\%$ and $27\%$ without the check).

\begin{table*}[t]
\caption{\label{tab:superseded}Model-check results of S28 and S29 (ion hardware unless stated; constant offsets). The offset was added to the model's fidelity, which inflated the check's detection power; the detection and pre-alarm rows are superseded by S30 (Table~\ref{tab:results}). Columns as in Table~\ref{tab:results}.}
\footnotesize
\setlength{\tabcolsep}{2.5pt}
\begin{ruledtabular}
\begin{tabular}{p{2.35cm}p{2.1cm}p{2.75cm}p{2.2cm}p{3.0cm}p{3.45cm}l}
Setting & Arm & Per-user risk & Delivered-pair risk $r$ & Throughput or cost & Certificate or model check & Key\\
\hline
No model error & check & -- & -- & -- & false alarms 3/1{,}024 runs (level 3.04\%) & S28\\
$\Delta=0.044$ & fixed-step response & -- & after alarm 0.80\% (bound 2.27\%; bar failed) & about half of oracle & alarm 128/128 (epoch 8--23); bad pairs cut 70--92\% & S28\\
Persistent noise, fault, thin hardware & check & -- & after fault 8--23\% & -- & 2 alarms in 1{,}792 runs, both false & S28\\
$\Delta=0.02$ & check & -- & 19\% in runs without alarm & -- & alarm in 56/128 runs & S29\\
All offset cells & tracking & -- & after alarm: none in 512 runs & up $24$--$34$\,pt & margin $0.006$--$0.011$ above $\Delta$; below it in 2/512 runs & S29\\
$\Delta=0.044$ & tracking & before alarm 45--48\% (ion2020), 19--20\% (ionproj); 53\%, 27\% without check & whole run 7.8\% & $\approx0.78$ & -- & S29\\
Alarm forced after 8 tests & tracking; fixed-step & -- & clean in 128/128; 1/128 runs (S30, offset in the truth: tracking failed in 1/128) & -- & -- & S29\\
$\Delta=0.044$; $0.08$ & fixed-step & -- & leaked in 4/128; 13/128 runs & -- & -- & S29\\
$\Delta=0.044$ & tomography, 16 (64, 256) points; exact margin & -- & -- & 0.35--0.53 ($\approx0.67$, $0.81$); 0.91--0.94 & -- & S29\\
Ablations & no bound; no gate & -- & no bound: 102/512 runs leaked & no gate: $-14$ to $-16$\,pt & -- & S29\\
\end{tabular}
\end{ruledtabular}
\end{table*}

\subsection{Provable-loop simulations (S30, S32, S34)}\label{app:s30}
The offset is placed in the true state as extra depolarizing noise that lowers the fidelity by exactly $\Delta_i$; the model is unchanged, so model fidelities never exceed 1. Scenarios (ion2020, ionproj; 90 epochs; 32 networks per cell; $p=0.1$ and $0.2$): constant offsets $0.02$, $0.044$, $0.08$; a constant $0.044$ from epoch 30 (late onset); a storage-proportional offset $\Delta_i=0.044\min(h_i/\bar H,3)$ with $\bar H=0.0105$\,s; per-link heterogeneous offsets (path mean of per-link Gamma draws, mean $0.044$, sd $0.02$); alarms forced at the 8th test (with no offset, with $0.044$, and before the late onset). Levels $\alpha=0.05$, $\alpha_{\rm hi}=\alpha'=0.025$ (so the bound of Theorem~\ref{thm:post} is $0.10$), $\beta=0.05$; $\theta_{\max}=0.10$ per unit $\psi$, with $\psi_i=1$ (constant shape) or $\psi_i=\min(h_i/\bar H,3)$ (storage shape, $\psi_{\max}=3$); bet grid as in Appendix~\ref{app:post}. Arms: the provable rule behind the mixture, plug-in and hybrid checks (constant shape), the hybrid with the storage shape, the always-on variant (Corollary~\ref{cor:always}), S29's tracking rule, a fixed margin of $0.10$ and the exact margin. The model check is pooled per network. After an alarm the check restarts whenever $\bar\theta$ falls by $0.01$.

\paragraph*{S32 and S34.} Both studies reuse this simulator and its code unchanged, with 32 networks per cell, paired across arms. S32 (3{,}328 runs, seeds 19331--19334) has three blocks. Block R (ion2020, ionproj; $p=0.1$ and $0.2$) compares the running minimum with the current-look bound of Remark~\ref{rem:reopen}, both behind the hybrid check, with no offset, a constant offset $0.044$, the late onset, the late onset with an alarm forced at the 8th test (before the onset), and a forced miscoverage: an alarm forced at the 8th test, after which every test of the first window before its first look (at least 150 tests) is recorded as a pass. Block E (ion hardware; $p=0.1$; no offset and offset $0.044$) uses thresholds $0.85$ and $0.92$, and calibration on $10\%$ instead of $30\%$ of heralded pairs (service rate unchanged), with $\theta_{\max}=0.10$ as set for $\tau=0.9$; its arms are the guard without a check, the gated provable margin and the always-on margin, with paired controls at the default settings. Block T (NV, SiV at their native threshold $0.8$; no offset and $0.044$; $p=0.1$ and $0.2$; the same three arms) is descriptive. S34 (4{,}928 runs, seeds 19341--19344) runs the guard without a check, the current-look bound behind the hybrid check ($p=0.1$ and $0.2$) and the always-on margin ($p=0.1$) on ion2020 and ionproj, with no offset, a constant offset $0.044$ in the true state, and $0.044$ with an alarm forced at the 8th test, on square grids of side $5$ (40 links; a reference with fresh seeds), $10$ (180 links) and $30$ (1{,}740 links). The load variants are the same total request rate as S30--S32 on every grid (204/340, 1{,}004/1{,}860 and 10{,}204/19{,}940 users for ion2020/ionproj on the three grids, so requests per user fall as the grid grows); matched exposure on $10\times10$ and $30\times30$, with demand restricted to a random subset of 204 or 340 users drawn once per network; and three times the request rate on $5\times5$ and $30\times30$. This gives 1{,}408, 1{,}408 and 2{,}112 runs on the three grids. Per-user measures of S34 use served users.

\subsection{Other topologies (S3D-1 to S3D-3)}\label{app:s3d}
These studies use a fork of the S34 simulator that generalises the square grid to $d$-dimensional lattices and to a square grid with both diagonals (hop distance $\max(|\Delta x|,|\Delta y|)$). The upstream modules are kept verbatim; before each confirmatory study, a gate job on the cluster reproduced recorded runs exactly in every column (S3D-1: 4 S34 runs; S3D-2: 6 S34 runs, 2 of them with users enumerated lazily; S3D-3: 3 S34 runs and 1 S3D-2 run). Models, arms and the check are those of S34, requests are served one at a time, and each cell has 32 networks (in S3D-2 and S3D-3, 8 in the cells that only check that an unchecked offset harms users). Seeds are fresh: 19351--19354, 19361--19364 and 19371--19374.

\emph{S3D-1} (1{,}344 runs): ion2020 and ionproj with near-neighbour demand (1 to $\Lmax$ hops) at the same total rate on a $5^3$ cube (125 nodes, 300 links), an $11\times11$ grid (matched nodes) and a $13\times13$ grid (matched links). \emph{S3D-2} (1{,}224 runs) and \emph{S3D-3} (408 runs): projected links (fut97--fut99, Table~\ref{tab:hw}: $R=1000$\,s$^{-1}$, $T_2=1$\,s, gate error $0.003$, readout error $0.001$) with link fidelity $0.97$, $0.98$ or $0.99$; requests are between uniformly random nodes, and those within a service radius of 5, 6 or 8 hops (the analytic reach at $\tau=0.9$ plus 2) are served, at 3 per second on every lattice. Lattices: $10^3$ (2{,}700 links), $32\times32$ (1{,}984), $37\times37$ (2{,}664) and, in S3D-3, $32\times32$ with diagonals (3{,}906). Safety arms, at $p=0.1$: the provable margin behind the hybrid check after an alarm (S3D-1: at offset $0.044$, with natural and forced alarms; S3D-2 and S3D-3: forced) and the always-on margin (no offset and $0.044$). Pass bars per cell: one-sided 95\% Clopper--Pearson bounds of at most $0.10$ on harmed networks and $0.01$ on bad pairs (Table~\ref{tab:s3d}). The bad-pair bound, as pre-registered, pools all guarded deliveries of a cell and treats them as independent. Deliveries of one network share its links, calibration data and history, so, like the run bound of Table~\ref{tab:cluster}, it is reported for the record and is not a calibrated bound on the bad-pair probability. The inference rests on the network level: in every cell no network of 32 had a harmed user, a one-sided 95\% bound of $8.9\%$ on the share of harmed networks, assuming the networks are independent replicates (their seeds are derived from the study, lattice, link level and index). The count of 0 bad pairs in $8.8\times10^6$ deliveries is descriptive.

\begin{table*}[!htbp]
\caption{\label{tab:s3d}Safety on other topologies. Cells: lattice $\times$ hardware or link level $\times$ arm (after an alarm; always-on), 32 networks each; a network is harmed if any of its runs in the cell gave a user a bad pair (after the alarm, or over the whole run). Every cell has bound $8.9\%$ on harmed networks.}
\begin{ruledtabular}
\begin{tabular}{p{1.4cm}p{4.6cm}p{3.6cm}p{1.6cm}p{1.2cm}p{3.6cm}}
Study & Lattices & Links & $\Umax$ & Cells & Bad / guarded deliveries\\
\hline
S3D-1 & $5^3$; $11\times11$; $13\times13$ & ion2020, ionproj & 4--12 & 12 & 0\,/\,3{,}493{,}387\\
S3D-2 & $10^3$; $32\times32$; $37\times37$ & projected, $0.97$--$0.99$ & 17--104 & 18 & 0\,/\,3{,}783{,}809\\
S3D-3 & $32\times32$ with diagonals & projected, $0.97$--$0.99$ & 37--96 & 6 & 0\,/\,1{,}516{,}258\\
\end{tabular}
\end{ruledtabular}
\end{table*}

\paragraph*{Other pre-registered hypotheses.} With no check, offset $0.044$ harmed $64$--$77\%$ of served users on every lattice, and no pre-registered falsification test rejected the certificate's nominal violation rate of $5\%$ (largest one-sided lower bound on the violating-run rate $0.014$). This does not mean that no violations occurred: a certificate at level $5\%$ may be violated in some runs, and it was, for example, in 4 of 96 runs, one per network, on the $32\times32$ grid at $p=0.1$ (one-sided 95\% upper bound $9.3\%$) and in 3 of 64 on the $5^3$ cube at $p=0.2$ (upper bound $11.7\%$, the largest of any cell). These data are consistent with the nominal rate but do not show that the rate is below it. Detection carried over for ionproj (alarm rate $0.97$ on the cube against $1.00$ on $11\times11$) but not for ion2020 ($0.69$ against $0.97$; S3D-1's detection prediction failed). In S3D-1, delivered fidelity on the cube differed from that on the grids by at most $0.003$. The remaining hypotheses compare lattices. With near-neighbour demand, the cube served more of the hardest requests than both grids in 1 of 4 comparisons (S3D-1's primary benefit hypothesis failed). With uniform demand, it served $2.3$--$3.5$ times the share of the $32\times32$ grid (passed), and $1.16$--$1.71$ times that of the grid with diagonals (passed); under the always-on margin, its lead over the grid with diagonals was not significant at link fidelity $0.97$ (that secondary hypothesis failed). These results are reported with the studies~\cite{Repo3D2026}.

\subsection{Conservative pooled bounds}\label{app:pooled}
Let $X_1,\dots,X_n$ be independent indicators with $\Pr(X_j=1)=p_j$ (network $j$ has a harmed user in some pooled run), $S=\sum_jX_j$ and $\bar p=n^{-1}\sum_jp_j$. For $k\leq n-2$ let $u(k)$ be the one-sided Clopper--Pearson upper bound at level $\alpha$, defined by $\Pr(\mathrm{Bin}(n,u(k))\leq k)=\alpha$. At the top end we use $u(n-1)=1-\alpha/n$ instead of the Clopper--Pearson value $(1-\alpha)^{1/n}$, which it exceeds, and $u(n)=1$; this changes no bound reported here, all of which have $k\leq2$.

\emph{Claim.} If $n\,u(k)\geq k+1$ for every $k\leq n-2$, then $\Pr(\bar p>u(S))\leq\alpha$ for every $(p_1,\dots,p_n)$.

\emph{Proof.} The function $u$ is increasing, so $\{\bar p>u(S)\}=\{S\leq k^\ast\}$ with $k^\ast=\max\{k:u(k)<\bar p\}$; the event is empty if no such $k$ exists. Two cases cover all $\bar p$.
(i) If $\bar p\leq1-\alpha/n=u(n-1)$, then $k^\ast\leq n-2$, so $k^\ast+1\leq n\,u(k^\ast)<n\bar p$, that is, $k^\ast\leq n\bar p-1$. In this range Hoeffding's theorem~\cite{Hoeffding1956} (Theorem~4) gives $\Pr(S\leq k^\ast)\leq\Pr(\mathrm{Bin}(n,\bar p)\leq k^\ast)$. The binomial distribution function is decreasing in its success probability and $u(k^\ast)<\bar p$, so this is at most $\Pr(\mathrm{Bin}(n,u(k^\ast))\leq k^\ast)=\alpha$.
(ii) If $\bar p>1-\alpha/n$, then $k^\ast=n-1$ and, by the union bound, $\Pr(S\leq n-1)=\Pr(\exists j:X_j=0)\leq\sum_j(1-p_j)=n(1-\bar p)<\alpha$. $\square$

The endpoint matters. With the Clopper--Pearson value at $k=n-1$, heterogeneous probabilities with $\bar p$ just above $(1-\alpha)^{1/n}$ give non-coverage slightly above $\alpha$ (for $n=64$ and two strata, we find up to $0.051$).

\emph{Where the condition holds.} At $\alpha=0.05$ we checked numerically that $n\,u(k)-(k+1)\geq0.64$ for every $k\leq n-2$ and each $n\in\{32,64,96,128\}$, the network counts of Table~\ref{tab:cluster}. With the endpoint above, the claim then holds for every value of the unknown $\bar p$. The pooled bounds of Table~\ref{tab:cluster} are therefore conservative bounds on the equal-weight average $\bar p$ over strata, not exact ones. Within a stratum, where the $p_j$ are equal, the Clopper--Pearson bound is exact. Both statements need the networks to be independent, which holds because each network's seed is derived from its study, hardware type, grid and index (Sec.~\ref{sec:methods-sim}).

\section{Real-data analyses}\label{app:real}
All eight analyses were pre-registered (hypotheses, bars, frozen scripts) before the reported statistic was computed; for the Innsbruck certificate the counts had been seen in Stage 1, so only the method was fixed in advance. No Delft data value of a stage was read before that stage's pre-registration had been committed and pushed (local commit and extraction times in Table~\ref{tab:keys}; the push order is in the terminal record in the ledger). The S35B pre-registration was written after the S35 results were known. Hashes are in the ledger~\cite{Repo2026}.

\paragraph*{ICFO 2021}~\cite{LagoRivera2021}.
\begin{itemize}
\item Time-tagged heralds from 11 runs at 2--25\,$\mu$s storage were decoded with one pre-declared fix for a time-tag glitch.
\item Over-dispersion of the echo probability per herald across blocks: dispersion ratio 0.73--5.6, significant in 7 of 11 runs at p-value $<0.05$ uncorrected and in 5 of 11 (2, 7, 7\#2, 12 and 25\,$\mu$s) after a Bonferroni correction over runs.
\item A Beta-mixture CS never excluded the full-run estimate in 10 of 11 runs.
\item The pre-registered primary visibility estimator \emph{failed} (unphysical $V=1.014$), because phase settings were not recorded. The authors' qualitative claims and two-photon suppression ($0.033$ vs $0.036(8)$) were reproduced.
\end{itemize}

\paragraph*{USTC 2026}~\cite{USTC2026}.
\begin{itemize}
\item Model: a pure-dephasing memory model built from single-ion coherence curves.
\item The model predicts the remote-entanglement fidelity change within error at every storage time (pre-registered pass), but is optimistic by $0.020$--$0.044$.
\item 200 sets of $\sim\!2{,}025$ events: the QBER dispersion ratio is $3.57$; the anytime CS excluded the full-run rate for $3.5\%$ of the run.
\item Caveat: unequal set sizes, which cannot be checked from the published data, would inflate the dispersion.
\end{itemize}

\paragraph*{Innsbruck}~\cite{Krutyanskiy2023}.
\begin{itemize}
\item Stage 1 reproduced the ion--photon fidelity: $0.920\pm0.004$ (best maximally entangled state) against $0.924$ in the authors' data repository ($0.929$ in the publication). Maximum-likelihood tomography of all nine settings gives a $\Phi^+$ fidelity of $0.913$.
\item The certificate (Stage 3) uses the ion--photon tomography data, the six diagonal scans, with detector-swapped pairs pooled.
\item Error counts: $XX$ $170/2355$, $YY$ $144/2406$, $ZZ$ $97/2411$.
\item Bounds: one-sided Chernoff--KL bounds at $\alpha/24$ per basis, union over 2 detector mappings $\times$ 4 Bell targets.
\item Result: $\hat\Fid=0.9139$, $\Fid_L=0.8867$.
\item Assumption H: the herald-conditioned state is the same across the six scans (stationarity), and detection is fair (no loss depending on the setting).
\item This analysis is not blind: the counts had been seen in Stage 1. Only the method was fixed in advance.
\end{itemize}

\paragraph*{Delft 2024}~\cite{Stolk2024}.
\begin{itemize}
\item Data: processed outcome counts per basis (raw counts $N$, readout-corrected probabilities and correlators, the authors' model correlators and a sweep over 16 herald windows of 5--20\,ns), CC BY 4.0. There are no time-ordered per-event records, so only fixed-sample analyses are possible; the basis is not randomised per pair, and events are assumed i.i.d.\ within each basis.
\item Before pre-registration we read only the file structure and the source code of the authors' notebooks, which identify the windows of their figures (10\,ns for Fig.~3, 15\,ns for Fig.~4) and mark the plotted fidelities (about $0.57$ and $0.533$); before S35B we had also seen the S35 results and the published Fig.~4 fidelity $0.534(15)$. Stage 1 (S35) analysed Fig.~3, which is post-selected entanglement (the qubits are measured right after spin--photon entanglement); its pre-registration mis-described Fig.~4, which was then analysed in Stage 2 (S35B) under its own pre-registration.
\item Gates (stop rules): outcome and basis labels, the parity convention (our correlators equal the authors' to $10^{-5}$) and the target states (our fidelities reproduce the authors' values to $10^{-5}$) passed in both stages.
\item Certificate (D1): per-basis error counts from the raw $N$ against the target Bell state, one-sided Clopper--Pearson bounds at $\alpha/3$ per basis, $\Fid_L=1-\sum_B e_B^{\rm ub}/2$. S35 (Fig.~3, $\Psi^\pm$ pooled; $1{,}436$/$1{,}535$/$1{,}415$ events in $X$/$Y$/$Z$): $\hat\Fid=0.505$, $\Fid_L=0.464$. S35B (Fig.~4, delivered $\Psi^-$; $450$/$452$/$449$): $\hat\Fid=0.474$, $\Fid_L=0.399$. Readout-corrected (authors' correction; normal bound): $0.558$ and $0.510$.
\item Model offset (D2): $\Delta=\Fid_{\rm model}-\Fid_{\rm data}$, with data and model uncertainties combined in quadrature; pass if $|\Delta|\leq2u+0.02$ and $\Delta-1.645u\leq0$. S35: $\Delta=-0.004\pm0.008$ (90\% interval $[-0.018,0.009]$, each end a one-sided 95\% bound; power $0.97$ at $\Delta=0.03$). The sweep's model file gives a model fidelity $0.008$ higher at 10\,ns ($\Delta=+0.004$, also within the bar). S35B, with the count-rate-weighted model of the authors' Fig.~4: $\Delta=-0.007\pm0.015$ (90\% interval $[-0.032,0.018]$; power $0.63$).
\item Descriptive sweeps (nested windows, no test): for the post-selected states $\Delta$ (sweep model file) increases, not monotonically, from $-0.004$ at 5\,ns to $+0.016$ at 20\,ns; for the delivered states $\Delta<0$ at every window ($-0.042$ to $-0.0035$).
\item Calibration on synthetic data at $200$--$3{,}000$ events per basis and state (the delivered-state data have about $225$ per state): the certificate's false-certification rate at $\Fid=0.5$ was at most $0.0008$; the offset test's false-failure rate with an exact model was $0.046$--$0.050$.
\end{itemize}
\paragraph*{Oxford 2022}~\cite{Nadlinger2022}.
\begin{itemize}
\item Data: the shot-by-shot record of the 1{,}920{,}000-round characterisation run of Ref.~\cite{Nadlinger2022} (its Fig.~3 and Fig.~S4): for each heralded pair, in time order, the herald time, both measurement settings and both outcomes. Bob chose the round type with a quantum random number generator after each herald (test probability $\gamma\approx1/2$; CHSH settings uniform; key rounds measure both qubits in the same basis). The two nodes are about 2\,m apart~\cite{Nadlinger2022}. According to the authors, every herald produced a row, including during ion loss and recalibration. The record was provided by its authors and is not redistributed.
\item Before pre-registration we read only the file's header comment, column names and row count, and the published values for this run ($S=2.677(6)$, QBER $1.44(2)\%$). The pre-registration was committed at 01:32:45 UTC (local commit time); the local repository was in sync with the remote at 01:35:22 UTC, before the locked analysis started at 01:35:26 UTC (2\,min\,41\,s after the commit). A rerun of the locked scripts on a second machine gave identical output.
\item Gates: the design matched (test fraction $0.5080$; each test setting $0.2495$--$0.2506$ of test rounds; key rounds all in Alice's setting 0); exactly one of the eight CHSH sign conventions reproduced the published value ($S=2.6768$; the next largest $|S|$ was $0.048$), which fixed the convention, and the key-round QBER was $1.439\%$.
\item Certificate: with $L_i$ the loss indicator of test round $i$ and $q_i$ its conditional probability, a pair measured at the nominal settings (two orthogonal axes for Alice, Bob's at $\pm\pi/4$ between them) has $\Fid_i\geq(\langle a_0b_+\rangle+\langle a_1b_-\rangle)/2=\sqrt2(1-2q_i)$, with $b_\pm=(b_0\pm b_1)/\sqrt2$, by positivity of its Bell-diagonal weights, for the maximally entangled state of the measurement frame (for each herald pattern, since Bob's phase is calibrated per pattern). With $T_i$ the test indicator and $p$ the test probability, $d_i=(1-p)T_iL_i-p(1-T_i)q_i$ has conditional mean zero and range 1, so Hoeffding supermartingales on a grid of 20 values of $\lambda$ (each at $\alpha/20$, $\alpha=0.05$) give, for all $t$ simultaneously, $pQ_K(t)\leq(1-p)\sum_{i\leq t}T_iL_i+\beta_t$ for the kept pairs' summed loss probability $Q_K$, with $\beta_t=\min_g[\ln(20/\alpha)/\lambda_g+\lambda_gt/8]$, $t$ counting all rounds and $\lambda_g$ on a logarithmic grid tuned to $t$ between $10^3$ and $4\times10^6$; then $\Fid_L(t)=\sqrt2(1-2Q_K^{\rm ub}(t)/K_t)$. Result: $\Fid_L=0.917$ over $944{,}576$ kept pairs; $\Fid_L$ first exceeded $0.9$ after $217{,}893$ kept pairs and stayed above it after $219{,}185$ (descriptive). With a union over the eight sign conventions ($\alpha/8$ each): $0.914$.
\item Test probability: the realised test fraction, $0.50803\pm0.00036$, is consistent with $130/256$ and not with $1/2$; the authors confirmed (private communication) that the round type was drawn from 8 random bits with test probability $130/256$, ten times the $13/256$ of their key run. Using $p=1/2$ below the true value increases each $d_i$ by $(p'-p)[T_iL_i+(1-T_i)q_i]\geq0$, so the pre-registered bound remains valid and is conservative. With $p=130/256$ (exploratory, with the confirmed test probability): $\Fid_L=0.932$, above $0.9$ at all times after $68{,}129$ kept pairs.
\item Consistency: at every time with at least 1{,}000 kept pairs, $\Fid_L(t)$ stayed below $1-e_Z/2$, a loose upper bound implied by the kept pairs' own key-basis error fraction $e_Z$ (final value $0.993$; the tighter $1-e_Z$ is about $0.986$).
\item Without trusted measurements, the self-testing bound $\Fid_{\rm ext}\geq\tfrac12+\tfrac12(S-\beta^*)/(2\sqrt2-\beta^*)$, $\beta^*=(16+14\sqrt2)/17$~\cite{Kaniewski2016}, is affine in $S$ and holds for every state, so it applies to the same anytime bound on the kept pairs' average $S$; it gives $0.837$ ($0.867$ with $p=130/256$). It still assumes uniform, independent settings and complete recording.
\item Fluctuations: over 96 blocks of 20{,}000 rounds (the header lists 96 acquisition chunks totalling 1{,}920{,}000 rounds; according to the authors the chunks are a software convenience, with calibrations about hourly), the CHSH loss rate (mean $0.165$; chunks $0.151$--$0.178$) gave $\chi^2=143.1$ (95 degrees of freedom, $p=0.001$), a dispersion ratio of $1.51$ and an excess relative fluctuation of $1.6\%$; the lag-1 autocorrelation of chunk rates was $-0.17$ (one-sided permutation $p=0.95$), so the predicted persistence was not found. The key-round QBER gave $\chi^2=117.6$ ($p=0.058$).
\item Calibration on synthetic data with the locked code: no violation of the anytime bound in any of 400 runs per scenario at $2\times10^5$ rounds and 20 at full size (stationary, drifting, stepped and slowly wandering sources), nor in 200 runs of $2\times10^4$ rounds with an adaptive adversary; the margin to the true bound at full size was $0.014$. The over-dispersion test had a false-positive rate of $0.051$ on a stationary link; a stationary end-to-end test file produced one such false positive ($p=0.038$), so the test's result is reported with its effect size.
\end{itemize}

\paragraph*{Harvard 2024 (S37)}~\cite{Knaut2024}.
\begin{itemize}
\item Data: per-shot records of the electron--electron $\mu$ sweep (5 runs), the nuclear-spin storage series (10--1000\,ms) and the telecom-fibre series (0--40\,km), with the authors' analysis notebooks; shared privately and not redistributed. The deployed-loop data were not analysed (no configuration in the shared notebook). Before pre-registration we read only the folder structure, the notebook source and the published values.
\item Pre-registration commit 6e37b80 (pushed 18:35:41 UTC); Stage 1 ran 61\,s later and Stage 2 at 18:41:45 UTC. The script was tested beforehand only on synthetic data in the real file layout.
\item Gates: our per-herald records reproduced the authors' Bell-count tables exactly in every run (G0); $\Fid(\Phi^-)=0.862\pm0.029$ and $\Fid(\Phi^+)=0.740\pm0.035$ against $0.86(3)$, $0.74(3)$ (G1); nuclear-spin 10\,ms, error-detected $0.774\pm0.050$ and raw $0.643\pm0.054$ against $0.77(5)$, $0.64(5)$ (G2). The authors' code counts a herald when a detector's click count is odd; we kept this rule.
\item H1 (memory model, error-detected $\Phi^-$, anchored at 250\,ms, $T_2=2.11$ and $2.10$\,s): offsets $-0.055$, $-0.078$, $-0.090$ at 500, 750, 1000\,ms; pooled $-0.077$ ($[-0.207,0.050]$; 20{,}000 bootstrap resamples); predicted pass.
\item H2 (over-dispersion; 10 blocks; Bonferroni over 13 valid tests): herald rate $p$ down to $4.5\times10^{-7}$ (min $p\times13=5.8\times10^{-6}$); $ZZ$ error rate $p=0.55$--$0.78$ in the three testable runs, 8 tests NA (too few errors); predicted pass.
\item H3/H4 (post-hoc test choice, $p=0.1$ and $0.2$, seeds 37001--38000): violations 0/1000 in 7 cells and 1/1000 in one; median certificates on the published run $0.293$/$0.527$ ($\Phi^-$) and $0.175$/$0.385$ ($\Phi^+$); pooled over the five runs $0.44$--$0.51$.
\end{itemize}

\paragraph*{IonQ 2026 (S38)}~\cite{Hartung2026}.
\begin{itemize}
\item Data: the public repository at commit 8a08f695fca8 (Apache-2.0): runs 64563 (Fig.~1 of Ref.~\cite{Hartung2026}) and the six Fig.~2 runs (XY8-1 to XY8-16), each 9 settings $\times$ 1{,}000 heralded pairs. Before pre-registration we read the README, the run list, the analysis code and the HDF5 structure only.
\item Pre-registration commit a4fe90b (pushed 19:21:34 UTC); Stage 1 ran 16\,s later and Stage 2 at 19:22:31 UTC.
\item Gates: our parser reproduced the authors' outcome counts and Fig.~2 fidelity formula in all 7 runs (G0); the authors' maximum-likelihood code gave $\Fid=0.8794\pm0.0073$ against $0.879(7)$, and the rate $1.032$\,kHz against $1.032(11)$ (G1).
\item H2: 28 tests (success per attempt; $XX$, $YY$, $ZZ$ error rates; 10 blocks each), min $p\times28=1$; prediction (no drift) held.
\item H3: 0 violations in all 21 cells (7 runs $\times$ $p=0.1,0.2,0.5$; 1{,}000 draws each, seeds 38001--39000).
\item H5: run 64563 at $p=0.2$, median certificate $0.818$ against the predicted $0.817$ (band $0.788$--$0.842$, simulated from $\Fid=0.879$ at $N=3{,}000$, seed 3800); $p=0.1$: $0.790$ (predicted $0.789$); $p=0.5$: $0.836$ (predicted $0.834$).
\item Descriptive (median certificate at $p=0.2$, linear fidelity, rate): XY8-1 $0.808$, $0.873$, $0.72$\,kHz; XY8-2 $0.806$, $0.871$, $1.04$\,kHz; XY8-3 $0.789$, $0.857$, $1.16$\,kHz; XY8-4 $0.784$, $0.853$, $1.21$\,kHz; XY8-8 $0.734$, $0.811$, $1.31$\,kHz; XY8-16 $0.654$, $0.742$, $1.35$\,kHz.
\end{itemize}

\paragraph*{Oxford Sr--Sr 2025 (S39)}~\cite{Main2025b}.
\begin{itemize}
\item Data: one characterisation run (RID 565472) of $200{,}000$ heralded Sr$^+$--Sr$^+$ events between the two Oxford nodes, provided by G.~Araneda; per event, the setting drawn at each node independently ($+X$, $+Y$, $-X$, $-Y$ each with probability $1/6$, $Z$ with $1/3$; NumPy pseudorandom), both outcomes, the herald pattern (four two-detector coincidences, brought to a common Bell-state frame by herald-dependent corrections), timestamps and attempt counts. According to the authors every recorded event is valid; the abort counter records re-cooling before a success. Before pre-registration we read the authors' data guide (which states the stored fidelities) and the file structure only.
\item Pre-registration commit (pushed 15:33:00 UTC); run 1, the confirmatory run, 24\,s later. An independent AI code review on synthetic data, without access to run 1's output, then corrected the sign of the reported phase (descriptive) and added input checks; the amendment was pushed at 15:53:04 UTC and rerun at 15:53:07 UTC. All gate and hypothesis verdicts are identical in both runs.
\item Gates: setting frequencies as documented at both nodes (equal codes at both nodes $0.2224$, independence predicts $0.2222$); exactly one convention reproduces the stored values: $F(\Psi^+)=0.9363$ against $0.9362$, per-pattern phase-optimised $0.958$--$0.967$ against $0.959$--$0.969$ (next candidate $0.50$).
\item H2 (drift in 10 blocks; 8 tests): nothing detected (smallest corrected $p=0.30$). The predicted drift of the success rate per attempt was not found, so H2 is scored as failed; the error rates were stable, as predicted.
\item H3, H4 ($p=0.05,0.1,0.2$; 1{,}000 draws each, seeds 39001--40000): 0 final violations; at 20 checkpoints in time order 1 of 1{,}000 draws exceeded the running reference at $p=0.05$, none at $0.1$ and $0.2$.
\item H5: median certificate $0.9273$ against the predicted $0.9275$ (band $0.9237$--$0.9309$, simulated from the stored $\Fid(\Psi^+)=0.9362$, which was known before the run); $0.9241$ ($0.9242$) at $p=0.1$, $0.9191$ ($0.9191$) at $p=0.05$.
\item Descriptive: correlators $XX$ $+0.891$, $YY$ $+0.888$, $ZZ$ $-0.967$, $XY$ $+0.289$, $YX$ $-0.304$; fidelity to the closest maximally entangled state $0.960$ at $\phi=-0.32$\,rad (leader first), stable over 10 time blocks ($-0.27$ to $-0.38$\,rad); a phase fixed from the first $10\%$ gives $0.961$ on the remaining $90\%$.
\end{itemize}

\paragraph*{Pairs needed (S37-D).} Synthetic computation, not a hypothesis test: each tested pair errs with probability $\tfrac23(1-\Fid)$; 1{,}000 draws per point, seed 3700 (Fig.~\ref{fig:pairs}). Pairs before the median certificate reaches $0.8$ with $20\%$ tested: about $2{,}500$ at $\Fid=0.879$ and $4{,}000$ at $0.862$; never at $0.740$.

\begin{table*}[t]
\caption{\label{tab:keys}Source keys.}
\begin{ruledtabular}
\begin{tabular}{p{2.2cm}p{14.8cm}}
G3 & Guard at scale: $30\times30$ grid, 900 nodes, 1{,}740 links; 256 runs (64 networks $\times$ 2 controllers $\times$ 2 priors); links at $\alpha/\Lmax$; exact model, stationary links; all pre-registered tests pass (runs.csv sha e9cd30683b249213).\\
S25 & Earlier simulation study; latency of the guards on ions.\\
S26 & Model-based layer at measured hardware levels (32 networks per cell); all 7 primary hypotheses passed, 2 secondary failed.\\
S27 & Both layers in a network, four conditions; both primary hypotheses passed.\\
S28 & Model check: detection and the first response (90 epochs, ion and thin hardware, $p=0.1$ and $0.2$; 5,248 runs); one primary hypothesis (post-alarm risk) failed.\\
S29 & Tracking response after an alarm (ion hardware; 4,352 runs), designed after S28's failure; all six primary hypotheses passed.\\
S30 & Provable closed loop in simulation (ion2020, ionproj; offset in the truth; Appendix~\ref{app:s30}); 8{,}384 runs, seeds 19321--19324; pre-registration commit f5f23f2 (19:48:04 UTC, local commit time), first shard 19:50:08 UTC; 5/5 primary and 6/6 secondary hypotheses pass.\\
S31 & Test-or-keep on IBM hardware, one job (Appendix~\ref{app:hw}); 2 of 3 binary hypotheses passed, detection failed.\\
S31b & Hardware rerun with the hybrid check (Proposition~\ref{prop:D}); seed 31032; ibm\_fez job db5ubf04qg6s73c40k20, 3\,s QPU; pre-registration commit 3520391 (19:37:59 UTC, local commit time; pushed 19:38:07), IBM job created 19:38:36 and finished 19:38:49 UTC; W1, W2, W3 (primary) and W5 (secondary) pass.\\
S3D-1 & Topology check: $5^3$ cube vs $11\times11$, $13\times13$ grids, ion presets (Appendix~\ref{app:s3d}); 1{,}344 runs; pre-registration commit fc3660d of Ref.~\cite{Repo3D2026}; H1 (safety) passed, H2 (benefit) failed, detection prediction failed for ion2020.\\
S3D-2 & Topology check at reach: $10^3$ cube vs $32\times32$, $37\times37$ grids, projected links; 1{,}224 runs; pre-registration commit 7e3dc97; all primary and secondary hypotheses passed.\\
S3D-3 & Topology check: $32\times32$ grid with diagonals; 408 runs; pre-registration commit 1fe2dba; both primary hypotheses passed, one secondary failed.\\
S32 & Current-look offset bound and robustness (threshold, calibration data, thin hardware; Appendix~\ref{app:s30}); 3{,}328 runs, seeds 19331--19334; pre-registration commit f647233 (22:45:35 UTC, local commit time), first shard 22:47:40 UTC; 6/6 primary and 4/4 secondary hypotheses pass (the certificate hypothesis at both $p$).\\
S33 & Fixed-$K$ versus anytime certificate (computation, not pre-registered; \texttt{s33/fixedk.py}): kept pairs to certify $0.9$ with a planned single bet versus Proposition~\ref{prop:C1}.\\
S34 & Closed loop on $5\times5$, $10\times10$ and $30\times30$ grids (same total rate, matched exposure, three times the rate; Appendix~\ref{app:s30}); 4{,}928 runs, seeds 19341--19344; pre-registration commit 1f5636c (00:52:51 UTC, local commit time), first shard 00:54:40 UTC; 3/3 primary and 5/5 secondary hypotheses pass (the certificate hypothesis at both $p$).\\
S36 & Oxford re-analysis (Appendix~\ref{app:real}): pre-registration commit d4d4a9d (01:32:45 UTC, local commit time; pushed), run 2\,min\,41\,s later, reproduced identically on a second machine; both primary hypotheses (certificate above $0.9$; over-dispersion) passed, the consistency check passed, and the predicted persistence was not found.\\
S35, S35B & Delft re-analyses (Appendix~\ref{app:real}): post-selected states (S35; pre-registration commit f9db617, 04:51:17 UTC, local commit time; extraction 14\,s later) and delivered states (S35B; commit 726549d, 05:06:41 UTC, local commit time; extraction 12\,s later); in each, the raw-count certificate failed (predicted) and the model-offset test passed (S35B underpowered).\\
S37 & Harvard SiV--SiV re-analysis (Appendix~\ref{app:real}): pre-registration commit 6e37b80 (pushed 18:35:41 UTC), run 61\,s later; all gates and all four predictions held.\\ %
S38 & IonQ ion--SiV re-analysis (Appendix~\ref{app:real}): pre-registration commit a4fe90b (pushed 19:21:34 UTC), run 16\,s later; all gates and all three predictions held.\\
S39 & Oxford Sr--Sr re-analysis (Appendix~\ref{app:real}): pre-registration pushed 15:33:00 UTC, confirmatory run 24\,s later; gates passed; H3, H4, H5 passed, H2 failed (no drift detected).\\ %
S41a & Bell pairs on IonQ Forte~1 with settings fixed in advance (Sec.~\ref{sec:hw}, Appendix~\ref{app:hw}); 30 jobs; pre-registration commit 1012b04 (pushed 2026-11-11 05:58:28 UTC), first job 3\,min later; job 13 re-run after an IonQ server error (lost job recorded); all gates and all five primary hypotheses (H3--H7) passed.\\
S41b & GHZ-3 and GHZ-4 states on IonQ Forte~1 (Sec.~\ref{sec:hw}); 20 jobs; pre-registration commit add3dd6 (pushed 2026-11-14 05:32:25 UTC), first job 2\,min later; all gates and all four primary hypotheses (H3b--H6b) passed.\\
S37-D & Pairs-needed design curve (synthetic computation, not pre-registered).\\
ICFO, USTC, Innsbruck & Real-data re-analyses (Appendix~\ref{app:real}).\\
\end{tabular}
\end{ruledtabular}
\end{table*}

\section{Hardware experiment}\label{app:hw}
\paragraph*{Design.} Each delivered pair is one shot of a two-qubit circuit that prepares $\Phi^+$ or, as an injected bad pair, $\Psi^-$. A schedule drawn in advance from a fixed seed fixes, for every shot, the state, the test/keep decision and the test basis. Calibration (phase 0) tests all 600 of its pairs in a random $X$/$Y$/$Z$ basis; the declared fidelity $\Fid_{\rm decl}$ is a one-sided 95\% Clopper--Pearson lower bound via Lemma~\ref{lem:F}. Phases A and B use $p=0.2$; phase B replaces pairs by $\Psi^-$ with probability $20\%$. Kept pairs are measured in the Bell basis (CX, H, measurement; outcome 00 means $\Phi^+$). The certificate (Proposition~\ref{prop:C1}, the eight-point $\nu$ grid of Appendix~\ref{app:cert}, $p=0.2$, $\alpha=0.05$, declared parity-flip bound $0.05$) is updated after every kept pair in phases A and B. The model check (Proposition~\ref{prop:MC}, $\beta=0.05$) uses the null bound $q=\tfrac23(1-\Fid_{\rm decl})$ on the tests of phases A and B; the declared fidelity is estimated from tests that include readout errors. The certificate and check code were unchanged from the simulations. The design was fixed with simulations before any hardware data, and a local noisy simulation of a different device was run as a dry run before the hardware job.

\paragraph*{Pre-registered outcomes and results of the first run (S31)} (single run, binary).
\begin{itemize}
\item W1, the certificate stays valid (never above the Bell-measured pass fraction of the kept pairs, at any kept pair): \emph{passed}. Final certificate $0.710$ against a pass fraction of $0.852$; the largest certificate-minus-truth difference was $-0.133$.
\item W2, no false alarm in phase A: \emph{passed} (206 tests).
\item W3, an alarm in phase B: \emph{failed}. Peak log-wealth $2.71$ against a threshold of $3.49$ (457 tests).
\item W4, descriptive: $\hat\Fid=0.8925$ from the 600 calibration tests (43 errors), $\Fid_{\rm decl}=0.8628$, $q=0.0914$; Bell pass fraction of kept pairs $0.967$ in phase A and $0.803$ in phase B; test error rates $7.2\%$ (phase 0), $5.8\%$ (A) and $15.5\%$ (B). The dry run passed W1 and alarmed in phase B (peak log-wealth $4.98$).
\end{itemize}
No re-analysis, re-tuning or re-run changes the W3 verdict.

\paragraph*{Rerun (S31b).} Same design and code, with a fresh schedule seed (31032) and the hybrid check of Proposition~\ref{prop:D} as the primary check; the fixed-bet mixture ran on the same outcomes as a paired comparison. Both checks used the null bound $q=\tfrac23(1-\Fid_{\rm decl})$, without the $s_{\max}$ term of Proposition~\ref{prop:MC}. After the first hybrid alarm, the offset bound of Definition~\ref{def:rule} ran on the post-alarm tests at level $0.025$, with $\psi_i=1$, no declared maximum ($\bar\theta$ starts at 1) and $y_i=\Fid_{\rm decl}-1+1.5E_i$: the declared model is the constant $\Fid_{\rm M}=\Fid_{\rm decl}$, with no link parameters, so no link confidence sequences enter. Pre-registered outcomes: W1 certificate valid (primary; passed; final $0.6526$ against $0.8403$, largest difference $-0.139$); W2 no hybrid alarm in phase A (primary; passed; 205 tests); W3 hybrid alarm in phase B (primary; passed; test 170 of 472); W5, secondary, $\bar\theta$ at the end at least the Bell-measured offset of the kept pairs after the alarm (passed; $0.312$ against $0.099$, 1{,}118 kept pairs). W5 assumes symmetric readout (Lemma~\ref{lem:onesided}); IBM readout is asymmetric, so W5 is a consistency check. Descriptive (W6): $\Fid_{\rm decl}=0.8851$ (35 errors in 600 tests), $q=0.0766$; test error rates $5.8\%$, $4.9\%$, $19.5\%$; Bell pass fraction $0.972$ (A) and $0.783$ (B); the fixed-bet mixture reached $-3.37$ at the end of phase A and alarmed at test 214. Pre-run power at the first run's rates: hybrid $78.9\%$, mixture $44.7\%$. A local noisy simulation run first as a dry run passed W1--W3 and W5.

\section{Provenance and pre-registration}\label{app:prov}
\paragraph*{Source keys.} Table~\ref{tab:keys} maps the keys of Table~\ref{tab:results} and the figure captions to the pre-registered studies of the repository~\cite{Repo2026}. Every confirmatory simulation and real-data analysis was pre-registered (hypotheses, pass bars, locked analysis script, code and task hashes), and every outcome, including failures, is reported; the scoreboard and audit trail are in the repository.

\begin{table*}[!htbp]
\caption{\label{tab:prereg}Three levels of pre-registration evidence. Push times are GitHub's server-side \texttt{PushEvent} records for the project repository (they record when a commit reached GitHub, not the repository's visibility at that time) (UTC; archived as \texttt{github\_push\_events\_2026-10-16.json}); run or data-read times are from the cluster's job records, the hardware result files or the analysis logs.}
\begin{ruledtabular}
\begin{tabular}{p{4.0cm}p{13.0cm}}
\multicolumn{2}{l}{\emph{Level 1: pre-registration public before the run or data read (server-side push time)}}\\
S26, S27 & commits fc22091, 63d20c6 pushed 11 Oct 01:42:05, 02:38:56; first shards 01:44:11, 02:40:40\\
S28, S29 & 023faed, a779735 pushed 06:24:59, 15:34:51; first shards 06:26:51, 15:39:34\\
S31, S31b (hardware) & b51306a, 3520391 pushed 16:58:44, 19:38:07; IBM jobs created 17:13:57 and 19:38:36 (IBM job records), 15\,min and 29\,s after the pushes\\
S30, S32 & f5f23f2, f647233 pushed 19:48:11, 22:45:44; first shards 19:50:08, 22:47:40\\
S34 & 1f5636c pushed 12 Oct 00:52:59; first shard 00:54:40\\
S35, S35B (Delft) & f9db617, 726549d pushed 13 Oct 04:51:20, 05:06:43; data extraction 04:51:31, 05:06:53\\
S36 (Oxford) & d4d4a9d pushed 14 Oct 01:32:47; analysis run 01:35:26\\
\multicolumn{2}{l}{\emph{Level 1b: pre-registration pushed before the run, to a repository~\cite{Repo3D2026} that was private at the time}}\\
S3D-1, S3D-2, S3D-3 & fc3660d, 7e3dc97, 1fe2dba pushed 18 Oct 05:13:42, 19 Oct 02:27:40 and 04:29:40 (GitHub repository activity log); gate jobs submitted 18 Oct 05:46:16, 19 Oct 02:28:23 and 04:30:32; first confirmatory shards 18 Oct 21:07:07, 19 Oct 02:41:55 and 04:42:12 (cluster job records; archived in \texttt{audit/} of Ref.~\cite{Repo3D2026})\\
\multicolumn{2}{l}{\emph{Level 2: pre-registration committed; no server-side record that it preceded the analysis}}\\
ICFO, USTC, Innsbruck stages 1--2, G3, S25 and earlier studies & committed before analysis according to the local record; the ICFO stage-1 commit was pushed on 10 Oct at 19:38:49, but its analysis time is not independently recorded\\
\multicolumn{2}{l}{\emph{Level 3: specified after partial exposure, or not pre-registered}}\\
Innsbruck certificate & counts seen in stage 1; only the method fixed in advance\\
S25 Block A; Studies v11, v16 & Block A started about 2\,min before its commit; v11 and v16 had partial exposure to confirmatory data\\
S35B; S31b & designed after S35's (different-data) results and after S31's summary rates, respectively; each then pre-registered at level 1\\
S33; all ``exploratory'' analyses & not pre-registered (S33 is a computation); labelled as such where reported\\
\end{tabular}
\end{ruledtabular}
\end{table*}

\paragraph*{Process.} The record covers 44 pre-registered studies and real-data stages up to S29, plus S30, S32, S34, S35, S35B, S36 and S3D-1 to S3D-3; S31 and S31b are reported separately (Appendix~\ref{app:hw}). Each confirmatory study fixed its hypotheses, pass bars, locked analysis script, code and task hashes before running. The pre-registration commits of S26, S27, S28, S29, S30, S32 and S34 precede their first cluster job by 2\,min 15\,s, 1\,min 51\,s, 2\,min 3\,s, 4\,min 51\,s, 2\,min 4\,s, 2\,min 5\,s and 1\,min 49\,s (commit times and the cluster's job records), and those of S35, S35B and S36 precede the first reading of their data by 14\,s, 12\,s and 2\,min 41\,s; the pushes of S3D-1, S3D-2 and S3D-3 precede their first confirmatory shards by 15\,h 53\,min, 14\,min 15\,s and 12\,min 32\,s (and their gate jobs, which only re-run recorded runs, by 32\,min 34\,s, 43\,s and 52\,s); GitHub's server-side push records, archived in the repository because the events interface keeps them for only 90 days, show that each of these commits had been pushed to the project's GitHub repository before the run or data read (Table~\ref{tab:prereg}); the repository was created on 10 October 2026 at 19:28 UTC, its public events feed records no later change of visibility, and Zenodo, which archives only public repositories, archived its first release at 19:43 UTC that day, before any of these studies (records archived as \texttt{github\_repo\_meta\_2026-10-16.json} and \texttt{github\_push\_events\_2026-10-16.json}). Earlier exceptions are recorded in the ledger: S25 Block A started about 2\,min before its commit, and Studies v11 and v16 had partial exposure to confirmatory data. The analyses of S26--S30, S32, S34 and S36 were re-run on a second machine with identical output (S28--S30: identical apart from one header line, which differs only in formatting). In a simulation programme where pilots are cheap and inform the design, pre-registration mainly adds transparency rather than protection against forking paths; S29, for example, was designed after S28's failure on the same simulator.

\paragraph*{Hardware run (S31).} IBM \texttt{ibm\_fez}, job \texttt{db5s7lamb58s738aufbg} (created 2026-10-11 17:13:57 UTC, after the pre-registration push at 16:58:44), one job of 3,900 shots on physical qubits [0, 1], 3\,s of QPU usage. Code \texttt{ibm\_testkeep.py} (hash 3b94867c13068e85); schedule hash b8db2032104754d0 (seed 31031). The pre-registration commit (b51306a) is time-stamped about 15\,min before the result files were written (local commit time).

\paragraph*{IonQ Forte runs (S41a, S41b).} IonQ Forte~1 via Qollab, 36 qubits, 1{,}000 shots per job, debiasing off. S41a: jobs submitted 2026-11-11 06:01 UTC to 2026-11-13 21:31 UTC; frozen analysis run once on all 30 jobs (stage-2 output sha256/16 da082837fa26c88a); job 13's first submission (id 01a1d512-67a3-78f0-b691-4d0c48282484) returned an IonQ API error with no counts, was set aside and listed by the analysis, and was re-run as pre-registered. S41b: jobs submitted 2026-11-14 05:34 UTC to 2026-11-15 12:51 UTC, no lost or repeated jobs; stage-2 output sha256/16 de05a4d5d9eaae34. Raw job files, checksums, the verbatim outputs and the frozen code are in the repository (\texttt{external\_checks/ionq\_forte\_hw\_2026}, \texttt{ionq\_forte\_ghz\_2026}). Descriptives: Bell clean $e_X,e_Y,e_Z=1.43,1.49,1.49\%$; damaged pairs $15.5,16.1,1.4\%$ (predicted $15.7,15.7$, unchanged); GHZ-3 $e_Z=2.8\%$, $e_s=2.4$--$3.0\%$; GHZ-4 $e_Z=3.6\%$, $e_s=3.4$--$4.1\%$; dispersion across job positions up to $\chi^2/\mathrm{df}=6.5$; mean pairwise correlation of errors in a shot $+0.017$ to $+0.026$.

\paragraph*{Scoreboard.} Up to S29, 63 primary hypotheses were tested: 55 passed and 8 failed. Seven failures predate the final studies and include two failures of our own real-data estimator (ICFO). The eighth is S28's post-alarm risk (Sec.~\ref{sec:methods-sim}), which S29 then addressed with a new, separately pre-registered design. Of the 36 secondary hypotheses, 29 passed or were confirmed, 6 failed and 1 was unclear. Failed hypotheses remain recorded as failures; later redesigns were pre-registered as new studies. The scoreboard counts only labelled hypotheses. A further 77 registered tests with a pass bar but no stated tier (19 failed their bar, 1 was not met and 2 not confirmed) and 5 gating checks are listed separately. The full list is in the repository. S30, S32 and S34 added 14 primary hypotheses (5, 6 and 3) and 15 secondary hypotheses (6, 4 and 5), all passed. Including them, 77 primary hypotheses were tested (69 passed, 8 failed) and 51 secondary (44 passed or confirmed, 6 failed, 1 unclear). With the hardware hypotheses of S31 (W1, W2 passed; W3 failed) and S31b (W1--W3 primary, passed; W5 secondary, passed; Appendix~\ref{app:hw}), the totals are 83 primary (74 passed, 9 failed) and 52 secondary (45 passed or confirmed, 6 failed, 1 unclear). The Delft re-analyses S35 and S35B added 4 primary hypotheses (both raw-count certificates failed, as predicted; both model-offset tests passed, one underpowered) and 2 secondary (both passed), for totals of 87 primary (76 passed, 11 failed) and 54 secondary (47 passed or confirmed, 6 failed, 1 unclear). S36 (Oxford) added 2 primary hypotheses (both passed) and 2 secondary (one passed; the persistence prediction failed), for totals of 89 primary (78 passed, 11 failed) and 56 secondary (48 passed or confirmed, 7 failed, 1 unclear). S3D-1 to S3D-3 added 6 primary hypotheses (5 passed; S3D-1's benefit hypothesis failed) and 12 secondary hypotheses and predictions (10 passed; S3D-1's detection prediction and S3D-3's comparison under the always-on margin failed), for totals of 95 primary (83 passed, 12 failed) and 68 secondary (58 passed or confirmed, 9 failed, 1 unclear). The re-analyses S37 (Harvard; 3 primary, all passed), S38 (IonQ; 3, all passed) and S39 (Oxford Sr--Sr; 4, 3 passed) bring the primary total to 105 (92 passed, 13 failed); their secondary results are listed in the repository. The IonQ Forte runs S41a (5 primary, all passed) and S41b (4, all passed) bring it to 114 (101 passed, 13 failed). The gates of S35, S35B and S36 are stop rules, not hypotheses, and are not counted.

\paragraph*{Corrections and review.} Independent AI agents reviewed the proofs, and the code of S26--S30, S32 and S34 before their runs; the analyses of S35, S35B and S36 were tested end to end on synthetic files before their runs, and the manuscript text on them was fact-checked by independent agents. Their reports, the errors they found and the corrections are in the repository. %
Corrected statements and reviewer checks are recorded in the repository's revision history.